\documentclass{acmtrans2m}

\newcommand{\commentout}[1]{}

\ifx\pdfoutput\undefined
  \usepackage{graphicx}
  \DeclareGraphicsExtensions{.eps}
\else
 \usepackage[pdftex]{graphicx}
 \DeclareGraphicsExtensions{.pdf,.eps,.ps}
\fi

\newtheorem{theorem}{Theorem}[section]
\newtheorem{lemma}[theorem]{Lemma}
\newtheorem{corollary}[theorem]{Corollary}
\newtheorem{proposition}[theorem]{Proposition}
\newdef{definition}[theorem]{{\sc Definition}}

\newdef{example}[theorem]{{\sc Example}}
\newtheorem{remark}[theorem]{Remark}
\newcommand{\thm}{\begin{theorem}}
\newcommand{\lem}{\begin{lemma}}
\newcommand{\pro}{\begin{proposition}}
\newcommand{\dfn}{\begin{definition}}
\newcommand{\rem}{\begin{remark}}
\newcommand{\xam}{\begin{example}}
\newcommand{\cor}{\begin{corollary}}
\newcommand{\prf}{\begin{proof}}
\newcommand{\ethm}{\end{theorem}}
\newcommand{\elem}{\end{lemma}}
\newcommand{\epro}{\end{proposition}}
\newcommand{\edfn}{\bbox\end{definition}}
\newcommand{\erem}{\bbox\end{remark}}
\newcommand{\exam}{\bbox\end{example}}
\newcommand{\ecor}{\end{corollary}}
\newcommand{\eprf}{\end{proof}}
\newcommand{\beqn}{\begin{equation}}
\newcommand{\eeqn}{\end{equation}}
\newcommand{\wbox}{\mbox{$\sqcap$\llap{$\sqcup$}}}
\newcommand{\bbox}{\wbox}
\newenvironment{RETHM}[2]{\trivlist \item[\hskip 10pt\hskip\labelsep{\sc #1\hskip 5pt\relax\ref{#2}.}]\it}{\endtrivlist}
\newcommand{\rethm}[1]{\begin{RETHM}{Theorem}{#1}}
\newcommand{\repro}[1]{\begin{RETHM}{Proposition}{#1}}
\newcommand{\relem}[1]{\begin{RETHM}{Lemma}{#1}}
\newcommand{\recor}[1]{\begin{RETHM}{Corollary}{#1}}

\newcommand{\erethm}{\end{RETHM}}
\newcommand{\erepro}{\end{RETHM}}
\newcommand{\erelem}{\end{RETHM}}
\newcommand{\erecor}{\end{RETHM}}

\newcommand{\rimp}{\Rightarrow}
\newcommand{\nthm}[1]{\begin{oldthm}{#1}}
\newcommand{\enthm}{\end{oldthm} \medskip}
\newenvironment{oldthm}[1]{\medskip\par\noindent{\bf Theorem #1:} \em \noindent}{\par}

\newcommand{\othm}{\rethm}
\newcommand{\eothm}{\erethm}

\newcommand{\opro}{\repro}
\newcommand{\eopro}{\erepro}

\newcommand{\union}{\cup}
\newcommand{\inter}{\cap}
\renewcommand{\phi}{\varphi}
\newenvironment{prog}{\begin{array}[t]{@{}l@{}}}{\end{array}}

\newcommand{\cL}{\mathcal{L}}

\newcommand{\len}[1]{|#1|}
\newcommand{\card}[1]{\#({#1})}

\newcommand{\Att}{\mathit{\bf Attractive}}
\newcommand{\Smart}{\mathit{\bf Smart}}
\newcommand{\Said}{\mathit{\bf Said}}
\newcommand{\true}{\mathit{\bf{true}}}
\newcommand{\false}{\mathit{\bf{false}}}
\newcommand{\Permitted}{\mathit{\bf Perm}}
\renewcommand{\Pr}{\mathit{\bf Pr}}

\newcommand{\vtab}{\phantom{Le}}

\newcommand{\cd}{d}
\newcommand{\scd}{S}
\newcommand{\cc}{e}
\newcommand{\scc}{E}

\newcommand{\lic}{\mathit{license}}

\newcommand{\grant}{\mathit{grant}}
\newcommand{\cond}{\mathit{cond}}
\newcommand{\conc}{\mathit{conc}}

\newcommand{\princ}{\mathit{prin}}

\newcommand{\rht}{\mathit{right}}
\newcommand{\rsrc}{\mathit{rsrc}}
\newcommand{\var}{\mathit{var}}
\newcommand{\princVarSet}{\mathit{prinVar}}
\newcommand{\rsrcVarSet}{\mathit{rsrcVar}}
\newcommand{\princVar}{\mathit{x_p}}

\newcommand{\rsrcVar}{\mathit{x_r}}
\newcommand{\primPrinc}{\mathit{p}}

\newcommand{\issue}{\mathtt{issue}}
\newcommand{\allCond}{\mathtt{AllConditions}}

\newcommand{\primitiveProp}{\mathit{primitiveProp}}
\newcommand{\primitivePrinc}{\mathit{primitivePrin}}

\newcommand{\keyHolder}{\mathtt{KeyHolder}}
\newcommand{\possessProperty}{\mathtt{PossessProperty}}

\newcommand{\allPrincipals}{\mathtt{AllPrincipals}}

\newcommand{\imp}{\rightarrow}

\newcommand{\XA}{\textbf{Auth}}

\newcommand{\CMet}{\textbf{Holds}}

\newcommand{\XProc}{\textbf{Query}}
\newcommand{\XProcTwo}{\textbf{Query2}}
\newcommand{\XProcFour}{\textbf{Query3}}

\newcommand{\CMetTwo}{\textbf{Holds2}}
\newcommand{\XATwo}{\textbf{Auth2}}
\newcommand{\CMetThree}{\textbf{Holds}}

\newcommand{\Princ}{\mathit{Principal}}

\newcommand{\Rht}{\mathit{Right}}
\newcommand{\Rsrc}{\mathit{Resource}}

\newcommand{\Val}{\mathsf{Val}}

\newcommand{\Eql}[2]{\mathsf{Val}(#1 \iff #2)}
\newcommand{\transwithE}[5]{#1^{#2, #3, #4, #5}}
\newcommand{\tranwithE}[1]{#1^{L, A, \scd, \scc}}
\newcommand{\tranwithEempty}[1]{#1^{L, A, \emptyset, \scc}}
\newcommand{\EStar}{{\cal E}^*}
\newcommand{\SStar}{{\cal S}^*}

\renewcommand{\iff}{\Leftrightarrow}

\newcommand{\pred}[1]{\mathbf{#1}}

\newcommand{\const}[1]{\mathbf{#1}}

\newcommand{\numP}{n}

\newcommand{\lenC}{h}

\newcommand{\EX}{X}

\newcommand{\andNd}{\mathit{and}}
\newcommand{\orNd}{\mathit{or}}
\newcommand{\nandNd}{\mathit{non-and}}
\title{A Formal Foundation for XrML}
\author{Joseph Y.~Halpern and Vicky Weissman \\ Cornell University}

\commentout{
\title{A Formal Foundation for XrM%
\thanks{Authors supported in part by NSF under grant
CTC-0208535, by ONR under grants  N00014-00-1-03-41 and
N00014-01-10-511, by the DoD Multidisciplinary University Research
Initiative (MURI) program administered by the ONR under
grant N00014-01-1-0795, and by AFOSR under grant F49620-02-1-0101.
A preliminary version of this paper appeared at the 17th IEEE Computer Security
Foundations Workshop in Pacific Grove, California, 2004.
}}

\author{
Joseph Y. Halpern\\
Cornell University\\
Ithaca, NY 14853\\
halpern@cs.cornell.edu
\and
Vicky Weissman\\
Cornell University\\
Ithaca, NY 14853\\
vickyw@cs.cornell.edu
}

\maketitle
\thispagestyle{empty}
}%

\begin{abstract}
XrML is becoming a popular language in industry for writing software
licenses.  The semantics for XrML is implicitly given by an algorithm
that determines if a permission follows from a set of licenses.  We
focus on a fragment of the language and use it to
highlight some problematic aspects of the algorithm.  We then correct
the problems, introduce formal semantics, and show that our semantics
captures the (corrected) algorithm.  Next, we consider the complexity of
determining if a permission is implied by a set of XrML licenses.  We
prove that the general problem is undecidable, but it is polynomial-time
computable for an expressive fragment of the language.  We extend XrML
to capture a wider range of licenses by adding negation to the language.
Finally, we discuss the key differences between XrML and MPEG-21, an
international standard based on XrML.
\end{abstract}

\category{H.2.7}{Database Management}{Database Administration}[Security \and integrity \and protection]
\category{K.4.4}{Computers and Society}{Electronic Commerce}[Security]
\terms{Security, Languages}
\keywords{Digital Rights Management}

\begin{document}

\begin{bottomstuff}
Authors' address: J.~Halpern and V.~Weissman, Cornell University,
Ithaca, NY 14853.\newline
Authors supported in part by NSF under
grants CTC-0208535, ITR-0325453, and IIS-0534064,
by ONR under grants  N00014-00-1-03-41 and
N00014-01-10-511, by the DoD Multidisciplinary University Research
Initiative (MURI) program administered by the ONR under
grant N00014-01-1-0795, and by AFOSR under
grants F49620-02-1-0101 and FA9550-05-1-0055.
A preliminary version of this paper appeared at the 17th IEEE Computer Security
Foundations Workshop in Pacific Grove, California, 2004.
\end{bottomstuff}

\maketitle
\section{Introduction}
The eXtensible rights Markup Language (XrML) is becoming an increasingly popular
language in which to write software licenses.  When first released in 2000, XrML
received the support of many technology providers, content owners, distributors,
and retailers, including Adobe Systems, Hewlett-Packard Laboratories, Microsoft,
Xerox Corp., Barnesandnoble.com, and Time Warner Trade Publishing.  In fact,
Microsoft, OverDrive, and DMDsecure have publicly announced their agreement to
build products and/or services that are XrML compliant.  Currently, XrML is being
used by international standard committees as the basis for application-specific
languages that are designed for use across entire industries.  For
example, the 
Moving Picture Experts Group (MPEG) has selected XrML as the foundation for their
MPEG-21 Rights Expression Language, henceforth referred to as MPEG-21
(see http://www.xrml.org).  It is clear that a number of industries are moving
towards a standard language for writing licenses and that many of these standard
languages are likely to be based on XrML.  To understand the new standards, we
need to understand XrML.

XrML does not have formal semantics.  Instead, the XrML specification \cite{XrML}
presents the semantics in two ways.  First is an English description of the
language.  Second is an English description of an algorithm that determines if a
permission follows from a set of licenses.  Unfortunately, the two versions of the
semantics do not agree.  To make matters worse, the algorithm has unintuitive
consequences that do not seem to reflect the language developers' intent.

To address these issues, we provide formal semantics for
a fragment of XrML.
We focus on a fragment because the entire language is somewhat unwieldy.
An
XrML license says that an agent grants a permission if certain
conditions hold.
Our
fragment includes only two types of permissions and only two types of conditions.
We give our fragment formal semantics by defining
a translation from licenses in
the fragment
to formulas in
first-order logic extended with a validity operator.  We argue that the translation
preserves the meaning of the XrML statements by proving that the algorithm included
in the XrML document, slightly modified to correct the unintuitive
behavior, 
matches our semantics.  More precisely, the algorithm says that a
permission follows from a 
set of licenses if and only if the translated permission is a logical
consequence of 
the translated licenses.  We then consider the complexity of determining if a
permission is implied by a set of licenses.  We show that the general problem is
undecidable, even for our fragment.
The problem is decidable in polynomial time if we restrict the fragment slightly.

A shortcoming of XrML is that it does not support negation.  For example, in XrML,
we cannot write ``customers may \emph{not} edit the software''.  The XrML developers
deal with this limitation, to some extent, by assuming that an action is forbidden
unless it is explicitly permitted.  As a result, a license writer does not need to say
that an action is forbidden, because the prohibition is already implied.  This
approach might be acceptable in various instances, but it is difficult to believe that
most license writers really want to forbid \emph{every} action that they do not
explicitly permit.  So, the approach does not capture the license writer's actual intent.
Moreover, it limits the class of licenses that can be expressed, because it removes the
distinction between forbidden and unregulated actions.  For example, in XrML, we cannot
say ``a hospital may petition for an exemption if it permits an action that the government
forbids''.  Similarly, a course instructor cannot say ``if the university does not object,
then Alice is permitted to audit the class''.  In this paper, we extend XrML to include
such statements and consider the effect of the addition on the language's tractability.

MPEG-21 is an international standard based on XrML.  When we first decided to give XrML
formal semantics, the MPEG committee had released a beta version of its language, which was
XrML with minor revisions, and was preparing the final release.  We chose to give semantics
to the beta language first (before analyzing the official XrML specification, as is done
here), because we hoped that any problems we found would be corrected in the final version
of MPEG-21.  This is, in fact, what occured.  After discussing our results with Thomas
DeMartini and Xin Wang of the MPEG Standards Committee, the committee released their ISO
standard \cite{RelFinal}; the shortcomings that we identified are addressed in the standard.
We conjecture that all of our results for XrML hold with minor changes for MPEG-21, although
we have not verified the details.

The rest of the paper is organized as follows.  In the next section we present
our
fragment of XrML.  In Section~\ref{s:xrmlAlg} we review XrML's algorithm for
answering queries.  After considering some examples in which the algorithm's behavior is
unintuitive and almost certainly unintended, we propose corrections that we believe captures
the designers' intent.  Formal semantics for
our fragment
are given in Section~\ref{s:semantics-xrml},
and the revised algorithm is shown to be sound and complete with respect to the semantics.  In
Section~\ref{s:trac} we show that the problem of determining if a permission follows from a set
of licenses is undecidable.  We also discuss a fragment of
XrML
that is both tractable
and relatively expressive.  In Section~\ref{s:core} we outline how our
results can be extended to
a substantial
fragment of XrML.
Negation is added to XrML in Section~\ref{s:extend}.
The analysis of this paper had an impact on practice.
MPEG-21 REL, an international standard based on XrML,
incorporates the developers' response to our concerns about XrML.
We describe MPEG-21 REL, and how it deals with our concerns,
in Section~\ref{s:MPEG-21}.
We conclude in Section~\ref{s:concl}.  All of the proofs are in the
appendix.

\section{Syntax}\label{s:XrMLsyntax}
XrML is an XML-based language; it follows XML-conventions.  Rather than present that syntax,
we use an alternative syntax that is more concise and, we believe, more intuitive.  In this
section, we introduce our syntax for a
fragment of XrML (the rest of the
language is discussed in Section~\ref{s:core}) and describe the key differences between the
syntax used in the XrML specification and that used here.

At the heart of XrML is the notion of a \emph{license}.  A license is a (principal, grant)
pair, where the license $(p, g)$ means $p$ issues (i.e., says) $g$.  For example, the
license (Alice, Bob is smart) means ``Alice says `Bob is smart'$\,$''.

A grant has the form $\forall x_1 \ldots \forall x_n(\mbox{condition} \imp \mbox
{conclusion})$, which intuitively means that the condition implies the conclusion
under all appropriate substitutions.  Conditions and conclusions are defined as follows.
\begin{itemize}
\item A condition has the form $\cd_1\land\ldots\land\cd_n$, where each $\cd_i$ is either
$\true$ or $\Said(p, \cc)$ for some principal $p$ and conclusion $\cc$.  Roughly speaking,
the condition $\true$ always holds and the condition $\Said(p, \cc)$ holds if $p$ issues a
grant that says $\cc$ holds if a condition $\cd$ holds, and $\cd$ does, in fact, hold.
\item A conclusion has either the form $\Permitted(p, r, s)$ or the form $\Pr(p)$, where
$\Pr$ is a property, $p$ is a principal, $r$ is a right (i.e., an action), and $s$ is a
resource.  The conclusion $\Permitted(p, r, s)$ means $p$ may exercise $r$ over $s$.  For
example, $\Permitted(\mathit{Bob}, \mathit{edit}, \mathit{budget\hspace{2 pt}report})$ means
Bob may edit the budget report.  The conclusion $\Pr(p)$ means $p$ has the property $\Pr$.
For example, the conclusion $\Att(\mathit{Bob})$ means Bob is attractive.
\end{itemize}
We abbreviate the grant $\forall x_1 \ldots \forall x_n(\true\imp \cc)$ as
$\forall x_1 \ldots \forall x_n\cc$.  Also, we try to consistently use $\cd$, possibly
subscripted, to denote a generic condition and $\cc$, possibly subscripted, to denote a
generic conclusion.

\begin{sloppypar}
Consider the following example.  Suppose that Alice issues the grant ``Bob is smart''
and Amy issues the grant ``if Alice says that Bob is smart, then he is attractive''.  We
can write the first license in our syntax as $(\mathit{Alice}, g_1)$, where
$g_1 = \Smart(\mathit{Bob})$ (recall that this is an abbreviation for
$\true\imp\Smart(\mathit{Bob})$), and we can write the second as
$(\mathit{Amy}, g_2)$, where $g_2 =  \Said(\mathit{Alice}, \Smart(\mathit{Bob})) \imp
\Att(\mathit{Bob})$.  Because $(\mathit{Alice}, g_1)$ is in the set of issued licenses,
$\Said(\mathit{Alice}, \Smart(\mathit{Bob}))$ holds.  It follows from this fact and the
license $(\mathit{Amy}, g_2)$ that $\Said(\mathit{Amy}, \Att(\mathit{Bob}))$ holds as
well.
\end{sloppypar}

The sets of principals, properties, rights, and resources depend on the particular application.
For example, a multimedia application might have a principal for each employee and each
customer; properties such as ``hearing impaired'' and ``manager''; rights such as ``edit'' and
``download''; and a resource for each object such as a movie.  We assume the application gives us
a finite set $\primitivePrinc$ of principals and a finite set $\primitiveProp$ of properties.  We
then define the components in our language as follows.
\begin{itemize}
\item The set $P$ of principals is the result of closing $\primitivePrinc$ under union.  (Here and
elsewhere we identify a principal $p\in \primitivePrinc$ with the singleton $\{p\}$ and write
$\{p_1, \ldots, p_n\}$ rather than $\{p_1\}\union\ldots\union \{p_n\}$.)  The interpretation of a
principal $\{p_1,\ldots, p_n\}$ depends on context; that is, the interpretation depends on whether
the principal appears as the first argument in a $\Said$ condition, in a conclusion, or in a
license.  We discuss this later in the paper (primarily in Section~\ref{s:trac}).
\item The set of properties is $\primitiveProp$.
We assume that
every property in $\primitiveProp$
takes a single argument and that argument is of sort $\Princ$.  For example, $\primitiveProp$ can
include the property $\mathit{\bf Employee}$, where $\mathit{\bf Employee}(x)$ means principal $x$
is an employee, but it cannot include the property $\mathit{\bf MotherOf}$, where
$\mathit{\bf MotherOf}(x, y)$ means principal $x$ is the mother of principal $y$, nor can it include
the property $\mathit{\bf Vehicle}$, where $\mathit{\bf Vehicle}(x)$ means resource $x$ is a vehicle
(e.g., a motorcycle, car, or truck).  The results in this paper continue to hold if we extend the
language to include properties that take multiple arguments of various sorts (i.e., principals,
rights, and resources).  It is also easy to show that closing $\primitiveProp$ under conjunction adds
no expressive power to the language.  Closing under negation does add expressive power; we return to
this issue in Section~\ref{s:extend}.
\item The only right in our language is $\issue$ and the only resources are grants.  Intuitively, if
a principal $p$ has the right to issue a grant $g$, and $p$ does issue $g$, then $g$ is a true
statement.  Including additional rights and resources in our language does not significantly affect
the discussion.
\end{itemize}

We formally define the syntax according to the following grammar.
\[
\begin{array}{lll}
\lic &::=& (\princ, \grant)\\
\grant &::=& \forall \var \ldots \forall \var (\cond\imp\conc)\\
\var &::=& \princVar ~|~ \rsrcVar\\
\cond &::=& \true ~|~ \Said(\princ, \conc) ~|~ \cond\land\cond \\
\conc &::=& \Pr(\princ) ~|~ \Permitted(\princ, \rht, \rsrc) \\
\princ &::=& \{\primPrinc\} ~|~ \{\princVar\} ~|~ \princ\union\princ \\
\rht &::=& \issue \\
\rsrc &::=& \grant ~|~ \rsrcVar,
\end{array}
\]
where $\Pr$ is an element of $\primitiveProp$, $\primPrinc$ is an element of
$\primitivePrinc$, $\princVar$ is an element of $\princVarSet$, which is the set of variables
ranging over primitive principles, and $\rsrcVar$ is an element of $\rsrcVarSet$, which is the
set of variables ranging over resources.  For the remainder of this paper we assume that the first
argument in a license is a singleton.  Because the XrML document treats the license
$(\{p_1, \ldots, p_n\}, g)$ as an abbreviation for the set of licenses
$\{(p, g)\mid p\in \{p_1, \ldots, p_n\}\}$, it is easy to modify our discussion to support all of the
licenses included in the grammar.

As mentioned at the beginning of this section, the grammar presented
here is not
identical to that described in the XrML document.  Certain components of XrML are
omitted from our language.  These are discussed in Section~\ref{s:core}.  The XrML
components that are included are represented using a syntax that we believe is more
intuitive.  The main differences between the syntax of our language and the syntax
of XrML are described below.
\begin{itemize}
\item Instead of assuming that the application provides a set of primitive principals, XrML assumes
that the application provides a set $K$ of cryptographic keys; the set of primitive principals is
$\{\keyHolder(k)\mid k\in K\}$. We could take $\primitivePrinc$ to be this set; however, our more
general approach leads to a simpler discussion.  Moreover, our results do not change if we restrict
primitive principals to those of the form $\keyHolder(k)$.
\item XrML does not have conclusions of the form $\Pr(p)$.  To capture properties, XrML uses a right
called $\possessProperty$ and considers the properties given by the application to be resources.  The
conclusion $\Pr(p)$ in our grammar corresponds to the conclusion
$\Permitted(p, \possessProperty, \Pr)$ in XrML.  We have two types of conclusions because we believe
the grammar should help distinguish the conceptually different notions of permissions and properties,
rather than confounding them.
\item Rather than writing $\allPrincipals(p_1, \ldots, p_n)$, $\allCond(c_1, \ldots, c_n)$, and
$\allCond()$, we use the more standard notations $\{p_1, \ldots, p_n\}$, $c_1\land\ldots\land c_n$,
and {\bf true}, respectively.  Rather than writing
$\pred{PrerequisiteRight}(p, \cc)$, we use the shorter and, we believe, more appropriate notation
$\Said(p, \cc)$.
\item As discussed previously, XrML abbreviates a set of licenses
$\{(p_i, g_j) \mid i \le n, j\le m\}$ as the single license
$(\{p_1, \ldots, p_n\}, \{g_1, \ldots, g_m\})$.  For ease of exposition, we do not do this.
\end{itemize}

\section{XrML's Authorization Algorithm}\label{s:xrmlAlg}
The XrML document includes a procedure that we call $\XProc$ to determine if a conclusion follows from
a set of licenses (and some additional input that is discussed below).  In this section we present and
analyze the parts of the algorithm that pertain to our fragment.

Before describing the algorithm, we note that some aspects of $\XProc$ are inefficient.  This is
acknowledged in the XrML document, which explains that $\XProc$ was designed with clarity as the primary
goal; it is the responsibility of the language implementors to create efficient algorithms with the same input/output behavior as $\XProc$.  (In Section~\ref{s:trac}, we show that it is highly unlikely that
such an efficient algorithm exists.)

\subsection{A Description of $\XProc$}\label{sec:XProcdesc}
The input to $\XProc$ is a closed conclusion $\cc$ (i.e., a conclusion with no free variables), a set
$L$ of licenses $(p,g)$ such that $p$ is variable-free, and a set $R$ of grants; $\XProc$ returns
$\true$ if $\cc$ is implied by $L$ and $R$, and returns $\false$ otherwise.  To explain the intuition
behind $L$ and $R$, we first note that the procedure treats a predefined set of principals as trusted.
If a trusted principal issues the grant $g$, then $g$ is in $R$ and it is assumed to be true.  If the
license $(p, g)$ is in $L$, then $p$ issued $g$ (i.e., $p$ says $g$) and $p$ is not an implicitly
trusted principal.  To clarify the inferences that are drawn from $R$ and $L$, suppose that the grant
$g$ is $\mathit{\bf QueenOfSiam}(\mathit{Alice})$, which means Alice is Queen of Siam, and the grant
$g'$ is $\Permitted(\mathit{Alice}, \issue, g)$, which means Alice may issue $g$.  If $g\in R$, then
we assume that Alice really is queen.  If $(\mathit{Alice}, g)$ is in $L$, then Alice says that she
is the queen, but we cannot conclude that she is royalty from this statement alone.  If
$(\mathit{Alice}, g)$ is in $L$ and $g'$ is in $R$, then we assume that Alice has the authority to
declare herself queen, because $g'\in R$; we assume that she exercises that authority, because
$(\mathit{Alice}, g)\in L$; and we conclude that Alice is queen, because this follows from the two
assumptions.

$\XProc$ begins by calling the $\XA$ algorithm.  $\XA$ takes $\cc$, $L$, and $R$ as input; it returns
a set $D$ of closed conditions (i.e., conditions with no free variables).  Roughly speaking, a closed
condition $\cd$ is in $D$ if $\cd$, $L$, and $R$ together imply $\cc$.  To determine if a condition
in $D$ holds, $\XProc$ relies on the $\CMet$ algorithm.  The input to $\CMet$ is a closed condition
$\cd$ and a set $L$ of licenses; $\CMet(d, L)$ returns true if the licenses in $L$ imply $\cd$, and
returns $\false$ otherwise.  If $\CMet(\cd, L)$ returns $\true$ for some $\cd$ in $D$, then $\XProc$
returns $\true$, indicating that $L$ implies $\cc$.  $\XProc$ is summarized in
Figure~\ref{tb:XrMLProc1}.

\begin{figure*}[htb]
\begin{center}
\begin{tabular}{|l|}\hline
\underline{$\XProc(\cc, L, R)$:}\\\\
$D := \XA(\cc, L, R)$\\
{\bf if} $\CMet(\cd, L) = \true$ for a condition $\cd\in D$\\
{\bf then} return $\true$\\
{\bf else} return $\false$\\
\hline
\end{tabular}
\end{center}
\caption{The $\XProc$ Algorithm}
\label{tb:XrMLProc1}
\end{figure*}

We now discuss $\XA$ and $\CMet$ in some detail.  To define $\XA$, we first consider the case where
$L = \emptyset$.  Define a \emph{closed substitution} to be a mapping from variables to closed
expressions of the appropriate sort.  Given a closed substitution $\sigma$ and an expression $t$, let
$t\sigma$ be the expression that arises after all free variables $x$ in $t$ are replaced by
$\sigma(x)$.  Roughly speaking, $\XA(\cc, \emptyset, R)$ returns the set $D$ of closed conditions such
that each condition in $D$, in conjunction with the grants in $R$, implies $\cc$.  That is, $\cd\in D$
iff there is a grant $g = \forall x_1\ldots\forall x_n(\cd_g\imp\cc_g)$ in $R$ and a closed
substitution $\sigma$ such that $\cd = \cd_g\sigma$ and $\cc_g$ implies $\cc$.  $\XA$ determines
whether $\cc_g$ implies $\cc$ in a somewhat nonstandard way.  In particular, it makes the
\emph{subset assumption}, which says that any property or permission attributed to a principal $p$ is
attributed to every principal that includes $p$.  In other words, if $p\subseteq p'$, then $\Pr(p)$
implies $\Pr(p')$ and $\Permitted(p, r, s)$ implies $\Permitted(p', r, s)$.  Thus,
\[
\begin{array}{lll}
\XA(\Pr(p), \emptyset, R) &=&
\{\cd\mid\mbox{for some grant }g = \forall x_1\ldots\forall
x_n(\cd_g\imp\Pr(p_g))\in R
\mbox{ and closed}\\
&&\mbox{substitution }\sigma, \cd_g\sigma
= \cd\mbox{ and } p_g\sigma \subseteq p\} \mbox{ and }
\end{array}
\]
\[
\begin{array}{lll}
\XA(\Permitted(p, r, s), \emptyset, R) &=&
\{\cd\mid\mbox{for some grant }
\forall x_1\ldots\forall x_n(\cd_g\imp\Permitted(p_g, r_g, s_g))\in R\\
&&\mbox{and closed substitution }\sigma,
\cd_g\sigma = \cd, p_g\sigma \subseteq p, r_g\sigma = r,\\
&&\mbox{and } s_g\sigma = s\}.
\end{array}
\]

\begin{sloppypar}
Suppose that $L\ne \emptyset$.  Then we reduce to the previous case by taking
$\XA(\cc,L,R) = \XA(\cc,\emptyset, R')$, where, intuitively, $R'$ is the set of legitimate grants;
that is, $R'$ consists of the grants in $R$ and the grants issued by someone who has the authority
to do so.  It seems reasonable to call $\XProc(\Permitted(p, \issue, g), L, R)$ to determine if a
principal $p$ has the authority to issue a grant $g$.  However, if $\XA$ calls
$\XProc(\Permitted(p, \issue, g), L, R)$ to construct $R'$, then the algorithm will not terminate,
because $\XProc$ calls $\XA$, leading to an infinite call tree.  So, instead of calling
$\XProc(\Permitted(p, \issue, g), L, R)$, the XrML algorithm determines if $p$ is permitted to
issue $g$ by checking if $\CMet(\cd,L) = \true$ for some $\cd$ in the set
$\XA(\Permitted(p, \issue, g), L - \{(p,g)\}, R)$.  We discuss the consequences of this solution
in Section~\ref{s:alg2}.  In summary,
\[
\begin{array}{lll}
R' &=& R\union R'', \mbox{ where}\\
R'' &=& \{g \mid \mbox{ for some licence}(p,g)\in L\mbox{ and condition }\cd,\\
&&\cd\in\XA(\Permitted(p, \issue, g), L-\{(p, g)\}, R)\mbox{ and} \CMet(\cd, L) = \true\}
\end{array}
\]
Pseudocode for $\XA$ is given in Figure~\ref{tb:auth}.
\end{sloppypar}

\begin{figure*}[htb]
\begin{center}
\begin{tabular}{|l|}\hline
\underline{$\XA(\cc, L, R)$:}\\\\
$D := \emptyset$\\
{\bf if} $L = \emptyset$\\
{\bf then} \\
\vtab\% Find $D$, the conditions under which $R$ implies $e$\\
\vtab{\bf if} $\cc = \Pr(p)$\\
\vtab\vtab{\bf for} each grant $\forall x_1 \ldots \forall x_n(\cd_g\imp\Pr(p_g)) \in R$\\
\vtab\vtab\vtab $D := D \union \{\cd\mid\cd_g\sigma = \cd\mbox{ and }p_g\sigma\subseteq p,
\mbox{ for some closed substitution }\sigma\}$\\
\vtab{\bf if} $\cc = \Permitted(p, r, s)$\\
\vtab\vtab{\bf for} each grant $\forall x_1 \ldots \forall x_n(\cd_g\imp\Permitted(p_g, r_g, s_g)) \in R$\\
\vtab\vtab\vtab $D := D \union \{\cd\mid\cd_g\sigma = \cd, p_g\sigma\subseteq p, r_g\sigma = r,
\mbox{ and }s_g\sigma = s, \mbox{ for some closed substitution }\sigma\}$\\
{\bf else} \\
\vtab\% Find $R'$\\
\vtab $R' := R$\\
\vtab {\bf for} each license $(p, g) \in L$\\
\vtab\vtab $L' := L - \{(p, g)\}$\\
\vtab\vtab $D' := \XA(\Permitted(p, \issue, g), L', R)$\\
\vtab\vtab {\bf if} $\CMet(\cd, L) = \true$ for a condition $\cd\in D'$\\
\vtab\vtab {\bf then} $R' := R'\union \{g\}$\\
\vtab\% Find $D$, the conditions under which $R'$ implies $\cc$\\
\vtab $D := \XA(\cc, \emptyset, R')$\\
return $D$\\
\hline
\end{tabular}
\end{center}
\caption{The $\XA$ Algorithm}
\label{tb:auth}
\end{figure*}

We define $\CMet(\cd, L)$ by induction on the structure of $\cd$.  If $\cd$ is {\bf true},
then $\CMet(\cd, L) = \true$.  If $\cd = \Said(p, \cc)$, then $\CMet(d, L) = \true$ iff
$p$ issues a grant $\forall x_1 \ldots\forall x_n (\cd_g\imp\cc_g)$ such that, for some
substitution $\sigma$, $e_g\sigma = e$ and $\CMet(\cd_g\sigma, L) = \true$.  In this context,
a principal $\{p_1, \ldots, p_n\}$ issues a grant $g$ if $p_i$ issues $g$ for some
$i = 1, \ldots, n$.  If $\cd = \cd_1\land\ldots\land\cd_n$, where each $\cd_i$ is $\true$ or
a $\Said$ condition, then $\CMet(\cd, L) = \bigwedge_{i = 1, \ldots, n}\CMet(\cd_i, L)$.
Pseudocode for $\CMet$ is given in Figure~\ref{tb:CMet}.

\begin{figure*}[htb]
\begin{center}
\begin{tabular}{|l|}\hline
\underline{$\CMet(\cd, L)$:}\\\\
{\bf if} $\cd = \true$ \\
{\bf then} return $\true$\\\\
{\bf if} $\cd = \Said(p, \cc)$\\
{\bf then}\\
\vtab $R_p = \{g \mid \mbox{for some principal }p', (p', g)\in L\mbox{ and } p'\in p\}$\\
\vtab $D := \{\cd' \mid \mbox{for some grant $\forall x_1 \ldots \forall x_n(\cd_g\imp\cc_g) \in R_p$ and}$\\
\vtab\phantom{ $D := $}$\mbox{closed substitution $\sigma$, } \cd_g\sigma = \cd'\mbox{ and }\cc_g\sigma = \cc\}$\\
\vtab {\bf if} $\CMet(\cd', L) = \true$ for a condition $\cd' \in D$\\
\vtab {\bf then} return $\true$\\
\vtab {\bf else} return $\false$\\\\
{\bf if} $\cd = \cd_1\land\ldots\land\cd_n$, where each $\cd_i$ is $\true$ or a $\Said$ condition\\
{\bf then} return $\bigwedge_{i = 1, \ldots, n}\CMet(\cd_i, L)$\\
\hline
\end{tabular}
\end{center}
\caption{The $\CMet$ Algorithm}
\label{tb:CMet}
\end{figure*}

\begin{sloppypar}
\xam\label{ex:XrML1}
In Section~\ref{s:XrMLsyntax}, we argued informally that Amy says Bob is attractive if the set
of licenses is $L = \{(\mathit{Alice}, g_1), (\mathit{Amy}, g_2)\}$, where
$g_1 = \Smart(\mathit{Bob})$ and
$g_2 = \Said(\mathit{Alice}, \Smart(\mathit{Bob}))\imp\Att(\mathit{Bob}).$  The formal algorithm
gives the same conclusion.  Specifically, $\CMet(\Said(\mathit{Amy}, \Att(\mathit{Bob})), L)$
sets $R_{Amy} = \{g_2\}$ and calls $\CMet(\Said(\mathit{Alice}, \Smart(\mathit{Bob})), L)$.
During this call $R_{Alice}$ is set to $\{g_1\}$ and $\CMet(\true, L)$ is called.  Because
$\CMet(\true, L) = \true$, $\CMet(\Said(\mathit{Alice}, \Smart(\mathit{Bob})), L)
 = \true$ and, thus, $\CMet(\Said(\mathit{Amy}, \Att(\mathit{Bob})), L) = \true$.

Suppose that a trusted principal says that Amy has the authority to issue $g_2$ (i.e., if Amy
says $g_2$, then $g_2$ holds).  Then we can conclude that Bob really is attractive, because
$\XProc(\Att(\mathit{Bob}), L, R) = \true$, where $R = \{\Permitted(\mathit{Amy}, \issue, g_2)\}$.
Specifically, $\XProc$ begins by calling $\XA(\Att(\mathit{Bob}), L, R)$.
$\XA(\Att(\mathit{Bob}), L, R)$, in turn, calls $\XA(\Att(\mathit{Bob}), \emptyset, R')$, where
$R' = \{g_2, \Permitted(\mathit{Amy}, \issue, g_2)\}$.
$\XA(\Att(\mathit{Bob}), \emptyset, R') = \{\Said(\mathit{Alice}, \Smart(\mathit{Bob}))\}$.  So,
Bob is attractive if the condition $\Said(\mathit{Alice}, \Smart(\mathit{Bob}))$ holds.  To
determine if the condition holds, $\XProc$ calls
$\CMet(\Said(\mathit{Alice}, \Smart(\mathit{Bob})), L)$.  We have already shown that
$\CMet(\Said(\mathit{Alice}, \Smart(\mathit{Bob})), L) = \true$; we evaluated this call during our
analysis of $\CMet(\Said(\mathit{Amy}, \Att(\mathit{Bob})), L)$.  So Bob is indeed attractive.
\exam
\end{sloppypar}

$\XProc$ as described here and in the XrML specification is somewhat ambiguous.  For example,
the specification does not say in which order the conditions in $D$ should be tested to see if at
least one condition in $D$ holds.  As a result, there are a number of possible executions of a call
$\XProc(\cc, L, R)$, depending on the implementation of $\XProc$.  It is easy to see that, for a
particular input, every execution that terminates returns the same output.  However, as we show in
Example~\ref{ex:terminate}, whether $\XProc$ terminates can depend on how it is implemented.  A
similar issue arises with $\XA$ and $\CMet$.  We talk about an execution of $\XProc$, $\XA$, or
$\CMet$ only if the choice of execution affects whether the algorithm terminates.  For example, we
write $\XProc(\cc, L, R) = \true$ if every execution of $\XProc(\cc, L, R)$ returns $\true$.

\subsection{An Analysis of $\XProc$}\label{s:alg2}
In this section we present five examples in which $\XProc$ gives unexpected results.
Example~\ref{ex:subset} reveals a mismatch between $\XProc$ and the informal language description;
the discrepancy exists because $\XA$ makes the subset assumption and the informal language
description does not.  Example~\ref{ex:lostL} demonstrates that a license $(p, g)$ should not be
removed from the set of licenses when determining if $p$ is permitted to issue $g$.
Examples~\ref{ex:terminate}, \ref{ex:terminateC}, and \ref{ex:terminateB}, show that a reasonable
implementation of $\XProc$ does not terminate on all inputs, for three quite different reasons:
Example~\ref{ex:terminate} shows that on some inputs $\CMet$ makes infinitely many identical calls,
Example~\ref{ex:terminateC} shows that on some inputs the call tree for $\XProc$ includes an
infinite path of distinct nodes; and Example~\ref{ex:terminateB} shows that on some inputs the call
tree for $\XProc$ includes a node with infinitely many distinct children.

\begin{sloppypar}
\xam\label{ex:subset}
Suppose that Alice is quietly walking beside her two giggling daughters, Betty and Bonnie.  Are the
three of them a quiet group?  Intuitively, they are not, because Betty and Bonnie are giggling.
According to $\XProc$, however, the answer is yes.  Since Alice is quiet and $\XA$ makes the subset
assumption, $\XProc$ concludes that the principal $\{\mathit{Alice,Betty, Bonnie}\}$ is quiet; that
is, $\XProc(\mathit{\bf Quiet}(\{\mathit{Alice, Betty, Bonnie}\}),
\emptyset,\{{\bf Quiet}(\mathit{Alice})\}) = \true.$
\exam
\end{sloppypar}

\begin{sloppypar}
\xam\label{ex:lostL}
Suppose that Alice says that she is smart, and if Alice says that she is smart, then she is permitted
to say that she is smart.  Is Alice smart?  Intuitively, she is, because Alice is permitted to say
that she is smart and she does so.  But consider $\XProc({\bf Smart}(\mathit{Alice}), L, R)$, where
$L = \{(\mathit{Alice}, g)\}$, $R = \{\Said(\mathit{Alice},
{\bf Smart}(\mathit{Alice}))\imp\Permitted(\mathit{Alice}, \issue, g)\}$, and
$g = {\bf Smart}(\mathit{Alice})$.  $\XProc({\bf Smart}(\mathit{Alice}), L, R)$ begins by calling
$\XA({\bf Smart}(\mathit{Alice}), L, R)$.  $\XA$ checks whether or not Alice is permitted to issue $g$.
It determines that Alice may not issue $g$, because the permission does not follow from $R$ and
$L - \{(\mathit{Alice}, g)\}$.  Since Alice is not permitted to issue $g$, $\XA$ sets $R' = R$ and
returns $\emptyset$.  Because $\XA$ returns $\emptyset$, $\XProc$ returns $\false$.
\exam
\end{sloppypar}

\xam\label{ex:terminate}
Suppose that Alice issues the grant ``if I say Bob is smart, then he is'' and Alice is permitted to
issue this grant.  Can we conclude that Bob is smart?  To answer the
question using $\XProc$, let
$\cc = {\bf Smart}(\mathit{Bob})$, $g = \Said(\mathit{Alice}, \cc)\rimp \cc$,
$L = \{(\mathit{Alice}, g)\}$, and $R = \{\Permitted(\mathit{Alice}, \issue, g)\}$.  We are interested
in the output of $\XProc(\cc, L, R)$.  $\XProc(\cc, L, R)$ begins by calling  $\XA(\cc, L, R)$, which
returns the set $D = \{\Said(\mathit{Alice}, \cc)\}$.  $\XProc$ then calls
$\CMet(\Said(\mathit{Alice}, \cc), L)$, which sets $R_{\mathit{Alice}} = \{g\}$ and calls
$\CMet(\Said(\mathit{Alice}, \cc), L)$ again.  It is easy to see that an infinite number of calls to
$\CMet(\Said(\mathit{Alice}, \cc), L)$ are made during the execution of $\XProc(\cc, L, R)$ and thus
the execution does not terminate.

It is tempting to conclude that a set $L$ of licenses and a set $R$ of grants imply a conclusion $\cc$
only if $\XProc(\cc, L, R)$ terminates and returns $\true$.  Unfortunately, whether $\XProc(\cc, L, R)$
terminates can depend on the order in which the calls to $\CMet$ are made.  To see why, consider a
slight modification of the previous example where we add the grant
$\{\mathit{\bf Smart}(\mathit{Bob})\}$ to $R$.  Intuitively, this means that an implicitly trusted
principal says that Bob is smart.  It now seems reasonable to expect that every execution of
$\XProc(\cc, L, R')$ returns $\true$, where $R' = R\union \{\cc\}$, and $\cc$, $L$, and $R$ are as
defined in the original example.  Surely the issued grants imply that Bob is smart, since a grant
issued by a trusted principal says just that!  However, only some of the executions terminate.  Every
execution of $\XProc$ begins by calling $\XA(\cc, L, R')$, and every execution of $\XA(\cc, L, R')$
returns $\{\Said(\mathit{Alice}, \cc), \true\}$.  If an execution of $\XProc$ next calls
$\CMet(\true, L)$, then that execution of $\XProc$ returns $\true$.  On
the other hand, if the 
execution calls $\CMet(\Said(\mathit{Alice}, \cc), L)$ and then waits for the call to return before
calling $\CMet(\true, L)$, then the execution does not terminate for the same reason that every
execution of $\XProc(\cc, L, R)$ does not terminate.
\exam

\begin{sloppypar}
\xam\label{ex:terminateC}
Suppose that Alice says ``for all grants $g$, if I say I am allowed to issue the grant
$\Permitted(\const{Alice}, \issue, g)$, then I am allowed to issue $g$'', and Alice is allowed to
issue that statement.  Is Alice allowed to issue the grant $\pred{Nap}(\const{Alice})$?  To answer
this question using $\XProc$, some abbreviations are useful.  For all grants $g$, we abbreviate the
condition
$\Said(\const{Alice}, \Permitted(\const{Alice}, \issue, \Permitted(\const{Alice}, \issue, g)))$ as
$\cd(g)$ and we abbreviate the grant $\Permitted(\const{Alice}, \issue, g)$ as $h(g)$.  We execute
$\XProc(\cc, L, R)$, where $\cc = \Permitted(\const{Alice}, \issue, \pred{Nap}(\const{Alice}))$,
$R = \{\Permitted(\const{Alice}, \issue, \forall x (d(x)\rimp\Permitted(\const{Alice}, \issue, x)))\}$,
and $L = \{(\const{Alice}, \forall x (\cd(x)\rimp\Permitted(\const{Alice}, \issue, x)))\}$.
$\XProc(\cc, L, R)$ begins by calling $\XA(\cc, L, R)$, which returns
$\{\cd(\pred{Nap}(\const{Alice}))\}$.  Next $\XProc$ calls $\CMet(\cd(\pred{Nap}(\const{Alice})), L)$,
which calls $\CMet(\cd(h(\pred{Nap}(\const{Alice}))), L)$, which calls
$\CMet(\cd(h(h(\pred{Nap}(\const{Alice})))), L)$, and so on.  It is not hard to see that, for all
integers $n>0$, $\CMet(\cd(h^n(\pred{Nap}(\const{Alice}))), L)$ is called, where $h^1(g) = h(g)$ and
$h^n(g) = h(h^{n-1}(g))$, for all grants $g$.  It follows that $\CMet$ does not terminate and, thus,
$\XProc$ does not terminate.
\exam
\end{sloppypar}

\xam\label{ex:terminateB}
Suppose that Alice may say that she is trusted if Bob says that Alice may issue some grant (any grant
at all).  May Alice say that she is trusted?  To answer this question using $\XProc$, we run
$\XProc(\cc, \emptyset, R),$ where
$\cc = \Permitted(\const{Alice}, \issue, \mathit{\bf Trusted}(\const{Alice}))$,
$R = \{\forall x (\cd(x)\imp \cc)\}$, and
$\cd(x) = \Said(\const{Bob}, \Permitted(\const{Alice}, \issue, x))$.  $\XProc$ begins by calling
$\XA(\cc, \emptyset, R),$ which returns $D = \{\cd(g)\mid g\mbox{ is a grant}\}$.  We show below that
$D$ is an infinite set, so every execution of $\XA$ that tries to compute $D$ does not terminate.
Even if $D$ is defined without explicitly listing all of its elements, $\XProc$ must determine if some
element in $D$ holds.  In fact, none do.  Thus, any  approach to testing if some condition in $D$ holds
by explicitly testing each condition will not terminate.

It remains to show that $D = \{d(g)\mid g \mbox{ is a grant}\}$ is an infinite set.  The key observation
is that infinitely many distinct grants can be expressed in the language, even if the vocabulary consists
of only one property $\Pr$ and one principal $p$.  To see why, define grants $g_n$, $n \ge 1$,
inductively by taking
$g_1 = \true\imp\Pr(p)$ and $g_{n+1} = \Said(\mathit{p}, \Permitted(p, \issue, g_{n}))\imp \Pr(p)$
for all $n>0$.  Since each of these grants is clearly distinct, $D$ is infinite.
\exam

\subsection{A Corrected Version of $\XProc$}\label{s:alg3}
In this section we revise $\XProc$ to correct the problems observed in Section~\ref{s:alg2}.  One of the
corrections is fairly straightforward.  We resolve the mismatch illustrated in Example~\ref{ex:subset} by
removing the subset assumption from $\XA$.  We note that the language is sufficiently expressive to force
the subset assumption, if desired, by including the following grants in $R$:
\[
\begin{array}{ll}
g = \forall x_1 \forall x_2 \forall x_3(\Permitted(x_1, \issue, x_2)\imp
\Permitted(x_1\union x_3, \issue, x_2))\\
g_i = \forall x_1 \forall x_2 (\Pr_i(x_1)\imp\Pr_i(x_1\union x_2)), \mathrm{ for}~ i = 1, \ldots, n,
\end{array}
\]
where $x_1$, $x_2$, and $x_3$ are variables of the appropriate sorts and $\Pr_1, \ldots, \Pr_n$ are the
properties in the language.  We now consider Examples~\ref{ex:lostL}, \ref{ex:terminate},
\ref{ex:terminateC}, and \ref{ex:terminateB}, in turn.

\begin{sloppypar}
The problem illustrated in Example~\ref{ex:lostL} lies in the definition of $R'$.
Recall that we define $\XA(e,L,R) = \XA(e,\emptyset,R')$.  Roughly speaking, $R'$
should consist of the set of grants in $R$ together with those issued by someone
who has the authority to do so.  In other words, $R'$ should be
$R\union \{g\mid \mbox{for some principal $p$ }, (p, g)\in L\mbox{ and }
\XProc(\Permitted(p, \issue, g), L, R) = \true\}$.  However, when computing
$\XProc(\Permitted(p, \issue, g), L, R)$, $\XA$ is given the argument $L-\{(p,g)\}$
rather than $L$.  Our solution is to do the ``right'' thing here, and compute
$\XProc(\Permitted(p,\issue,g),L,R)$.  But now we have to deal with the problem of
termination, since a consequence of our change is that $\XProc(e,L,R)$ terminates
only if the set $L = \emptyset$.  To ensure termination, we modify $\XA$ so that no
call is evaluated twice.  Specifically, the revised $\XA$ takes a fourth argument
$\scc$ that is the set of closed conditions that have been the first argument to a
previous call; $\XA(\cc, L, R, \scc)$ returns $\emptyset$ if $\cc\in\scc$.  Because
the revised $\XA$ calls $\XProc$, which calls $\XA$, we modify $\XProc$ to take
$\scc$ as its fourth argument.  A closed condition $\cc$ is implied by a set $L$ of
licenses and a set $R$ of grants if the modified $\XProc$ algorithm returns $\true$
on input $(\cc, L, R, \emptyset)$.  Pseudocode for the revised version of $\XProc$,
which we call $\XProcTwo$, and for the revised version of $\XA$, which we call
$\XATwo$, are given in Figures~\ref{tb:XrMLProc2} and \ref{tb:auth2}, respectively.
$\XProcTwo$ refers to the algorithm $\CMetTwo$, which is $\CMet$ modified to correct
the behavior seen in Example~\ref{ex:terminate} (discussed below).
\end{sloppypar}

\begin{figure*}[htb]
\begin{center}
\begin{tabular}{|l|}\hline
\underline{$\XProcTwo(\cc, L, R, \scc)$:}\\\\
$D := \XATwo(\cc, L, R, \scc)$\\
{\bf if} $\CMetTwo(\cd, L, \emptyset) = \true$ for a condition $\cd\in D$\\
{\bf then} return $\true$\\
{\bf else} return $\false$\\
\hline
\end{tabular}
\end{center}
\caption{The $\XProcTwo$ Algorithm}
\label{tb:XrMLProc2}
\end{figure*}

\begin{figure*}[htb]
\begin{center}
\begin{tabular}{|l|}\hline
\underline{$\XATwo(\cc, L, R, \scc)$:}\\\\
{\bf if} $\cc\in\scc$\\
{\bf then} return $\emptyset$\\
{\bf else}\\
\vtab $\scc' := \scc \union \{\cc\}$\\
\vtab $R' := R$\\
\vtab {\bf for} each license $(p, g) \in L$\\
\vtab\vtab {\bf if} $\XProcTwo(\Permitted(p, \issue, g), L, R, \scc') = \true$\\
\vtab\vtab {\bf then} $R' := R'\union \{g\}$\\
\vtab $D := \emptyset$\\
\vtab{\bf for} each grant $\forall x_1 \ldots \forall x_n(\cd_g\imp\cc_g) \in R'$\\
\vtab\vtab $D := D \union\{\cd\mid\cd_g\sigma = \cd\mbox{ and }  \cc_g\sigma = \cc,
\mbox{ for some closed substitution }\sigma\}$\\
\vtab return $D$\\
\hline
\end{tabular}
\end{center}
\caption{The $\XATwo$ Algorithm}
\label{tb:auth2}
\end{figure*}

The type of nontermination seen in Example~\ref{ex:terminate} occurs because $\XProc$ tries to verify
that a condition of the form $\Said(p, \cc)$ holds by checking if $\Said(p, \cc)$ holds.  To correct
the problem, we modify $\CMet$ to take a third argument $S$ that is the set of $\Said$ conditions that
have been the first argument to a previous call; that is, $S$ is the set of $\Said$ conditions that
are currently being evaluated.  If the revised $\CMet$ is called with a first argument $\cd$ that is
in $S$ (which means that the call was made when trying to determine whether $\cd$ holds), then the
algorithm returns $\false$, thereby halting the cycle.  Pseudocode for the revised version of $\CMet$,
which we call $\CMetTwo$, is given in Figure~\ref{tb:CMet2}.

\begin{figure*}[htb]
\begin{center}
\begin{tabular}{|l|}\hline
\underline{$\CMetTwo(\cd, L, S)$:}\\\\
{\bf if} $\cd = \true$ \\
{\bf then} return $\true$\\\\
{\bf if} $\cd = \cd_1\land\ldots\land\cd_n$\\
{\bf then} return $\bigwedge_{i = 1, \ldots, n}\CMetTwo(\cd_i, L, S)$\\ \\
{\bf if} $\cd = \Said(p, \cc)$ and $\cd\in S$\\
{\bf then} return $\false$\\\\
{\bf if} $\cd = \Said(p, \cc)$ and $\cd\not\in S$\\
{\bf then}\\
\vtab $S' = S\union \{\cd\}$\\
\vtab $R_p = \{g \mid \mbox{ for some principal }p', (p', g)\in L\mbox{ and } p'\in p\}$\\
\vtab $D := \{\cd'\mid \mbox{for some grant $\forall x_1 \ldots \forall x_n(\cd_g\imp\cc_g) \in R_p$
and}$\\
\vtab\phantom{ $D := $}$\mbox{closed substitution $\sigma$, }\cd_g\sigma = \cd'\mbox{ and }
\cc_g\sigma = \cc\}$\\
\vtab {\bf if} $\CMetTwo(\cd', L, S') = \true$ for a condition $\cd'\in D$\\
\vtab {\bf then} return $\true$\\
\vtab {\bf else} return $\false$\\
\hline
\end{tabular}
\end{center}
\caption{The $\CMetTwo$ Algorithm}
\label{tb:CMet2}
\end{figure*}

It is easy to see that the problem illustrated by Example~\ref{ex:terminate} does not
occur during the execution of $\CMetTwo$.  Moreover, the following theorem shows that
$\CMetTwo$ is correct in the sense that every execution of $\CMet$ and $\CMetTwo$
have the same input/output behavior on the inputs for which both executions terminate
and, if an execution of $\CMet$ terminates for a particular input $(\cd, L)$, then
some execution of $\CMetTwo(\cd, L,\emptyset)$ terminates as well.
\pro\label{t:correct2}
For all closed conditions $\cd$ and sets $L$ of licenses,
\begin{itemize}
\item[(a)] every execution of $\CMet(\cd, L)$ that terminates returns the same output,
\item[(b)] every execution of $\CMetTwo(\cd, L, \emptyset)$ that terminates returns
the same output,
\item[(c)] if an execution of $\CMet(\cd, L)$ terminates by returning the truth value $t$,
then an execution of $\CMetTwo(\cd, L, \emptyset)$ terminates by returning $t$.
\end{itemize}
\epro

Now consider Examples~\ref{ex:terminateC} and \ref{ex:terminateB}.  To address the type of
nontermination seen in these examples, we might hope to find an algorithm $\XProcFour$ that
returns the same output as $\XProcTwo$ on inputs for which an execution of $\XProcTwo$
terminates and returns $\false$ on all other inputs.  Returning $\false$ when no execution
of $\XProcTwo$ terminates gives an intuitively reasonable answer; moreover, this approach
is essentially what is done in MPEG-21 REL (see Section~\ref{s:MPEG-21} for details).
Unfortunately, as we show shortly (see Theorem~\ref{t:undec2}) this approach will not work
in general; there is no algorithm $\XProcFour$ with these properties, since whether
$\XProcTwo$ terminates on a given input is undecidable.

Since we cannot ``fix'' $\XProcTwo$, the best we can do is define some restrictions such
that, if the restrictions hold for a particular query, then the problems seen in
Examples~\ref{ex:terminateC} and~\ref{ex:terminateB} do not occur for that query.  We now
describe some conditions that are sufficient and that we suspect often hold in practice.

To describe our approach for avoiding the problem seen in Example~\ref{ex:terminateC}, let
$g$ and $g'$ be the grants $\forall x_1\ldots\forall x_n(\cd_g\imp\cc_g)$ and
$\forall x_1\ldots\forall x_m(\cd_{g'}\imp\cc_{g'})$ respectively.  The license $(p,g)$
\emph{affects} the license $(p', g')$ if and only if there are closed substitutions $\sigma$
and $\sigma'$ such that a condition of the form $\Said(p'', \cc_g\sigma)$ is mentioned in
$\cd_{g'}\sigma'$ and $p\subseteq p''$.  For example, consider the license set
$L = \{(\const{Alice}, g_1), (\const{Amy}, g_2)\}$, where $g_1 = \pred{Smart}(\const{Bob})$
and $g_2 = \forall x(\Said(\const{Alice}, \pred{Smart}(x))\rimp\pred{Attractive}(x))$.  The
license $(\const{Alice}, g_1)$ affects the license $(\const{Amy}, g_2)$ because the conditions
are satisfied if $\sigma$ is a closed substitution and $\sigma'$ is a closed substitution such
that $\sigma'(x) = \const{Bob}$.  A set $L$ of licenses is \emph{hierarchical} if there exists
a strict partial order $\prec$ on the licenses in $L$ such that, for all license
$\ell, \ell'\in L$, if $\ell$ affects $\ell'$ then $\ell\prec\ell'$.  Continuing our example,
$L$ is hierarchical because the ordering $(\const{Alice}, g_1) \prec (\const{Amy}, g_2)$
satisfies the requirements.  Observe that no hierarchical license set includes the license
$(\const{Alice}, \Said(\const{Alice}, \cc)\rimp\cc)$ because this license affects itself.  The
license set in Example~\ref{ex:terminateC} is not hierarchical for essentially the same reason.
It is not hard to see that by restricting the set of queries $(\cc, L, R, \scc)$ to those in
which $L$ is hierarchical, we avoid the type of circularity that causes the problem seen in
Example~\ref{ex:terminateC}.  In the next result and elsewhere, we use $\card{X}$ to denote the
cardinality of a set $X$.

\pro\label{p:fixExB}
If $\cd$ is a closed condition, $L$ is a hierarchical set of licenses, $S$ is a set of closed
$\Said$ conditions, and $T$ is the call tree of an execution of $\CMetTwo(\cd, L, S)$, then the
height of $T$ is at most $2\card{L}+1$.
\epro

We further restrict the language to avoid the problem seen in Example~\ref{ex:terminateB}.
To understand our restriction, recall that $\XA(\cc,L,R)$ first extends $R$ to $R'$ by
adding all the grants that are issued by someone who has the authority to do so.  Since
all the grants in $R'-R$ are in $L$, the set $R'$ must be finite.  Then $\XA$ creates the
possibly infinite set $R_\Sigma$ consisting of all substitution instances of grants in
$R'$, and returns $\{\cd \mid \cd\imp \cc \in R_{\Sigma}\}$.  (For simplicity here, we
are assuming that $\XA$ does not use the subset assumption; the subset assumption does
not affect our discussion.)  Since $\XA$ considers only the grants in $R_{\Sigma}$ whose
conclusion matches the first input to $\XA$, we could certainly replace $R_{\Sigma}$ by
$R_{\Sigma}'$, where
$$
\begin{array}{ll}
R_\Sigma' = &\{\cd_g\sigma\imp\cc \mid \forall x_1 \ldots \forall x_n (\cd_g\imp\cc_g)
\in R', \sigma \mbox{ is a closed substitution, and }\cc_g\sigma = \cc\}.
\end{array}
$$
Because $\cc$ is closed, $R_\Sigma'$ is finite if, for every grant $g$ in $R'$, if the
condition of $g$ mentions a free variable $x$, then either $x$ ranges over a finite set
or $x$ appears in the conclusion of $g$.  Our solution is simply to restrict the language
so that every grant has this property.  Since, in our fragment, there are infinitely many
resources (grants) and only finitely many principles, this amounts to restricting the
language so that if $\forall x_1 \ldots \forall x_n (\cd_g\imp\cc_g)$ is a grant, then
every free variable of sort $\Rsrc$ that appears in $\cd_g$ also appears in $\cc_g$.
We call a grant \emph{restrained} if it has this property; we call a license $(p, g)$
restrained if $g$ is restrained.  Thus, for example,
$\forall x \forall y (\Said(\emptyset, \Permitted(x, \issue, y))\imp
\Permitted(\mathit{Alice}, \issue, y))$ is restrained, but neither
$$
\begin{array}{ll}
\forall y \forall z (\Said(\emptyset, \Permitted(\mathit{Alice}, \issue, y))\imp
\Permitted(\mathit{Alice}, \issue, z))
\end{array}
$$
nor the grant $\forall x (d(x) \rimp\cc)$ in Example~\ref{ex:terminateB} is restrained.
It is easy to see that, for all restrained grants
$g = \forall x_1 \ldots \forall x_n (\cd_g\imp\cc_g)$ and closed conclusions $\cc$, if $n$
is the number of primitive principals in the language and $|g|$ is the length of $g$, then
there are at most $n^{|g|}$ grants of the form $\cd_g\sigma\imp\cc_g\sigma$ such that
$\sigma$ is a closed substitution and $\cc_g\sigma = \cc$.  Thus, by considering only
restrained grants and licenses, we solve the problem raised in Example~\ref{ex:terminateB}.

\section{Formal Semantics}\label{s:semantics-xrml}
In this section we provide formal semantics for the XrML fragment described in
Section~\ref{s:XrMLsyntax}.
We show that the semantics is correct 
in the sense that 
it captures the output of the (corrected) query algorithm, $\XProcTwo$.
We then consider two, arguably more intuitive,
semantics and show that neither captures $\XProcTwo$..

\subsection{A Correct Translation}\label{s:correctSemantics}
To give formal semantics to our fragment, we translate licenses in the grammar to formulas
in a modal many-sorted first-order logic.
The logic has three sorts: $\Princ$, $\Rht$, and $\Rsrc$.  The vocabulary includes
the following symbols, where $\primitivePrinc$ is the application-provided set of primitive
principals and $\primitiveProp$ is the application-provided set of properties:
\begin{itemize}
\item a constant $p$ of sort $\Princ$ for every principal $p \in \primitivePrinc$;
\item a constant $\issue$ of sort $\Rht$;
\item a ternary predicate $\Permitted$ that takes arguments of sort $\Princ$, $\Rht$, and
$\Rsrc$;
\item a unary predicate $\Pr$ that takes an argument of sort $\Princ$ for each property
$\Pr \in \primitiveProp$;
\item a function $\union: \Princ\times\Princ\longrightarrow\Princ$;
\item a function $f_g: s_1\times\ldots\times s_n\longrightarrow\Rsrc$ for each grant $g$ in
the language; if $x_1, \ldots, x_n$ are the free variables in $g$, then $x_i$ is of sort
$s_i$, for $i = 1, \ldots, n$.  If $g$ is closed, then the corresponding function is a
constant that we denote as $c_g$; and
\item a modal operator $\Val$ that takes a formula as its only argument.
\end{itemize}
Intuitively, $\Pr(p)$ means principal $p$ has property $\Pr$, and $\Val(\phi)$ means formula
$\phi$ is valid.  Notice that every principal in the grammar corresponds to a term in the
language, because $\union$ is a function symbol.

The semantics of our language is just the standard semantics for first-order logic, extended
to deal with $\Val$.  We restrict attention to models for which $\union$ satisfies the
following standard properties:
\begin{itemize}
\item[U1.] $\forall x((x\union x) = x)$
\item[U2.] $\forall x_1 \forall x_2 ((x_1 \union x_2) = (x_2 \union x_1))$
\item[U3.] $\forall x_1 \forall x_2 \forall x_3 ((x_1 \union (x_2 \union x_3)) = ((x_1 \union x_2) \union x_3))$
\item[U4.] $\forall x((x\union \emptyset) = x)$
\end{itemize}
We call such models \emph{acceptable}.  $\Val(\phi)$ is true in a model $m$ if $\phi$ is true
in all acceptable models.  If a formula $\phi$ is true in all acceptable models, then we say
that $\phi$ is \emph{acceptably valid}.  Thus, $\Val(\phi)$ is true in an acceptable model iff
$\phi$ is acceptably valid.

The translation takes four finite sets as parameters.  They are a set $L$ of licenses, a set
$A$ of closed resources, a set $\scd$ of closed $\Said$ conditions, and a set $\scc$ of closed
conclusions.
Roughly speaking, $L$ is the set of licenses that have been issued
and $A$
is the set of resources that are relevant to a particular application
For all XrML queries, $\scd = \emptyset$ and $\scc = \emptyset$.
(The reader is encouraged to take $\scd = \scc = \emptyset$ when first
trying to understand the details of the semantics.)
The input parameter $\scd$
allows users to specify a set of $\Said$ conditions that do not hold, regardless of $L$.  We
also use the parameter to insure that the translation of a $\Said$ condition does not enter an
infinite loop.  The input parameter $\scc$ corresponds to the fourth argument of $\XProcTwo$.
(Recall that an XrML query asks if a conclusion $\cc$ follows from a set $L$ of licenses and
set $R$ of grants; the answer is ``yes'' if $\XProcTwo(\cc, L, R, \emptyset)$ returns $\true$.)
By including $E$, we can give a translation that
agrees with
the $\XProcTwo$ algorithm.
The translation is defined below, where $\tranwithE{s}$ is the translation of the string $s$ given
input $L$, $A$, $\scd$, and $\scc$.

\begin{itemize}
\item If $\Permitted(p, \issue, g)\in \scc$ or $(p, g)\not\in L$, then $\tranwithE{(p, g)} = \true$.
\item If $\Permitted(p, \issue, g)\not\in \scc$ and $(p, g)\in L$, then $\tranwithE{(p, g)} =
\Permitted(p, \issue, c_g) \rimp\tranwithE{g}$.  Note that we assume $g$ is closed, because this
assumption is built into $\XProc$.
\item $\tranwithE{(\cd_g\imp\cc_g)} = ((\bigwedge_{\cc\in\scc}\neg\Eql{\tranwithE{\cc}}
{\tranwithE{\cc_g}})\land \tranwithE{\cd_g})\rimp\tranwithE{\cc_g}$.
\item $\tranwithE{(\forall x \phi)} = \bigwedge_{t\in T}(\tranwithE{\phi[x/t])}$, where $T = A$ if
$x$ is of sort $\Rsrc$, and $T = P$ if $x$ is of sort $\Princ$.  (Recall that $P$ is the set of
principals.)
\item $\tranwithE{\true} = \true$.
\item If $\tranwithE{\Said(p, \cc)} \in S$, then $\tranwithE{\Said(p, \cc)} = \false$.
\item If $\tranwithE{\Said(p, \cc)} \not\in S$, then
$\tranwithE{\Said(p, \cc)} = \Val((\bigwedge_{g\in R_p}\transwithE{g}{L}{A}{\scd'}{\emptyset})\rimp
\transwithE{\cc}{L}{A}{\scd'}{\emptyset}),$ where 
$R_p = \{g \mid (p', g)\in L \mbox{ for a }p'\in p\}$
and $\scd' = S \union\{\Said(p, \cc)\}$.
\item $\tranwithE{(\cd_1\land \cd_2)} = \tranwithE{\cd_1}\land \tranwithE{\cd_2}$.
\item $\tranwithE{\Permitted(p, r, s)} = \Permitted(p, r, s^*)$, where $s^* = s$ if $s$ is a variable
of sort $\Rsrc$, $s^* = c_s$ if $s$ is a closed grant, and $s^{*} = f_s(x_1, \ldots, x_n)$ if $s$ is an
open grant with free variables $x_1, \ldots, x_n$.
\item $\tranwithE{\Pr(p)} = \Pr(p)$.
\item for every principal $p$, $\tranwithE{\{p\}} = p$.
\end{itemize}
This translation has two features that seem somewhat inelegant.  The
first is that, in dealing with 
a universal quantifier, variables are replaced by the
constants over which they range; the second is the use of the $\Val$
operator.  In the next section, we explain in more detail why we
translated in this way.  For now, we show that, in a precise sense, our
translation captures the intended interpretation of the language.

Note that $\tranwithE{\Said(p, \cc)}$ does not depend on $\scc$.  This matches our intuition that the
meaning of a $\Said$ condition depends only on what principals have said, rather than on what is
actually true.  By adding $\Said(p, \cc)$ to $S$, we ensure that the meaning of the condition does not
depend on itself.  Finally, observe that $\tranwithE{\Said(p, \cc)}$ is defined in terms of the
translation of potentially more complex expressions.  Nevertheless, the following result shows that
the translation is well defined.
\thm\label{t:trans1}
For all strings $s$ in the language and all finite sets $L$ of licenses,
$A$ of closed resources, $\scd$ 
of closed $\Said$ conditions, and $\scc$ of closed conclusions,
$\tranwithE{s}$ is well defined. 
\ethm

\begin{sloppypar}
We believe that our semantics captures the intended meaning of XrML expressions, as implied by the
specification.  To make this precise, we show that $\XProcTwo$ agrees with the semantics on all queries.
Specifically, we show that for all terminating executions $\EX$ of $\XProcTwo(\cc, L, R, \scc)$, $\EX$
returns $\true$ iff $\bigwedge_{\ell\in L}\tranwithEempty{\ell}\land \bigwedge_{g\in R}\tranwithEempty{g}
\rimp\tranwithEempty{\cc}$ is acceptably valid, where $A = A(\cc, L, R, \scc, \EX)$ is the set of
closed resources that appear in the first argument of a call to $\XProcTwo$, $\XATwo$, or $\CMetTwo$
during execution $\EX$.  Intuitively, $A$ is the set of resources relevant to answering the query
$(\cc, L, R, \scc)$.  For example, suppose that, during a particular execution $\EX$ of
$\XProcTwo(\cc, L, R, \scc)$,
$\CMetTwo(\Said(p, \Permitted(p', \issue, \Permitted(p'', \issue, g))), L, S)$ is called.  Then
$A(\cc, L, R, \scc, \EX)$ includes $\Permitted(p'', \issue, g)$ and $g$.  Notice that if $\EX$ is a
terminating execution, then $A(\cc, L, R, \scc, \EX)$ is finite.
\thm\label{t:correct1}
Suppose that $(\cc, L, R, \scc)$ is a query and $\EX$ is a terminating execution of
$\XProcTwo(\cc, L, R, \scc)$.  Then $\EX$ returns $\true$ iff
$$
\bigwedge_{\ell\in L}\tranwithEempty{\ell}\land \bigwedge_{g\in R}\tranwithEempty{g}\rimp
\tranwithEempty{\cc}
$$
is acceptably valid, where $A = A(\cc, L, R, \scc, \EX)$.
\ethm
\end{sloppypar}
\subsection{Two Alternative Translations}\label{s:temptingTrans}
We now discuss why we captured universal quantification by replacing
variables by constants and the need for the $\Val$ operator.  We do so
by giving two arguably more natural alternative translations that do not
have these ``features'', and showing where they go wrong.  While this
does not show that there is no correct translation that translates
universal quantification as universal quantification, and does not use
$\Val$, it does show why finding such a translation
is nontrivial.

For all strings $s$ in our fragment, let $s^{L, A, S, E}_1$ be a
translation of $s$, where $L$, $A$,
$S$, and $E$ are as defined in Section~\ref{s:correctSemantics}.  The
formula $s^{L, A, S, E}_1$ is 
identical to $s^{L, A, S, E}$ except that
$(\forall x \phi)^{L, A, S, E}_1 = \forall x(\phi^{L, A, S, E}_1)$.
Notice that the new translation often leads to more concise formulas and
does not depend on the input 
parameter $A$.  
Unfortunately, this translation does not interact well with our use of
$\Val$ when it comes to universally quantified formulas involving
$\Said$.
The following example shows
why we rejected this translation.

\xam
Suppose that Alice may issue any grant.  Alice issues the grants ``if I say some principal $p$ is
great, then $p$ is also good'', ``if I say Bob is good, then Charlie is great'', and ``Bob is great.''
Can we conclude that Charlie is good?

To answer our question using $\XProcTwo$, 
let $L = \{({\mathit Alice}, g_A), ({\mathit Alice}, g_B), ({\mathit
Alice}, g_C)\}$ and consider
\mbox{$\XProcTwo({\bf Good}({\mathit Charlie}),
\{({\mathit Alice}, g_A), L, R, \emptyset)$}, where
\[
\begin{array}{lll}
g_A &=& \forall x (\Said(\mathit{Alice}, {\bf Great}(x))\imp {\bf Good}(x)),\\
g_B &=& \Said(\mathit{Alice}, {\bf Good}(\mathit{Bob}))\imp {\bf Great}(\mathit{Charlie}),\\
g_C &=& {\bf Great}({\mathit Bob})\\
R &=& \{\forall x(\Permitted({\mathit Alice}, \issue, x))\}
\end{array}
\]
It is not hard to see that the algorithm returns $\true$ (i.e., Charlie is good), which is the
intuitively correct answer.  Roughly speaking, the algorithm deduces that Charlie is good if Alice
says he is great; Alice says Charlie is great if Alice says Bob is
good; Alice says Bob is good if
Alice says Bob is great; and Alice does indeed say Bob is great.

\begin{sloppypar}
To answer our question using the revised translation, we 
need to 
determine the
validity of the formula 
$$
(\bigwedge_{\ell\in L}{\ell}^{L, \emptyset, \emptyset, \emptyset}_1 \land
\forall x(\Permitted({\mathit Alice}, \issue, x))) \rimp {\bf
Good}(\mathit{Charlie}). 
$$
It is easy to see that this formula equivalent to 
$$(g_A)^{L, \emptyset, \emptyset, \emptyset}_1  \land (g_B)^{L,
\emptyset, \emptyset, \emptyset}_1 \land (g_C)^{L, \emptyset, \emptyset,
\emptyset}_1 \rimp   {\bf Good}(\mathit{Charlie}).$$
Clearly, $(g_C)^{L, \emptyset, \emptyset, \emptyset}_1 = 
(g_C)^{L, \emptyset, \emptyset, \emptyset} = 
{\bf Great}(\mathit{Bob})$.  In the original translation, we combine
${\bf Great}(\mathit{Bob})$ with $(g_A)^{L, \emptyset, \emptyset,
\emptyset}$ to conclude  
${\bf Great}(\mathit{Bob})$, then combine ${\bf Great}(\mathit{Bob})$ with 
$(g_A)^{L, \emptyset, \emptyset, \emptyset}$ to derive ${\bf
Good}(\mathit{Charlie})$; that is, the formula corresponding to the
query is valid under the original translation.  Unfortunately,
the latter two steps fail with the
revised translation.  As suggested above, the problem lies in the
interaction of quantified formulas and $\Said$ in $g_A$.  
Consider the first step.  Note that $(g_A)^{L, \emptyset, \emptyset,
\emptyset}$ is equivalent to a conjunction of which one conjunct is
$(\Said(\mathit{Alice}, {\bf Great}(\mathit{Bob}))\imp {\bf
Good}(\mathit{Bob}))$.  When combined with ${\bf Great}(\mathit{Bob})$,
we can indeed conclude ${\bf Good}(\mathit{Bob})$.  On the other hand,
with the revised translation, $(g_A)^{L, \emptyset, \emptyset,
\emptyset}_1$ is 
$$\forall x(\Val((g_B)^{L,
\emptyset, \emptyset, \emptyset}_1 \land (g_C)^{L, \emptyset, \emptyset,
\emptyset}_1 \rimp   {\bf Great}(\mathit{x})) \rimp {\bf Good}(x)).$$
The $\Val$ formula is vacuously false, so $(g_A)^{L, \emptyset, \emptyset,
\emptyset}_1$ is vacuosuly true, and does not help in concluding
${\bf Good}(\mathit{Bob})$.  Thus, the formula corresponding to the
query is not valid under the revised translation; we do not get the
intuitively correct answer.
\bbox
\end{sloppypar}
\end{example}

Next, suppose that we modify our original translation so that
the
$\Val$
operator
is not used.
In particular, we fix the input parameter $\scc$ to be the empty set and
remove the validity
operator from the translation of $\Said$ conditions.  For all strings $s$, let
$s^{L, A, S, \emptyset}_2$ be the translation of $s$ that is identical to
$s^{L, A, S, \emptyset}$ except that, if $s$ is of the form ``$\cd_g\imp\cc_g$'', then
$s^{L, A, S, \emptyset}_2 = \cd_{g2}^{L, A, S, \emptyset}\rimp\cc_{g2}^{L, A, S, \emptyset}$ and,
if $s$ is of the form $\Said(p, \cc)$ and $s\not\in S$, then
$\Said(p, e)^{L, A, S, \emptyset}_2 =
(\bigwedge_{g\in R_p} g^{L, A, S', \emptyset}_2)\rimp e^{L, A, S', \emptyset}_2$,
where $R_p = \{g\mid (p', g)\in L \mbox{ for a }p'\in p\}$ and $S' = S\union\{\Said(p,e)\}$.
Observe that every translated string is a variable-free formula in first-order logic.
The following example
illustrates a problem
with this translation.
Roughly speaking, the problem is that, according to
the translation, every statement that follows from the given licenses and grants is said by every
principal.
\xam
Suppose that Alice cheated on an exam and, if Alice admits that she cheated, then she is trusted.
Is Alice trusted?  Intuitively, the answer is ``no'' because Alice has not confessed.

\begin{sloppypar}
To answer the question using $\XProcTwo$, we execute
$\XProcTwo({\bf Trusted}({\mathit Alice}), \emptyset, R, \emptyset),$
where
$R = \{{\bf Cheated}({\mathit Alice}), {\bf Said}({\mathit Alice}, {\bf Cheated}({\mathit Alice}))\imp
{\bf Trusted}({\mathit Alice})\}$.  It is not hard to see that $\XProcTwo$ returns $\false$,
indicating that Alice is not trusted.  Specifically, the algorithm determines that Alice is trusted
 only if Alice said she cheated and Alice has not done this.
\end{sloppypar}

To answer the question using the revised translation, we determine the
validity of the formula 
$
({\bf Cheated}({\mathit Alice})\land ((\true\rimp {\bf Cheated}({\mathit Alice}))\rimp{\bf Trusted}({\mathit Alice})))\rimp{\bf Trusted}({\mathit Alice})
$.
Standard manipulations show that the formula is logically equivalent to
$
({\bf Cheated}({\mathit Alice})\land ({\bf Cheated}({\mathit Alice})\rimp
{\bf Trusted}({\mathit Alice})))\rimp{\bf Trusted}({\mathit Alice}),
$
which is valid. So, if we use the revised translation, we conclude that
Alice is trusted.
If we use the translation in Section~\ref{s:correctSemantics}, then we determine that
Alice is not trusted.  This is because $\Val$ ``isolates'' the $\Said$ condition
from the statements implied by the given grants and the issued licenses.
As a result,
${\bf Said}({\mathit Alice}, {\bf Cheated}({\mathit Alice}))$ holds only if 
the grants issued by Alice, in isolation, imply ${\bf Cheated}({\mathit
Alice})$; 
that is, ${\bf Said}({\mathit Alice}, {\bf Cheated}({\mathit Alice}))$
holds only if  
$\Val(\true\rimp{\bf Cheated}({\mathit Alice}))$ is $\true$. 
Since $\true\rimp{\bf Cheated}({\mathit Alice})$ is not an acceptably valid formula, 
we conclude that 
${\bf Said}({\mathit Alice}, {\bf Cheated}({\mathit Alice}))$ does not hold and, thus,
${\bf Trusted}({\mathit Alice})$ does not hold.
\exam

\section{Complexity}\label{s:trac}
To answer a query $(\cc, L, R, \scc)$, we need to determine whether an execution of
$\XProcTwo(\cc, L, R, \scc)$ returns $\true$.  We claimed earlier that the problem of answering queries
is, in general, undecidable.  We now formalize this claim.  Recall that a grant $g$ is restrained if
every variable of sort $\Rsrc$ mentioned in the antecedent of $g$ is mentioned in the conclusion of $g$.
We say that \emph{a grant $g$ is in a set $L$ of licenses} if $(p,g)\in L$ for some principal $p$.  A
grant $g$ is in $R\union L$, for some set $R$ of grants, if $g$ is in $R$ or $g$ is in $L$.
\thm\label{t:undec2}
Determining whether some execution of $\XProcTwo(\cc, L, R, \scc)$ returns $\true$ is undecidable for
the set of queries $(\cc, L, R, \scc)$ such that at most one grant in $R\union L$ is not restrained.
\ethm
Let $\cL_0$ be the set of queries $(\cc, L, R, \scc)$ such that every grant in $R\union L$ is restrained.
In this section, we examine the computational complexity of answering queries for fragments of $\cL_0$.

We first show that the problem of answering queries for the full language $\cL_0$ is NP hard for two
quite different reasons.  The first stems from the fact that, if there are $n$ primitive principals, we
can construct $2^n$ principals using the $\union$ operator.  The second is that, to answer a query, we
might need to determine if exponentially many closed $\Said$ conditions hold.

We use the following definitions to state our results.  $\cL_1$ is the set of queries that do not
mention the $\union$ operator.  A grant $g$ is \emph{$n$-restricted} if the number of variables of sort
$\Princ$ that are mentioned in the antecedent of $g$ and not in the conclusion of $g$ is at most $\numP$.
$\cL_2^{\numP}$ is the set of queries $(\cc, L, R, \scc)$ such that all grants in $R \union L$ are
$n$-restricted.  A call $\CMetTwo(\cd, L, S)$ is \emph{$h$-bounded} if the call tree for every execution
of $\CMetTwo(\cd, L, S)$ has height at most $\lenC$.  Note that Proposition~\ref{p:fixExB} shows that
if $L$ is a hierarchical set of licenses, then $\CMetTwo(\cd, L, S)$ is $(2\card{L}+1)$-bounded.
$\cL_3^{\lenC}$ is the set of queries $(\cc, L, R, \scc)$ such that if an execution of
$\XProcTwo(\cc, L, R, \scc)$ calls $\CMetTwo(\cd, L, S)$, then $\CMetTwo(\cd, L, S)$ is $h$-bounded.
The next result shows that deciding if at least one execution of $\XProcTwo$ returns $\true$ is hard, even
if we restrict to queries in $\cL_0$ that satisfy any two of the following: the union operator is not
mentioned (i.e., restrict to $\cL_1$), the query is $n$-restricted for some fixed $n$, or all calls made
during an execution of the query are $h$-bounded for some fixed $h$.  (We show shortly that the set of
queries in $\cL_0$ that satisfy all three restrictions is tractable.)

For a formula $\phi$, let $\len{\phi}$ be the length of $\phi$ when viewed as a string of symbols.  For a
set $S$, let $\len{S}$ be the length of $S$; that is $\len{S} = \Sigma_{s\in S} |s|$.  Finally, we
abbreviate $\primitivePrinc$, the set of primitive principals, as $P_0$.

\thm\label{t:NPHardAll}
The problem of deciding
if some
execution of $\XProcTwo(\cc, L, R, \scc)$ returns $\true$ for $(\cc,L,R, \scc)\in\cL_0\cap\cL\cap\cL'$
is  NP-hard for $\cL, \cL' \in \{\cL_1, \cL_2^{0}, \cL_3^2\}$.
\ethm

If we make all three restrictions (that is, restrict to queries in
$\cL_0 \inter \cL_1 \inter \cL_2^n \inter \cL_3^h$, for some fixed $n$ and $h$), then determining whether
a query returns $\true$ is decidable in polynomial time.  However, as we might expect in light of
Theorem~\ref{t:NPHardAll}, the degree of the polynomial depends on $n$ and $h$, and the polynomial involves
constants that are exponential in $n$ and $h$.  Note that, for queries in
$\cL_0 \inter \cL_1 \inter \cL_2^n \inter \cL_3^h$, all executions of $\XProcTwo$ terminate and return the
same answer.  Termination is fairly easy to show since every call tree of an execution of
$\XProcTwo(\cc, L, R, \scc)$ has a finite branching factor if $(\cc, L, R, \scc) \in \cL_0$, and has finite
height if $(\cc, L, R, \scc)\in \cL_3^h$.  The fact that all executions of $\XProcTwo(\cc, L, R, \scc)$
return the same output for all queries
$(\cc, L, R, \scc) \in \cL_0 \inter \cL_1 \inter \cL_2^n \inter\cL_3^h$ follows easily from
Proposition~\ref{t:correct2}(b).

\thm\label{t:NPHardAll1}
For fixed $n$ and $h$, if $(\cc, L, R, \scc) \in\cL_0\cap\cL_1\cap\cL_2^{\numP}\cap\cL_3^{\lenC}$
then determining whether $\XProcTwo(\cc, L, R, \scc)$ returns $\true$ takes time
$O(\len{L}\len{E} + (\len{R} + \len{L})(\len{L}^{h-1}(\len{L} + \len{R}+ \len{\cc})^2))$.
\ethm
\noindent The big-O notation is hiding some rather complex (and uninformative) terms that are functions of
$n$ and $h$; we spell these out in the appendix.

In practice, we believe that queries are often in $\cL_0$ and, as shown in Proposition~\ref{p:fixExB}, if
we restrict to queries where the set $L$ of licenses has size at most $h$ and is hierarchical (which we
expect in practice will often be the case), than all call trees that arise are guaranteed to have height
at most $2h+1$.  Thus, in practice, we expect that we can restrict to queries in $\cL_2^n$ and $\cL_3^h$
for relatively small values of $n$ and $h$.  Moreover, even for larger values of $n$ and $h$ (say, as
large as 10), as long as the union operator does not appear, we expect that queries can be answered
efficiently, because the upper bound is quite conservative.

How reasonable is it to restrict to queries in $\cL_1$ that do not mention the $\union$ operator?  We
believe that XrML without the $\union$ operator is sufficiently expressive for many applications.  To
examine the effect of not using the $\union$ operator, note that principals appear as the first argument
in a license, in a $\Said$ condition, and in a conclusion.
\begin{itemize}
\item According to the XrML documentation, the license $(\{p_1,\ldots, p_n\}, g)$ is an abbreviation for
the set of licenses $\{(p, g)\mid p\in \{p_1, \ldots, p_n\}\}$.  It follows that we can restrict the
first argument of licenses to primitive principals and variables without sacrificing any expressive
power.  (In fact, we can restrict the first argument of licenses to only primitive principals, because
$\XProc$ assumes that if $(p, g)$ is a license in $L$, then $p$ is variable-free.)
\item We
can replace all conditions of the form
$\Said(\{p1, \ldots, p_n\},\cc)$, where $p_1, \ldots, p_n$ are primitive principals, by a condition
$\Said(\{p_1, \ldots, p_n\}^*,\cc)$, where $\{p_1, \ldots, p_n\}^*$ is a new primitive principal, and then
expand the set $L$ of issued licenses by adding a new license $(\{p_1, \ldots, p_n\}^*, g)$ for every
license $(p, g)$ already in $L$, where $p \in \{p_1, \ldots, p_n\}$.  It is not hard to show that this
results in at most a quadratic increase in the number of grants.  Thus, as long as the first argument to
$\Said$ is variable-free, we can express it without using $\union$.
\begin{sloppypar}
\item To understand the impact of our restriction on conclusions, we need to consider the meaning of
statements such as $\mathit{\bf Trust}(\mathit\{Alice, Bob\})$ and
$\Permitted(\mathit\{Alice, Bob\}, issue, g)$.  According to the XrML document,
$\mathit{\bf Trust}(\mathit\{Alice, Bob\})$ means Alice and Bob together (i.e., when viewed as a single
entity) is trusted; $\Permitted(\mathit\{Alice, Bob\}, issue, g)$ means Alice and Bob is permitted to issue
$g$.  However, the XrML document does not explain precisely what it means for Alice and Bob to be viewed as
a single entity.  Indeed, it seems to treat this notion somewhat inconsistently (recall the inconsistent use
of the subset assumption).  There are other difficulties with sets.  Notice that if
$\{\mathit{Alice}, \mathit{Bob}\}$ is permitted to issue a grant, then presumably $g$ holds if
$\{\mathit{Alice}, \mathit{Bob}\}$ issues $g$.  However, according to the XrML documentation, the license
$(\{\mathit{Alice}, \mathit{Bob}\}, g)$ is simply an \emph{abbreviation} for the set of licenses
$\{(\{\mathit{Alice}\}, g), (\{\mathit{Bob}\}, g)\}$.  So it is unclear whether a principal that is not a
singleton can issue a license.  Furthermore, if principals that are not singletons can issue grants and
$\{\mathit{Alice}, \mathit{Bob}\}$ is permitted to issue a grant $g$, then it seems reasonable to conclude
that $g$ holds if $g$ is issued by both Alice and Bob, but it is not clear whether $g$ holds if it is issued
by only Alice (or by only Bob).
\end{sloppypar}

There may well be applications for which these notions have an obvious and
clear semantics.  But we suspect that such applications typically include
only a relatively small set of groups of interest.
In that case, it may be possible to simply take these groups
to be new primitive principals, and express the relationship between the
group and its elements in the language.  (This approach has the added
advantage of forcing license writers to be clear about the semantics of
groups.)
\end{itemize}
In short, we are optimistic that many applications do not need the union function.

\section{The Entire XrML Language}\label{s:core}
XrML has several components that are not in our fragment.  Most have been excluded simply for ease of
exposition.
That is, our work can be extended in a straightforward way to a
much larger fragment of XrML.
In this section we list the main omissions, briefly discussing each one.
Giving formal semantics to the entire XrML language remains an open problem.

\begin{itemize}
\item XrML supports \emph{patterns}, where a pattern restricts the terms over which a variable
ranges. For example, if the variable $x$ is restricted to the pattern ``ends in Simpson'', then
$x$ ranges over the terms that meet this syntactic constraint (e.g., $x$ ranges over
$\{Homer Simpson, Marge Simpson, \ldots\}$).
Our semantics includes the patterns that correspond to properties in our fragment.
Continuing the example, we could capture the pattern ``ends in Simpson''
by having the property  $\mathit{\bf Simpson}$ in the
language and having the set of grants determine which terms have the property.

XrML
also
allows a pattern to be a set of patterns.  We can express a set of patterns as a
conjunction of patterns.  Since we can express conjunctions of
properties in our fragment, we can
also capture sets
of the corresponding patterns.

Patterns can be written in any language that the writer chooses.  The default is to
write patterns as XPath expressions.  First-order logic is not
well-suited to capturing
XPath expressions; the situation may be even worse with other languages.
Therefore we do not believe our semantics can be easily extended to
include all patterns.
The significance of this limitation is not yet clear.

\item XrML supports \emph{delegable grants}.  A delegable grant $g$ can be viewed as a conjunction of a
grant $g'$ in our fragment and a set $G$ of grants that, essentially, allow other principals to issue $g'$.
For example, the delegable grant ``Doctor Alice may view Charlie's medical file and she may also give the
right to view the file
to her colleague, Doctor Bob'' can be viewed as the conjunction of the grant
``Doctor Alice may view Charlie's medical file'' and the grant ``Alice is permitted to issue the grant
`Doctor Bob may view Charlie's medical file'\, ''.

The XrML specification
also
supports more general types of delegation.  For
example, in XrML, we can
say ``Doctor Alice may view Charlie's medical file and may delegate this
right to anyone under
any condition that she specifies.''
The extent to which our semantics can capture delegation, as defined in the XrML specification,
is an open problem.
\item XrML supports \emph{grantGroups}, where a grantGroup is a set of grants.  We can extend our syntax
to support grantGroups by closing the set of grants (as currently defined) under the union operator.
Note that our proposed treatment of grantGroups is quite similar to our current treatment of principals.
\item XrML has variables that range over conditions.  It is
not clear how this capability
is intended to be used in practice.  Our hope is that the practical
applications will translate easily
to our fragment.  Examining this issue is left as an open problem.
\begin{sloppypar}
\item XrML includes rights, resources, and conditions that are not in our fragment.  There should be no
difficulty in extending our translation to handle these new features, and proving an analogue of
Theorem~\ref{t:correct1}.  But we might not be able to answer queries in the extended language.  The
problem is that XrML allows resource terms to be formed by applying functions other than $\union$.  For
example, MPEG-21 REL extends XrML by defining a \emph{container} resource that is a sequence of resources.
This naturally translates to a function $\mbox{container:}\Rsrc\times\Rsrc\longrightarrow\Rsrc$, so that
the container $\langle s_1, s_2, s_3\rangle$ is translated as
$\mbox{container}(s_1, \mbox{container}(s_2, s_3))$.  Allowing such functions makes the problem of deciding
if a conclusion follows from a set of XrML licenses and grants undecidable, for much the same reason that
the validity problem for negation-free Datalog with function symbols is undecidable \cite{NS}.
\end{sloppypar}
\item XrML allows an application to define additional principals, rights, resources, and conditions within
the XrML framework.  Obviously, we cannot analyze terms that have yet to be defined; however, we do not
anticipate any difficulty in extending the translation to deal with these terms and getting an analogue of
Theorem~\ref{t:correct1}.
\item XrML allows licenses to be encrypted and supports abbreviations via the \emph{Inventory} component.
However, the XrML procedure for determining if a permission follows from a set of licenses assumes that all
licenses are unencrypted and all abbreviations have been replaced by the statements for which they stand.
In other words, these features are engineering conveniences that are not part of understanding or reasoning
about licenses.
\end{itemize}

\section{Negation}\label{s:extend}
We believe that many license writers will find it important to deny permissions explicitly and to state
conclusions based on whether a permission is granted, denied, or neither granted nor denied by a particular
principal.  For example, Alice's mother might want to say ``Alice is not permitted to enter the adult
website'', a teacher might want to say ``if the university does not object, then Alice is permitted to audit
the class'', and a lawyer might want to say ``if the hospital permits an action that the government forbids,
then the hospital is not compliant''.

\begin{sloppypar}
We can write these statements in XrML by using special ``negated predicates''.  For example, we can write
$\pred{Prohibited}(\const{Alice}, \const{enter}, \const{adult~website})$ to capture
``Alice is not permitted to enter the adult website'%
\footnote{Since XrML allows the application to define only additional principals, rights, resources, and
conditions, we cannot add $\pred{Prohibited}$ to XrML without extending the framework, but the extension is
so minor that we ignore it here; moreover, there are no implications as far as complexity goes.},
$\pred{NotSaid}(\const{University}, \pred{Prohibited}(\const{Alice}, \const{audit}, \const{class}))$
to capture ``the university does not say that Alice is not permitted to audit the class'' (i.e., the
university does not object to Alice auditing the class), and $\pred{NotCompliant}(\const{Hospital})$ to
capture ``the hospital is not compliant''.  We remark that this approach of using ``negated predicates'' has
appeared before in the literature \cite{JSS,MS04}; it is essentially the technique used by XACML \cite{XACML},
another popular license language.
\end{sloppypar}

Adding negated predicates to XrML is straightforward; reasoning about statements in the extended language is
not.  One problem is that we have to handle statements that are intuitively inconsistent.  For example,
consider the grants $\Permitted(\const{Alice}, \issue, g)$ and $\pred{Prohibited}(\const{Alice}, \issue, g)$,
which say that Alice is permitted and prohibited to issue the grant $g$.  It is not clear what we should
conclude from these grants.  In particular, it is not clear if Alice should be allowed to issue $g$.  (The
languages that include negated predicates typically require the policy writer to specify how inconsistencies
should be resolved.)

Other problems arise if we extend XrML so that the set of conditions includes $\Pr(p)$ and $\pred{NotPr}(p)$,
in addition to $\Said(p,\cc)$ and $\true$.
\xam\label{ex:diff1}
Suppose that a company allows employees to access their server and allows nonemployees access if they sign a
nondisclosure agreement.  If Alice cannot prove that she is an employee, can she still get access to the
server by signing a nondisclosure agreement?  Intuitively, she should be able to, because Alice is either an
employee, in which case she has permission, or she is not an employee, in which case she still has permission
because she signed the waiver.  However, if we express the query in the obvious way (using negated predicates),
then Alice is not permitted, because
\[
\begin{array}{l}
\pred{SignedWaiver}(\const{Alice})\land
\forall x (\pred{Employee}(x) \rimp\Permitted(x, \const{access}, \const{server}))\land \\
\forall x (\pred{NotEmployee}(x)\land\pred{SignedWaiver}(x) \rimp
\Permitted(x, \const{access}, \const{server}))\rimp\\
\Permitted(\const{Alice}, \const{access}, \const{server})
\end{array}
\]
is not valid.
\exam

To address the unintuitive behavior shown in Example~\ref{ex:diff1}, we could replace the negated predicates
by a negation operator, which is the standard approach in logic.  Let XrML$^\neg$ be XrML extended so that
the set of conditions includes  $\neg\Said(p, \cc)$ as well as $\Said(p,\cc)$, and the set of conclusions
includes $\neg\Pr(p)$ and $\neg\Permitted(p, r, s)$, as well as $\Pr(p)$ and $\Permitted(p,r,s)$.  There is
no problem extending the semantics of XrML to XrML$^\neg$.  Moreover, by replacing $\pred{NotEmployee}$ in
Example~\ref{ex:diff1} by $\neg \pred{Employee}$, we get the intuitively correct answer.  The downside of
allowing negation is intractability.  Recall that $\cL_0\cap\cL_1\cap\cL_2^0\cap\cL_3^2$ is a small fragment
of XrML: the licenses in this fragment do not mention the $\union$ operator, every variable in the antecedent
of a grant appears in its conclusion, and the execution tree for all calls to $\CMetTwo$ has height at most
two.  Theorem~\ref{t:NPHardAll} shows that queries in $\cL_0\cap\cL_1\cap\cL_2^0\cap\cL_3^2$ are tractable;
however, as we now show, adding negation to this relatively small language makes it intractable.
\thm\label{t:ext1}
Let $(\cc, L, R, \scc)$ be a tuple in $\cL_0\cap\cL_1\cap\cL_2^0\cap\cL_3^2$ extended to include negated
$\Said$ conditions and negated conclusions.  The problem of deciding whether
$$
\bigwedge_{\ell\in L} \tranwithE{\ell} \land \bigwedge_{g\in R} \tranwithE{g}\rimp\tranwithE{\cc}
$$
is valid is NP-hard.  This result holds even if $\cc$, all of the licenses in $L$, and all of the
conclusions in $\scc$ are in XrML, all but one of the grants in $R$ is in XrML, and the one grant that is
in XrML$^\neg$ -- XrML is of the form $\forall x_1\ldots\forall x_n(\neg \cc)$.
\ethm

We are currently investigating whether there is a tractable fragment of XrML$^\neg$ that is sufficiently
expressive to capture the grants and licenses that are of practical importance.  We expect that some
ideas from our work on Lithium \cite{HW03} will prove useful in this regard.

\section{MPEG-21 REL}\label{s:MPEG-21}
MPEG-21 is an international standard that is based on XrML.  In \cite{HW04}, we give semantics to a beta
version of MPEG-21.  All of the problems discussed in Section~\ref{s:alg2} are present in the beta version.
We reported these issues to Xin Wang and Thomas DeMartini of the MPEG-21 working group before the final
version was released, and our concerns were addressed in the final version (although not exactly as
specified in Section~\ref{s:alg3}).

The key differences between XrML and MPEG-21 are as follows.
\begin{itemize}
\item MPEG-21
consistently makes the subset assumption;
a principal $\{p_1, \ldots, p_n\}$ has
all of the properties and permissions of principal $p_i$, for $i = 1, \ldots, n$.
\item A $\Said$ condition takes a \emph{trustRoot} $s$ and a conclusion $\cc$.  No definition of trustRoot
is given in the specification; rather, it is assumed that the application will associate with every
trustRoot $s$, set $L$ of licenses, and set $R$ of grants a set $G(s, L, R)$ of grants.  $\Said(s, \cc)$
holds if the set $L$ of issued licenses and $G(s, L, R)$ together imply $\cc$, where $R$ is the set of
grants that implicitly hold.
\item Rather than defining an algorithm, MPEG-21 says that $L$ and $R$ imply $\cc$ if there is a
\emph{proof tree} that shows the result holds.  Roughly speaking, a proof tree $t$ shows that $L$ and $R$
imply $\cc$ if (a) $t$ includes a grant $g$ that implies $\cc$ if certain conditions hold; (b) for each of
these conditions, $t$ includes a proof tree showing that the condition does, in fact, hold, and (c) either
$g$ is in $R$ or, for some principal $p$, $(p,g)$ is in $L$ and $t$ includes a proof tree showing that $p$
is permitted to issue $g$.
\end {itemize}
We believe that the translation and corresponding proof of correctness given in
Section~\ref{s:correctSemantics}
can be modified in a straightforward way to apply to MPEG-21.  If this is
indeed the case, then an appropriately modified $\XProcTwo$ can be used
to answer queries about licenses and
grants that are written in MPEG-21.

\section{Concluding Remarks}\label{s:concl}
XrML is a popular language that does not have formal semantics.  Since there are no formal semantics,
we cannot argue that the XrML algorithm is incorrect, but its behavior on certain input does seem
unreasonable.  To address the problem, we modified the algorithm, provided formal semantics for
an interesting fragment of XrML,
and showed that the modified algorithm
corresponds to our semantics in a precise sense.

We have examined only a fragment of XrML.  A key reason for XrML's popularity is that the framework
is extensible; applications can define new components (i.e., principals, rights, resources, and
conditions) to suit their needs.  We do not believe there will be be any difficulty in giving
semantics to the extended language.  The real question is whether we can find useful {\em tractable\/}
extensions.  As we have already seen, functions pose no semantic difficulties, but adding them makes
the problem of answering queries in XrML undecidable.  Another obvious and desirable feature is
negation.  Currently, XrML does not support negation in either the condition or conclusion of grants.
This is a significant expressive weakness.  Without negation, license writers cannot forbid an action
explicitly nor can they say that a conclusion holds if a permission is denied or unregulated by a
particular principal.  While it is easy to extend XrML to include negation, doing so without placing
further restrictions on the language makes it intractable.  We suspect
that we can use our earlier
work \cite{HW03} to find a fragment of XrML with negation that is tractable and substantially more
expressive.

Of course, it remains an open question whether XrML (or some extension
of it) is the ``best'' policy language to use to for rights managment
(and, more generally, trust management).  Many
languages have been proposed to do this, including XACML \cite{XACML},
ODRL \cite{ODRL}, numerous variants of Datalog
\cite{DeT,LGF03,LMW,MS04,Tre}, SPKI/SDSI \cite{HM01a,LM06,spki1,spki2},
and our own language Lithium \cite{HW03}.
As the references above indicate, a number of these have even been given
semantics using first-order or modal logic.
Comparing the strengths and weaknesses of all these approaches (and the
semantic methods used to capture them) remains an open direction for
future research.

Our work emphasizes the need for collaboration
between language developers and
the formal methods community.  Our analysis of XrML demonstrates that a
language without formal semantics is prone to ambiguities and inconsistencies, even if
that language is carefully crafted and reviewed by industry.  The good news is that collaborations
are possible.  The XrML developers that we contacted answered our questions and listened to our
concerns.  When they designed the next version of XrML, which is the ISO Standard MPEG-21 REL,
they did not make the same mistakes.

\subsection*{Acknowledgements}
Many thanks to Xin Wang and Thomas DeMartini, who answered our questions about the intended meaning
of various MPEG-21 components.

\appendix
\section{Proofs}
\opro{t:correct2}
For all closed conditions $\cd$ and sets $L$ of licenses,
\begin{itemize}
\item[(a)] every execution of $\CMet(\cd, L)$ that terminates returns the same output,
\item[(b)] every execution of $\CMetTwo(\cd, L, \emptyset)$ that terminates returns the same output,
\item[(c)] if an execution of $\CMet(\cd, L)$ terminates by returning the truth value $t$, then
an execution of $\CMetTwo(\cd, L, \emptyset)$ terminates by returning $t$.
\end{itemize}
\eopro
\prf
Parts (a) and (b) are immediate from the description of the $\CMet$ and $\CMetTwo$.
To prove part (c), say that a call tree for $\CMet(\cd,L)$ is \emph{non-repeating}
if it is not the case that there exists a path $p$ in the call tree and two nodes
$n_1$ and $n_2$ on the path such that both nodes are labeled by the same call to
$\CMet$.  If $\CMet(\cd, L)$ terminates, then it has a finite call tree.  Moreover,
it is easy to see that if there is a finite call tree for $\CMet(\cd,L)$, then
there is a nonrepeating call tree:  If there is a call to $\CMet(\cd',L')$ at two
nodes on a path, we simply replace the subtree below the first call to
$\CMet(\cd',L')$ by the subtree below the last call to $\CMet(\cd',L')$.  A
non-repeating call tree for $\CMet(\cd, L)$ is essentially a call tree for
$\CMetTwo(\cd, L, \emptyset)$; the same calls are made at every step (the third
component has to change appropriately).
\eprf

For the proofs of Proposition~\ref{p:fixExB} and Lemma~\ref{complexityestimate}, we
rely on the observation that, if $T$ is the call tree for an execution of
$\CMetTwo(\cd, L, S)$, then $T$ can be viewed as an and-or tree, where a node labeled
$\CMetTwo(\cd', L, S')$ is an $\andNd$ node if $\cd'$ is a conjunction with at least
two conjuncts, an $\orNd$ node if $\cd'$ is a $\Said$ condition and
$\CMetTwo(\cd', L, S')$ makes at least one recursive call, and a leaf if $\cd'$ is
$\true$ or if $\cd'$ is a $\Said$ condition and $\CMetTwo(\cd', L, S')$ makes no
recursive calls.  For future reference, note that each node in $T$ can be assigned a
truth value in an obvious way.  An $\andNd$ node is assigned ``true'' if all its
children are; an $\orNd$ node is assigned ``true'' if at least one child is; a leaf
labeled $\CMetTwo(\true, L, S')$ is assigned ``true''; and a leaf labeled
$\CMetTwo(\Said(p,\cc), L, S')$ is assigned ``false''.

\opro{p:fixExB}
If $\cd$ is a closed condition, $L$ is a hierarchical set of licenses, $S$ is a set
of closed $\Said$ conditions, and $T$ is the call tree of an execution of
$\CMetTwo(\cd, L, S)$, then the height of $T$ is at most $2\card{L}+1$.
\eopro
\prf
Because $L$ is hierarchical, there exists a strict partial order $\prec$ on licenses
such that, if $\ell$ and $\ell'$ are licenses in $L$ and $\ell$ affects $\ell'$, then
$\ell\prec\ell'$.  A node $v$ in $T$ is a \emph{$\nandNd$} node if $v$ is an $\orNd$
node or a leaf.  It follows from the description of $\CMetTwo$ that every $\andNd$
node has at least two children and every child of an $\andNd$ node is a $\nandNd$ node.
So, if a path in $T$ from the root to a leaf has $n$ $\nandNd$ nodes, then that path
has at most $2n$ total nodes; thus, it suffices to show that every path in $T$ has at
most $\card{L}+1$ $\nandNd$ nodes.  If $L = \emptyset$, then it is immediate from the
description of $\CMetTwo$ that $T$ has height at most $1$.  Suppose that $L\ne\emptyset$.
Then, for every path $t$ in $T$, either $t$ includes at most $2$ $\nandNd$ nodes, in
which case $t$ mentions at most $\card{L} + 1$ $\nandNd$ nodes, or $t$ includes $2$
$\nandNd$ nodes $v_i$ and $v_j$ such that an $\orNd$ node precedes $v_i$, which precedes
$v_j$, and no $\orNd$ node is between $v_i$ and $v_j$.  If $v_i$ has a label of the form
$\CMetTwo(\cd_i, L, S_i)$ and $v_j$ has a label of the form $\CMetTwo(\cd_j, L, S_j)$,
then it follows from the description of $\CMetTwo$ that there are licenses $(p_i, g_i)$
and $(p_j, g_j)$ in $L$ and closed substitutions $\sigma_i$ and $\sigma_j$ such that the
antecedent of $g_i$ under $\sigma_i$ mentions $\cd_i$; the antecedent of $g_j$ under
$\sigma_j$ mentions $\cd_j$; and $(p_j, g_j)$ affects $(p_i, g_i)$.  Thus,
$(p_j, g_j) \prec (p_i, g_i)$.  It follows that $t$ has at most $\card{L}+1$ $\nandNd$
nodes.
\eprf

\dfn
Suppose that $(\cc, L, R, \scc)$ is a query,
$\EX$ is an execution of $\XProcTwo(\cc, L, R, \scc)$, and
$A = A(\cc, L, R, \scc, \EX)$.
Define
\[
\begin{array}{lll}
\EStar(\cc, L, R) &=& \{\Permitted(p, \issue, g)\mid (p,g)\in L\}\union\{\cc\}.\\
\SStar(\cc, L, R, \scc, \EX) &=& \{\Said(p, \pred{Pr}(p'))\mid p, p'\in P\mbox{ and }
\pred{Pr}\in\primitiveProp\}\union\\
&&\{\Said(p, \Permitted(p', \issue, g))\mid p, p'\in P\mbox{ and }g\in A\}.
\end{array}
\]
\edfn

\othm{t:trans1}
For all strings $s$ in the language and all finite sets $L$ of licenses, $A$ of closed
resources, $\scd$ of closed $\Said$ conditions, and $\scc$ of closed conclusions,
$\tranwithE{s}$ is well defined.
\eothm
\prf
Let $S_L$ be the set of $\Said$ conditions that are mentioned in issued grants; that is,
$\Said(p, \cc)\in S_L$ iff there is a license $(p', g) \in L$ such that $g$ mentions
$\Said(p, \cc)$.  Let $S_s$ be the set of $\Said$ conditions mentioned in $s$.  Finally,
let $S_{L,s} = S_L\union S_s$.  We define a lexicographic order on the tuples $(s, S)$
such that $(s, S) < (s', S')$ iff either (a) $\card{S_{L,s}-S} < \card{S_{L,s}-S'}$ or
(b) $\card{S_{L,s}-S} = \card{S_{L,s}-S'}$ and $\len{s}<\len{s'}$.  The proof is by
induction on this ordering.  If $\card{S_{L,s}-S} = 0$ and $\len{s}=1$, then
$\tranwithE{s} = s$, so the translation is well defined.  The inductive step is trivial
except when $s = \Said(p, \cc)$ and $s\not\in S$.

Suppose that $s$ is of the form $\Said(p, \cc)$ and $s\not\in S$.  Recall that
$$
\tranwithE{\Said(p, \cc)} = \Val(\bigwedge_{g\in R_p}\transwithE{g}{L}{A}{\scd'}{\emptyset}
\rimp \transwithE{\cc}{L}{A}{\scd'}{\emptyset}),
$$
where $R_p = \{g \mid (p', g)\in L \mbox{ for a }p'\in p\}$ and
$\scd' = S \union\{\Said(p, \cc)\}$.  Because $L$ is a finite set, $R_p$ is a finite set
and because $\cc$ is a conclusion, $\transwithE{\cc}{L}{A}{\scd'}{\emptyset}$ is well
defined.  So, to prove that $\tranwithE{\Said(p, \cc)}$ is well defined, it suffices to
show that $\transwithE{g}{L}{A}{\scd'}{\emptyset}$ is well defined for all $g\in R_p$.
Suppose that $s\not\in S_L$.  Then $\card{S_{L,g} - S'} = \card{S_{L} - S'}$ since
$S_{L, g} = S_L$; $\card{S_{L} - S'} = \card{S_L - S}$ since $s\not\in S_L$;
$\card{S_L - S} < \card{S_L - S \union \{s\}}$ since $s \not\in S_L$; and
$\card{S_L - S \union \{s\}} = \card{S_{L,s}- S}$ since $s\not\in S$.  So, putting
the pieces together, $\card{S_{L,g} - S'} < \card{S_{L,s}- S}$ and, by the induction
hypothesis, $\transwithE{g}{L}{A}{\scd'}{\emptyset}$ is well defined.
Suppose that $s\in S_L$.  Then $\card{S_{L,g} - S'} = \card{S_L - S'}$ since
$S_{L, g} = S_L$; $\card{S_L - S'} < \card{S_L - S}$ since $s\in S_L - S$; and
$\card{S_L - S} = \card{S_{L, s} - S}$ since $s \in L$.  Again, putting the pieces
together, $\card{S_{L,g} - S'} < \card{S_{L,s}- S}$, so
$\transwithE{g}{L}{A}{\scd'}{\emptyset}$ is well defined by the induction hypothesis.
\eprf

We next prove Theorem~\ref{t:correct1}.  We actually prove a stronger result, given as
Theorem~\ref{t:generalizeCorrect1}; Theorem~\ref{t:generalizeCorrect1}(c) is
Theorem~\ref{t:correct1}.  The next five lemmas 
provide a deeper understanding of the properties of the $\XProcTwo$,
$\XATwo$, and $\CMetTwo$ algorithms and the translation, and are used
in the proof of Theorem~\ref{t:generalizeCorrect1}.

\lem\label{l1}
Suppose that $(\cc, L, R, \scc)$ is a query.  Then during an execution $\EX$ of
$\XProcTwo(\cc, L, R, \scc)$
\begin{itemize}
\item[(a)] every call made to $\XProcTwo$, $\XATwo$, and $\CMetTwo$ takes $L$ as its
second argument;
\item[(b)] every call made to $\XProcTwo$ and $\XATwo$ takes $R$ as its third argument;
\item[(c)] if $\XProcTwo(\cc', L, R, \scc')$ is called, then $\cc'\in \EStar(\cc, L, R)$;
\item[(d)] if $\XATwo(\cc', L, R, \scc')$ is called, then $\cc'\in \EStar(\cc, L, R)$;
and
\item[(e)] if $\CMetThree(\cd, L, S)$ is called, then every conjunct of $\cd$ is in
$\SStar(\cc, L, R, \scc, \EX)\union \{\true\}$.
\end{itemize}
\elem
\prf
Parts (a) through (d) follow immediately from the descriptions of $\XProcTwo$, $\XATwo$,
and $\CMetTwo$.  For part (e), suppose that $\CMetThree(\cd, L, S)$ is called.  Because
$\cd$ is a closed condition, every conjunct of $\cd$ is either $\true$ or of the form
$\Said(p, \cc')$, where $p$ is a closed principal and $\cc'$ is a closed conclusion.
If $\cc'$ is of the form $\pred{Pr}(p')$, then $\Said(p, \cc')$ is clearly in
$\SStar(\cc, L, R, \scc, \EX)$.  Otherwise, $\cc'$ is of the form
$\Permitted(p', \issue, g)$.  Because $\cc'$ is an input to a call made during $\EX$ and
$g$ is mentioned in $\cc'$, $g\in A(\cc, L, R, \scc, \EX)$.
\eprf

\lem\label{l5}
Suppose that $(\cc, L, R, \scc)$ is a query such that $\cc\in\scc$, $A$ is a set of
closed resources, and $S$ is a set of closed $\Said$ conditions.  Then
$\bigwedge_{\ell\in L}\tranwithE{\ell}\land \bigwedge_{g\in R}\tranwithE{g}\rimp\tranwithE{\cc}$
is not acceptably valid (and hence not valid).
\elem
\prf
Let $m$ be an acceptable model that satisfies $\tranwithE{\cc'}$ iff $\cc' \ne \cc$.  Recall
that, for a grant $g = \forall x_1 \ldots \forall x_n (\cd_g\imp\cc_g)$, $\tranwithE{g}$ is a
conjunction of formulas of the form
$$
(\bigwedge_{\cc\in\scc}\neg\Eql{\tranwithE{\cc}}{\tranwithE{(\cc_g\sigma)}}\land
\tranwithE{(\cd_g\sigma)})\rimp\tranwithE{(\cc_g\sigma)},
$$
where $\sigma$ is a closed substitution.  If $\cc\in \scc$, then $m$ satisfies $\tranwithE{g}$
because, for all substitutions $\sigma$, either $\tranwithE{(\cc_g\sigma)} \ne \tranwithE{\cc}$,
in which case $m$ satisfies $\tranwithE{(\cc_g\sigma)}$, or
$\tranwithE{(\cc_g\sigma)} = \tranwithE{\cc}$, in which case
$\bigwedge_{\cc\in\scc}\neg\Eql{\tranwithE{\cc}}{\tranwithE{(\cc_g\sigma)}}$ is equivalent to
$\false$.  Since $m$ satisfies every grant, $m$ satisfies
$\bigwedge_{\ell\in L}\tranwithE{\ell}\land \bigwedge_{g\in R}\tranwithE{g}$.  By construction,
$m$ does not satisfy $\tranwithE{\cc}$, so $m$ does not satisfy
$\bigwedge_{\ell\in L}\tranwithE{\ell}\land \bigwedge_{g\in R}\tranwithE{g}\rimp\tranwithE{\cc}$.
\eprf

\lem\label{l3}
Suppose that $(\cc, L, R, \scc)$ is a query, $A$ is a set of closed resources, and $S$ is a set
of closed $\Said$ conditions.  Then
(a) $\tranwithE{\cc'} = \transwithE{\cc'}{L}{A}{S}{(\scc\union\{\cc\})}$ for every closed
conclusion $\cc'$ in the language,
(b) $\tranwithE{g}\rimp\transwithE{g}{L}{A}{S}{(\scc\union\{\cc\})}$ is valid for every grant
$g$ in the language, and
(c) $\tranwithE{\ell}\rimp\transwithE{\ell}{L}{A}{S}{(\scc\union\{\cc\})}$ is valid for every
license $\ell$ in the language.
\elem
\prf
Part (a) follows immediately from the translation.

For part (b), let $g = \forall x_1 \ldots \forall x_n (\cd_g\imp\cc_g)$.  It is easy to see that
$\tranwithE{g}\rimp\transwithE{g}{L}{A}{\scd}{(\scc\union\{\cc\})}$ is valid if, for all closed
substitutions $\sigma$,
$\transwithE{\cd_g\sigma}{L}{A}{\scd}{(\scc\union\{\cc\})}\rimp\tranwithE{\cd_g\sigma}$ is valid.
The latter statement holds because the translation of a condition does not depend on the final
input argument (i.e., the set of conditions), so
$\transwithE{\cd_g\sigma}{L}{A}{\scd}{(\scc\union\{\cc\})} =  \tranwithE{\cd_g\sigma}$.

For part (c), let $\ell = (p, h)$.  If $\Permitted(p, \issue, h)\in\scc\union\{\cc\}$ or
$(p, h)\not\in L$, then $\transwithE{\ell}{L}{A}{\scd}{(\scc\union\{\cc\})} = \true$, so
$\tranwithE{\ell}\rimp\transwithE{\ell}{L}{A}{\scd}{(\scc\union\{\cc\})}$ is valid.  If
$\Permitted(p, \issue, h)\not\in\scc\union\{\cc\}$ and $(p, h)\in L$, then
$\tranwithE{\ell} = \Permitted(p, \issue, c_h)\rimp\tranwithE{h}$ and
$\transwithE{\ell}{L}{A}{\scd}{(\scc\union\{\cc\})} = \Permitted(p, \issue, c_h)\rimp
\transwithE{h}{L}{A}{\scd}{(\scc\union\{\cc\})}$.  It follows that
$\tranwithE{\ell}\rimp\transwithE{\ell}{L}{A}{\scd}{(\scc\union\{\cc\})}$ is valid if
$\tranwithE{h}\rimp\transwithE{h}{L}{A}{\scd}{(\scc\union\{\cc\})}$ is valid.  The latter formula
is valid by part (b).
\eprf

\dfn
For a set $A$ of closed resources, an \emph{$A$-closed} substitution $\sigma$ is a closed substitution
such that, for all variables $x$ of sort $\Rsrc$, $\sigma(x)\in A$.
\edfn

\lem\label{l7}
Suppose that $G$ is a set of grants, $L$ is a set of licenses, $A$ is a set of closed resources,
$\scd$ is a set of closed $\Said$ conditions, $\scc$ is a set of grants, and $\cc$ is a closed
conclusion.  Then $\bigwedge_{g\in G}\tranwithE{g}\rimp\tranwithE{\cc}$ is acceptably valid iff
$\tranwithE{g}\rimp\tranwithE{\cc}$ is acceptably valid for some $g\in G$.  Moreover, for any
grant $g$, $\tranwithE{g}\rimp\tranwithE{\cc}$ is acceptably valid iff $\cc\not\in \scc$ and,
for some $A$-closed substitution $\sigma$, the formula $\tranwithE{\cd_g\sigma}$ is acceptably
valid and $\cc_g\sigma= \cc$.
\elem
\prf
We first show that $\bigwedge_{g\in G}\tranwithE{g}\rimp\tranwithE{\cc}$ is acceptably valid iff
$\tranwithE{g}\rimp\tranwithE{\cc}$ is acceptably valid for some $g\in G$.  The ``if'' direction
is trivial.  For the ``only if'' direction, suppose by way of contradiction that
$\bigwedge_{g\in G}\tranwithE{g}\rimp\tranwithE{\cc}$ is acceptably valid and
$\tranwithE{g}\rimp\tranwithE{\cc}$ is not acceptably valid for all $g\in G$.  Let $m$ be an
acceptable model such that, for all closed conclusions $\cc'$, $m$ satisfies $\tranwithE{\cc'}$
iff $\cc'\ne\cc$.  Since $\bigwedge_{g\in G}\tranwithE{g}\rimp\tranwithE{\cc}$ is acceptably valid,
there is a $g = \forall x_1\ldots\forall x_n(\cd_g\imp\cc_g) \in G$ such that $m$ does not satisfy
$\tranwithE{g}$.  By the translation, it follows that there is an $A$-closed substitution $\sigma$
such that $\cc_g\sigma\not\in \scc$, $\tranwithE{\cd_g\sigma}$ holds in $m$, and
$\cc_g\sigma \ne \cc$.  Because, for all conditions $\cd'$, $\tranwithE{\cd'}$ can be written as
$\Val(\phi)$ for an appropriate formula $\phi$, $\tranwithE{\cd_g\sigma}$ is acceptably valid since
it holds in an acceptable model.  It follows that $\tranwithE{g}\rimp\tranwithE{\cc}$ is acceptably
valid, which contradicts the assumption.

It remains to show that $\tranwithE{g}\rimp\tranwithE{\cc}$ is acceptably valid for a grant
$g = \forall x_1\ldots\forall x_n(\cd_g\imp\cc_g)$ iff $\cc\not\in \scc$ and, for some $A$-closed
substitution $\sigma$, the formula $\tranwithE{\cd_g\sigma}$ is acceptably valid and
$\cc_g\sigma= \cc$.  The ``if'' direction is immediate from the translation.  For the ``only if''
direction, suppose by way of contradiction that $\tranwithE{g}\rimp\tranwithE{\cc}$ is acceptably
valid and either $\cc\in \scc$ or, for each $A$-closed substitution $\sigma$, either
$\tranwithE{\cd_g\sigma}$ is not valid or $\cc_g\sigma\ne \cc$.  Let $m$ be the acceptable model
defined above; that is, for all conclusions $\cc'$, $m$ satisfies $\tranwithE{\cc'}$ iff
$\cc'\ne\cc$.  We can get a contradiction by showing that $m$ satisfies $\tranwithE{g}$.  If
$\cc \in \scc$, then $m$ satisfies $\tranwithE{g}$ since either $\cc_g\sigma \in \scc$ (because
$\cc_g\sigma = \cc$), or $\cc_g\sigma$ holds in $m$ (because $\cc_g\sigma \ne \cc$).  Otherwise,
by assumption, either $\tranwithE{\cd_g\sigma}$ is not acceptably valid or $\cc_g\sigma\ne \cc$,
for each $A$-closed substitution $\sigma$.  Note that, because $\tranwithE{\cd_g\sigma}$ (like
every formula of the form $\tranwithE{\cd}$ for some condition $\cd$) is equivalent to a formula
of the form $\Val(\phi)$, then if it is not acceptably valid, it is not true in any acceptable
model and, in particular, not in $m$.  It then easily follows from the translation that $m$
satisfies $\tranwithE{g}$.  This gives us the desired contradiction.
\eprf

\dfn
Let $(\cc, L, R, \scc)$ be a query, let $\EX$ be a terminating execution of
$\XProcTwo(\cc, L, R, \scc)$, and let $A = A(\cc, L, R, \scc, \EX)$.  Then
\[
\begin{array}{lll}
G(\cc, L, R, \scc, \EX)
&=& R\union \{h\mid\mbox{ for some principal $p$, }(p,h)\in L \mbox{ and }\\
&&((\bigwedge_{\ell\in L}\transwithE{\ell}{L}{A}{\emptyset}{(\scc\union\{\cc\})})\land
(\bigwedge_{g\in R}\transwithE{g}{L}{A}{\emptyset}{(\scc\union\{\cc\})}))\rimp
\Permitted(p, \issue, c_h)\\&&\mbox{ is acceptably valid}\}.
\end{array}
\]
\edfn

\lem\label{l2}
Suppose that $(\cc, L, R, \scc)$ is a query, $\EX$ is a terminating execution of
$\XProcTwo(\cc, L, R, \scc)$, and $A = A(\cc, L, R, \scc, \EX)$.  Then
$
\bigwedge_{\ell\in L}\transwithE{\ell}{L}{A}{\emptyset}{\scc}\land\bigwedge_{g\in R}
\transwithE{g}{L}{A}{\emptyset}{\scc}\rimp\transwithE{\cc}{L}{A}{\emptyset}{\scc}
$
is acceptably valid iff there is a grant $h\in G(\cc, L, R, \scc, \EX)$ such that
$\transwithE{h}{L}{A}{\emptyset}{\scc}\rimp\transwithE{\cc}{L}{A}{\emptyset}{\scc}$
is acceptably valid.
\elem
\prf
For the ``if'' direction, suppose that $h$ is a grant in $G(\cc, L, R, \scc, \EX)$ such that
$\transwithE{h}{L}{A}{\emptyset}{\scc}\rimp\transwithE{\cc}{L}{A}{\emptyset}{\scc}$ is
acceptably valid.  If $h\in R$, then
$
\bigwedge_{\ell\in L}\transwithE{\ell}{L}{A}{\emptyset}{\scc}\land\bigwedge_{g\in R}\transwithE{g}{L}{A}{\emptyset}{\scc}\rimp\transwithE{\cc}{L}{A}{\emptyset}{\scc}
$
is acceptably valid.  If $h\in G(\cc, L, R, \scc, \EX) - R$, then there is a principal $p$ such
that
\begin{itemize}
\item[(1a)] $(p, h)\in L$,
\item[(1b)] $\Permitted(p, \issue, h)\not\in\scc$, and
\item[(1c)]
$
\bigwedge_{\ell\in L}\transwithE{\ell}{L}{A}{\emptyset}{(\scc\union\{\cc\})}\land
\bigwedge_{g\in R}\transwithE{g}{L}{A}{\emptyset}{(\scc\union\{\cc\})}\rimp\Permitted(p, \issue, c_h)
$
is acceptably valid.
\end{itemize}
Let $\phi = \bigwedge_{\ell\in L}\transwithE{\ell}{L}{A}{\emptyset}{\scc}\land\bigwedge_{g\in R}
\transwithE{g}{L}{A}{\emptyset}{\scc}$.  It follows from (1a) that
$\phi\rimp\transwithE{(p,h)}{L}{A}{\emptyset}{\scc}$ is acceptably valid.  It follows from (1a), (1b),
and the translation that $\phi\rimp(\Permitted(p, \issue, c_h)\rimp\transwithE{h}{L}{A}{\emptyset}{\scc})$
is acceptably valid.  It follows from Lemma~\ref{l3} and (1c) that $\phi\rimp\Permitted(p, \issue, c_h)$
is acceptably valid, so $\phi\rimp\transwithE{h}{L}{A}{\emptyset}{\scc}$ is acceptably valid.  By
assumption $\transwithE{h}{L}{A}{\emptyset}{\scc}\rimp\transwithE{\cc}{L}{A}{\emptyset}{\scc}$ is
acceptably valid, so $\phi\rimp\transwithE{\cc}{L}{A}{\emptyset}{\scc}$ is acceptably valid.

For the ``only if'' direction, suppose that there is no grant $g\in G(\cc, L, R, \scc, \EX)$ such that
$\transwithE{g}{L}{A}{\emptyset}{\scc}\rimp\transwithE{\cc}{L}{A}{\emptyset}{\scc}$ is acceptably valid.
Let $m$ be an acceptable model that does not satisfy $\transwithE{\cc}{L}{A}{\emptyset}{\scc}$ and the
formulas in $\{\transwithE{\Permitted(p, \issue, h)}{L}{A}{\emptyset}{\scc}\mid (p,h)\in L,
\Permitted(p, \issue, h)\not\in\scc, \mbox{ and } h\not\in G(\cc, L, R, \scc, \EX)\}$.  Because $m$ does
not satisfy $\transwithE{\cc}{L}{A}{\emptyset}{\scc}$, it suffices to show that $m$ satisfies
$\bigwedge_{\ell\in L}\transwithE{\ell}{L}{A}{\emptyset}{\scc}\land\bigwedge_{g\in R}\transwithE{g}{L}{A}{\emptyset}{\scc}$.  We do this by showing that (1) $m$ satisfies
$\transwithE{(p, h)}{L}{A}{\emptyset}{\scc}$ for every license $(p, h)$ such that
$h\not\in G(\cc, L, R, \scc, \EX)$, and (2) $m$ satisfies $\transwithE{g}{L}{A}{\emptyset}{\scc}$ for
every grant $g\in G(\cc, L, R, \scc, \EX)$.

\begin{sloppypar}
For part (1), observe that if $\Permitted(p, \issue, h)\in \scc$ or $(p,h)\not\in L$, then
$\transwithE{(p,h)}{L}{A}{\emptyset}{\scc} = \true$, so $\transwithE{(p,h)}{L}{A}{\emptyset}{\scc}$ holds
in $m$.  If $\Permitted(p, \issue, h)\not\in\scc$ and $(p, h)\in L$, then
$\transwithE{(p, h)}{L}{A}{\emptyset}{\scc} = \Permitted(p, \issue, c_h)\rimp\transwithE{h}{L}{A}{\emptyset}{\scc}$
and, by construction, $m$ does not satisfy $\Permitted(p, \issue, c_h)$; so
$\transwithE{(p, h)}{L}{A}{\emptyset}{\scc}$ is again true in $m$.

For part (2), let $g = \forall x_1 \ldots\forall x_n(\cd_g\imp\cc_g)\in G(\cc, L, R, \scc, \EX)$,
and recall that $\transwithE{g}{L}{A}{\emptyset}{\scc}$ is the conjunction of formulas of the form
$$
(\bigwedge_{\cc\in\scc}\neg\Eql{\transwithE{\cc}{L}{A}{\emptyset}{\scc}}
{\transwithE{(\cc_g\sigma)}{L}{A}{\emptyset}{\scc}}\land
\transwithE{(\cd_g\sigma)}{L}{A}{\emptyset}{\scc})\rimp\transwithE{(\cc_g\sigma)}{L}{A}{\emptyset}{\scc},
$$
where $\sigma$ is an $A$-closed substitution.  Clearly, $m$ satisfies
$\transwithE{g}{L}{A}{\emptyset}{\scc}$ iff, for every $A$-closed substitution $\sigma$, $m$
satisfies $((\bigwedge_{\cc'\in\scc}\neg\Eql{\transwithE{\cc'}{L}{A}{\emptyset}{\scc}}
{\transwithE{(\cc_g\sigma)}{L}{A}{\emptyset}{\scc}}\land
\transwithE{(\cd_g\sigma)}{L}{A}{\emptyset}{\scc})
\rimp\transwithE{(\cc_g\sigma)}{L}{A}{\emptyset}{\scc}$.
It is easy to see that the latter statement holds if, for all $A$-closed substitutions $\sigma$,
either $\cc_g\sigma\in\scc$, $\transwithE{(\cd_g\sigma)}{L}{A}{\emptyset}{\scc}$ is not true in $m$,
or $\transwithE{(\cc_g\sigma)}{L}{A}{\emptyset}{\scc}$ is true in $m$.  We claim that this is indeed
the case.  To prove the claim, suppose by way of contradiction that $\cc_g\sigma\not\in\scc$,
$\transwithE{(\cd_g\sigma)}{L}{A}{\emptyset}{\scc}$ is true in $m$, and
$\transwithE{(\cc_g\sigma)}{L}{A}{\emptyset}{\scc}$ is not true in $m$.
Since $\transwithE{(\cc_g\sigma)}{L}{A}{\emptyset}{\scc}$ is not true in $m$, either
$\cc_g\sigma = \cc$ or $\cc_g\sigma \in \{\Permitted(p, \issue, h) \mid (p, h)\in L,
\Permitted(p, \issue, h)\not\in\scc, \mbox{ and } h\not\in G(\cc, L, R, \scc, \EX)\}$.
\end{sloppypar}

If $\cc_g\sigma = \cc$, then we claim that $\transwithE{g}{L}{A}{\emptyset}{\scc}
\rimp\transwithE{\cc}{L}{A}{\emptyset}{\scc}$ is acceptably valid.  To
see this note that 
$\transwithE{g}{L}{A}{\emptyset}{\scc}\rimp(\bigwedge_{\cc'\in\scc}\neg\Eql{\transwithE{\cc'}{L}{A}{\emptyset}{\scc}}{\transwithE{(\cc_g)\sigma}{L}{A}{\emptyset}{\scc}}\land \transwithE{(\cd_g\sigma)}{L}{A}{\emptyset}{\scc}\rimp\transwithE{(\cc_g\sigma)}{L}{A}{\emptyset}{\scc})$ is acceptably valid.
Since $\cc_g\sigma\not\in\scc$, the formula
$\bigwedge_{\cc'\in\scc}\neg\Eql{\transwithE{\cc'}{L}{A}{\emptyset}{\scc}}{\transwithE{(\cc_g\sigma)}{L}{A}{\emptyset}{\scc}}
= \true$; so,
$\transwithE{g}{L}{A}{\emptyset}{\scc}\rimp(\transwithE{(\cd_g\sigma)}{L}{A}{\emptyset}{\scc}\rimp\transwithE{(\cc_g\sigma)}{L}{A}{\emptyset}{\scc})$
is acceptably valid.  Since $\transwithE{(\cd_g\sigma)}{L}{A}{\emptyset}{\scc}$ is true in $m$ by
assumption, and, as we have observed, every formula of the form $\transwithE{\cd}{L}{A}{\emptyset}{\scc}$
is equivalent to $\Val(\phi)$ for some formula $\phi$, $\transwithE{(\cd_g\sigma)}{L}{A}{\emptyset}{\scc}$
is acceptably valid and, as a result,
$\transwithE{g}{L}{A}{\emptyset}{\scc}\rimp\transwithE{(\cc_g\sigma)}{L}{A}{\emptyset}{\scc}$ is
acceptably valid.  By assumption, $\cc_g\sigma = \cc$, so
$\transwithE{g}{L}{A}{\emptyset}{\scc}\rimp\transwithE{\cc}{L}{A}{\emptyset}{\scc}$ is acceptably valid.
Since $g\in G(\cc, L, R, \scc, \EX)$ and, by assumption, none of the grants in $G(\cc, L, R, \scc, \EX)$
imply $\transwithE{\cc}{L}{A}{\emptyset}{\scc}$, we have a contradiction.

\begin{sloppypar}
Finally, suppose that $\cc_g\sigma\ne\cc$ and $\cc_g\sigma=\Permitted(p,\issue, h)$, where $(p,h)\in L$,
$\Permitted(p, \issue, h)\not\in\scc$, and $h\not\in G(\cc, L, R, \scc, \EX)$.  We now prove that
$\transwithE{g}{L}{A}{\emptyset}{(\scc\union\{\cc\})}\rimp\Permitted(p, \issue, c_h)$ is acceptably valid,
so $h\in G(\cc, L, R, \scc, \EX)$, which contradicts the assumptions.  We begin by noting that
$\transwithE{g}{L}{A}{\emptyset}{(\scc\union\{\cc\})}\rimp((\bigwedge_{\cc'\in\scc\union\{\cc\}}\neg\Eql{\transwithE{\cc'}{L}{A}{\emptyset}{(\scc\union\{\cc\})}}
{\transwithE{(\cc_g\sigma)}{L}{A}{\emptyset}{(\scc\union\{\cc\})}}\land
\transwithE{(\cd_g\sigma)}{L}{A}{\emptyset}{(\scc\union\{\cc\})})\rimp
\transwithE{(\cc_g\sigma)}{L}{A}{\emptyset}{(\scc\union\{\cc\})})$ is acceptably valid.  By assumption,
$\cc_g\sigma \not\in\scc\union\{\cc\}$, so $\transwithE{g}{L}{A}{\emptyset}{(\scc\union\{\cc\})} \rimp
(\transwithE{(\cd_g\sigma)}{L}{A}{\emptyset}{(\scc\union\{\cc\})}\rimp\transwithE{(\cc_g\sigma)}{L}{A}{\emptyset}{(\scc\union\{\cc\})})$ is acceptably valid.  Since $\cc_g\sigma = \Permitted(p, \issue, h)$,
$\transwithE{\cc_g\sigma}{L}{A}{\emptyset}{(\scc\union\{\cc\})} =
\Permitted(p, \issue,c_h)$, so
$\transwithE{g}{L}{A}{\emptyset}{(\scc\union\{\cc\})}
\rimp(\transwithE{(\cd_g\sigma)}{L}{A}{\emptyset}{(\scc\union\{\cc\})}\rimp\Permitted(p,
\issue,c_h))$ is acceptably valid.  It remains to be shown that
$\transwithE{(\cd_g\sigma)}{L}{A}{\emptyset}{(\scc\union\{\cc\})}$ is
acceptably valid.  Because the translation of a condition does not
depend on the set of conclusions, it suffices to show that
$\transwithE{\cd_g\sigma}{L}{A}{\emptyset}{\scc}$ is acceptably valid.
But, as we observed above, this follows immediately from the assumption
that $\transwithE{\cd_g\sigma}{L}{A}{\emptyset}{\scc}$ is true in $m$. 
\end{sloppypar}
\eprf

\begin{sloppypar}
\thm\label{t:generalizeCorrect1}
Suppose that $(\cc, L, R, \scc)$ is a query, $\EX$ is a terminating execution of
$\XProcTwo(\cc, L, R, \scc)$, and $A = A(\cc, L, R, \scc, \EX)$.  Then for all calls of the form
$\CMetTwo(\cd, L, S)$, $\XATwo(\cc', L, R, \scc')$, or $\XProcTwo(\cc', L, R, \scc')$ made during
execution $\EX$, including the initial call,
\begin{itemize}
\item[(a)] $\CMetTwo(\cd, L, S)$ returns $\true$ iff $\transwithE{\cd}{L}{A}{S}{\scc'}$ is acceptably
valid, where $\scc'$ is an (arbitrary) set of closed conclusions;
\item[(b)] $\XATwo(\cc', L, R, \scc')$ returns the set $D$ of closed conditions, where
$D = \{\cd\mid \cc'\not\in\scc'$  and, for some grant
$\forall x_1\ldots\forall x_n(\cd_g\imp\cc_g)\in G(\cc', L, R, \scc', \EX)$ and closed substitution
$\sigma$, $\cd_g\sigma = \cd$ and  $\cc_g\sigma = \cc'\};$ and
\item[(c)] $\XProcTwo(\cc', L, R, \scc')$ returns $\true$ iff
$
\bigwedge_{\ell\in L}\transwithE{\ell}{L}{A}{\emptyset}{\scc'}\land \bigwedge_{g\in R}
\transwithE{\ell}{L}{A}{\emptyset}{\scc'}\rimp\transwithE{\cc'}{L}{A}{\emptyset}{\scc'}
$
is acceptably valid.
\end{itemize}
\ethm
\end{sloppypar}

\prf
We prove part (a) by induction on $\card{\SStar(\cc, L, R, \scc, \EX) - \scd}$, with a subinduction
on the structure of $\cd$.   Suppose that $\card{\SStar(\cc, L, R, \scc, \EX) - \scd} = 0$.  If
$\cd = \true$, then $\CMetTwo(\cd, L, S) = \true$ and $\transwithE{\cd}{L}{A}{S}{\scc'} = \true$.
Suppose that $\cd$ is of the form $\Said(p, \cc')$.  Then, by Lemma~\ref{l1},
$\cd \in \SStar(\cc, L, R, \scc)$.  By assumption, $\card{\SStar(\cc, L, R, \scc) - \scd} = 0$, so
$\cd\in \scd$.  It follows that $\CMetTwo(\cd, L, S) = \false$ and
$\transwithE{\cd}{L}{A}{S}{\scc'} = \false$.  Finally, if $d$ is a conjunction, then the result is
immediate from the induction hypothesis.  For the induction step, the argument used for the base
case applies if $\cd = \true$ or if $\cd$ is a conjunction of conditions.  Suppose that $\cd$ has
the form $\Said(p, \cc')$.  If $\cd\in \scd$, then $\CMetTwo(\cd, L, S) = \false$ and
$\transwithE{\cd}{L}{A}{S}{\scc'} = \false$.  If $\cd\not\in\scd$ then, by the description of
$\CMetTwo$, $\CMetTwo(\cd, L, S) = \true$ iff there is a grant
$g = \forall x_1\ldots\forall x_n(\cd_g\imp\cc_g) \in R_p$ and an $A$-closed substitution $\sigma$
such that $\CMetTwo(\cd_g\sigma, L, S\union\{\cd\}) = \true$ and $\cc_g\sigma = \cc'$.  By the
induction hypothesis, $\CMetTwo(\cd_g\sigma, L, S\union\{\cd\}) = \true$ iff
$\transwithE{\cd_g\sigma}{L}{A}{(S\union\{\cd\})}{\scc'}$ is acceptably valid.  By the translation,
the latter statement holds iff $\transwithE{\cd_g\sigma}{L}{A}{(S\union\{\cd\})}{\emptyset}$ is
acceptably valid.  So, by Lemma~\ref{l7}, $\CMetTwo(\cd, L, S) = \true$ iff
$(\bigwedge_{g\in R_p}\transwithE{g}{L}{A}{(S\union\{\cd\})}{\emptyset})\rimp
\transwithE{\cc}{L}{A}{(S\union\{\cd\})}{\emptyset}$ is acceptably valid.  It is immediate from the
translation that the latter statement holds iff $\transwithE{\cd}{L}{A}{S}{\scc'}$ is acceptably
valid.

We prove parts (b) and (c) by simultaneous induction on $\card{\EStar(\cc, L, R) - \scc'}$.  If
$\card{\EStar(\cc, L, R) - \scc'} = 0$, then $\cc'\in\EStar(\cc, L, R)$ by Lemma~\ref{l1}, so
$\cc'\in\scc'$.  Because $\cc'\in\scc'$, $\XATwo(\cc', L, R, \scc') = \emptyset$, so part (b) holds.
For part (c), $\XProcTwo$ begins by calling $\XATwo(\cc', L, R, \scc')$, which returns the empty
set, and then $\XProcTwo$ returns $\false$.  Since $\cc'\in\scc'$, it follows from Lemma~\ref{l5}
that
$
\bigwedge_{\ell\in L}\transwithE{\ell}{L}{A}{\emptyset}{\scc'}\land
\bigwedge_{g\in R}\transwithE{g}{L}{A}{\emptyset}{\scc'}\rimp\transwithE{\cc'}{L}{A}{\emptyset}{\scc'}
$
is not acceptably valid, so the invariant holds.

Now consider the inductive step.  For part (b), suppose that $\XATwo(\cc', L, R, \scc')$ is called
during the execution of $\XProcTwo(\cc, L, R, \scc)$.  If $\cc'\in\scc'$, then part (b) holds by
the same argument as in the base case.  If $\cc'\not\in\scc'$, then $\XATwo$ returns a set $D$ of
closed conditions such that $\cd\in D$ iff there is a grant
$\forall x_1 \ldots \forall x_n(\cd_h\imp\cc_h) \in S_L$ and a closed substitution $\sigma$ such
that $\cd_h\sigma = \cd$ and $\cc_h\sigma = \cc$, where
\[
\begin{array}{ll}
S_L =& R\union \{h\mid \mbox{for some principal $p$, } (p,h)\in L \mbox{ and, during execution $\EX$, }\\
&\XProcTwo(\Permitted(p, \issue, h), L, R, (\scc'\union\{\cc'\}))\mbox{ returns } \true\}.
\end{array}
\]
It clearly suffices to show that $S_L = G(\cc', L, R, \scc', \EX)$.  By Lemma~\ref{l1},
$\cc'\in\EStar(\cc, L, R)$ and, by assumption, $\cc\not\in\scc'$.  So it follows from the induction
hypothesis that
\[
\begin{array}{ll}
S_L = &R\union \{h\mid \mbox{for some principal $p$, } (p,h)\in L \mbox{ and }\\
&\bigwedge_{\ell\in L}\transwithE{\ell}{L}{A}{\emptyset}{(\scc'\union\{\cc'\})}\land
\bigwedge_{g\in R}\transwithE{g}{L}{A}{\emptyset}{(\scc'\union\{\cc'\})}\rimp
\Permitted(p, \issue, c_h)\mbox{ is }\\
&\mbox{acceptably valid}\},
\end{array}
\]
which is $G(\cc', L, R, \scc', \EX)$.

For part (c), observe that if $\cc'\in\scc'$ then we can use the same reasoning as in the base case
to show that the invariant holds.  If $\cc'\not\in\scc'$ then, during execution $\EX$,
$\XProcTwo(\cc', L, R, \scc')$ returns $\true$ iff there is a closed condition $\cd$ in the output
of $\XATwo(\cc', L, R, \scc')$ such that $\XProcTwo$ calls $\CMetTwo(\cd, L, \emptyset)$, which
returns $\true$.  By part (b), $\XATwo(\cc', L, R, \scc')$ returns a set of conditions that includes
$\cd$ iff there is a grant
$g = \forall x_1 \ldots \forall x_n(\cd_g\imp\cc_g)\in G(\cc', L, R, \scc', \EX)$ and a closed
substitution $\sigma$ such that $\cd_g\sigma = \cd$ and $\cc_g\sigma = \cc'$.  Moreover, since
$\CMetTwo(\cd, L, \emptyset)$ is called during execution $\EX$ of $\XProcTwo(\cc, L, R, \scc)$,
$\sigma$ is $A$-closed.  By part (a), $\CMetTwo(\cd, L, \emptyset) = \true$ iff
$\transwithE{\cd}{L}{A}{\emptyset}{\scc'}$ is acceptably valid.  So
$\XProcTwo(\cc', L, R, \scc')$ returns $\true$ iff there is a grant
$g = \forall x_1 \ldots \forall x_n(\cd_g\imp\cc_g)\in G(\cc', L, R, \scc', \EX)$ and an $A$-closed
substitution $\sigma$ such that $\transwithE{\cd}{L}{A}{\emptyset}{\scc'}$ is acceptably valid and
$\cc_g\sigma = \cc'$.  By assumption, $\cc'\not\in\scc'$; so, by Lemma~\ref{l7},
$\XProcTwo(\cc', L, R, \scc')=\true$ iff
$\transwithE{g}{L}{A}{\emptyset}{\scc'}\rimp\transwithE{\cc'}{L}{A}{\emptyset}{\scc'}$ is acceptably
valid for some $g\in G(\cc', L, R, \scc', \EX)$.  It follows from Lemma~\ref{l2} that the latter
statement holds iff
$
\bigwedge_{\ell\in L}\transwithE{\ell}{L}{A}{\emptyset}{\scc'}\land\bigwedge_{g\in R}\transwithE{g}{L}{A}{\emptyset}{\scc'}\rimp\transwithE{\cc'}{L}{A}{\emptyset}{\scc'}
$
is acceptably valid.
\eprf

\othm{t:undec2}
Determining whether some execution of $\XProcTwo(\cc, L, R, \scc)$ returns $\true$ is undecidable for
the set of queries $(\cc, L, R, \scc)$ such that at most one grant in $R\union L$ is not restrained.
\eothm
\begin{sloppypar}
\prf
We reduce the Post correspondence problem (PCP) \cite{PCP} to the problem of determining
whether some execution of
$\XProcTwo(\cc, L, R, \emptyset)$ returns $\true$ for a query $(\cc, L, R, \emptyset)$,
where all but one grant in $R\union L$ is restrained.
Let $\Sigma$ be an alphabet; let $s_1, \ldots, s_n$ and $t_1, \ldots, t_n$ be strings over $\Sigma$;
and, for all strings $s$ and $s'$, let $s\cdot s'$ be the concatenation of $s$ and $s'$.  We want to
determine if there are integers $i_1, \ldots, i_k \in \{1, \ldots, n\}$ such that
$s_{i_1}\cdot\ldots\cdot s_{i_k} = t_{i_1}\cdot\ldots\cdot t_{i_k}$.

To encode the problem as a query, assume that the language includes the primitive principal
$p_{\sigma}$ for each symbol $\sigma \in \Sigma$, the
primitive principal $p$, and the property $\pred{Pr}$.  For every string $s$ over $\Sigma$, define
a function $G_s$ from grants to grants by induction on the length of $s$.  If $s$ has length one
($s\in\Sigma$), then $G_s(g) = \Permitted(p_s, \issue, g)$.  If $s = \sigma s'$, then
$G_s = G_\sigma\circ G_{s'}$.  For all grants $g_1$ and $g_2$, define $G(g_1, g_2)$ to be the grant
$\Said(p, \Permitted(p, \issue, g_1))\imp\Permitted(p, \issue, g_2)$.

We claim that there are integers $i_1, \ldots, i_k \in \{1, \ldots, n\}$ such that
$s_{i_1}\cdot\ldots\cdot s_{i_k} = t_{i_1}\cdot\ldots\cdot t_{i_k}$ iff an execution of
$\XProcTwo(\pred{Pr}(p), L, R, \emptyset)$ returns $\true$, where
\[
\begin{array}{lll}
L &=& \{(p, \Permitted(p, \issue, G(G_{s_i}(\pred{Pr}(p)), G_{t_i}(\pred{Pr}(p)))))\mid i= 1, \ldots, n\}\union\\
&&\{(p, \forall x_1 \forall x_2(\Said(p, \Permitted(p, \issue, G(x_1, x_2)))\imp\\
&&\Permitted(p, \issue, G(G_{s_i}(x_1), G_{t_i}(x_2)))))\mid i = 1, \ldots, n\}
\end{array}
\]
and $R = \{\forall x(\Said(p, \Permitted(p, \issue, G(x, x)))\imp\pred{Pr}(p))\}$.

Recall that an execution of $\XProcTwo(\cc, L, R, \emptyset)$ returns $\true$ iff
an execution of $\XATwo(\cc, L, R, \emptyset)$ returns a set $D$ of conditions such
that an execution of $\CMetTwo(\cd, L, \emptyset)$ returns $\true$ for some
condition $\cd\in D$.  It is easy to see that every execution of
$\XATwo(\cc, L, R, \emptyset)$ returns the set
$D = \{\Said(p, \Permitted(p, \issue, G(g, g)))\mid g \mbox{ is a closed grant}\}$.
Moreover, if $\cd$ is of the form $\Said(p, \Permitted(p, \issue, G(g, g)))$, where
$g$ is a closed grant, then it is not hard to see that an execution of
$\CMetTwo(\cd, L, \emptyset)$ returns $\true$ iff there are integers
$i_1, \ldots, i_k \in \{1, \ldots, n\}$ such that
$g = G_{s_{i_1}}(G_{s_{i_2}}(\ldots G_{s_{i_k}}(\pred{Pr}(p))\ldots ))$ and
$g = G_{t_{i_1}}(G_{t_{i_2}}(\ldots G_{t_{i_k}}(\pred{Pr}(p))\ldots ))$.  The latter
statements holds iff there are integers $i_1, \ldots, i_k \in \{1, \ldots, n\}$ such
that $s_{i_1}\cdot\ldots\cdot s_{i_k} = t_{i_1}\cdot\ldots\cdot t_{i_k}$.
\eprf
\end{sloppypar}

\othm{t:NPHardAll}
The problem of deciding
if some execution of $\XProcTwo(\cc, L, R, \scc)$
returns $\true$ for $(\cc,L,R, \scc)\in\cL_0\cap\cL\cap\cL'$ is  NP-hard for
$\cL, \cL' \in \{\cL_1, \cL_2^{0}, \cL_3^2\}$.
\eothm
\prf
For the NP hardness results, it suffices to show that the problem of deciding whether
$\XProcTwo(\cc, L, R, \scc) = \true$ is NP-hard if (a)
$(\cc,L,R, \scc) \in \cL_0\inter\cL_2^{0}\inter \cL_3^2$,
(b) $(\cc,L,R, \scc) \in \cL_0\inter\cL_1 \inter \cL_3^2$, and
(c) $(\cc,L,R, \scc) \in \cL_0\inter\cL_1 \inter \cL_2^{0}$.

For part (a), we show that we can reduce the Hamiltonian path problem to the problem of determining whether
$\XProcTwo(\cc,L,R, \scc) = \true$, for some $(\cc, L, R, \scc) \in \cL_0\inter\cL_2^{0}\cap\cL_3^{2}$.  Given a graph $G(V,E)$, where $V = \{v_1, \ldots, v_n\}$, we take $v_1, \ldots, v_n$ to be primitive principles.  We also assume that the language has primitive properties {\bf Node}, {\bf Edge}, and {\bf Path}.  For each node $v \in V$, let $g_v$ be the grant ${\bf Node}(v)$ (recall that this is an abbreviation for $\true \imp {\bf Node}(v)$).  For each edge $e = (v,v') \in E$, let $g_{(v,v')}$ be the grant ${\bf Edge}(\{v,v'\})$ (recall that $\{v,v'\}$ is an abbreviation for $\{v\}\union\{v'\}$).
Finally, let $g$ be the grant $\forall x_1 \ldots \forall x_n (\cd_1\land \cd_2 \imp \mathit{\bf Path}(\{x_1, \ldots,x_n\}))$, where
\[
\begin{array}{lll}
\cd_1 &=& \bigwedge_{1\le i\le n} \Said(\mathit{Alice}, \mathit{\bf Node}(x_i)) \mbox{ and } \\
\cd_2 &=& \bigwedge_{1 \le i\le n-1}\Said(\mathit{Alice},
\mathit{\bf Edge}(\{x_i, x_{i+1}\})).
\end{array}
\]
Let $L = \{(\mathit{Alice}, g_v) \mid v \in V\} \union \{(\mathit{Alice}, g_e) \mid e \in E\}$ and let
$R = \{g\}$.  It is not hard to show that $\XProcTwo(\mathit{\bf Path}(\{v_1, \ldots, v_n\}), L, R, \emptyset) = \true$
iff $G$ has a Hamiltonian path. To see this, observe that $\XATwo(\mathit{\bf Path}(\{v_1, \ldots, v_n\}), L, R, \emptyset)$ returns $\{\cd_1\sigma \land \cd_2\sigma \mid \sigma(x_i) = v_\pi(i), i = 1, \ldots, n$, where $\pi$ is some permutation of $\{1, \ldots, n\}\}$.  The condition $\cd_2\sigma$ holds iff there is a path $x_1\sigma, \ldots, x_n\sigma$.  Thus, $\XProcTwo(\mathit{\bf Path}(\{v_1, \ldots, v_n\}), L, R, \emptyset) = \true$ iff there is a Hamiltonian path in $G$.  Moreover, it is clear that $(\mathit{\bf Path}(\{v_1, \ldots, v_n\}), L, R, \emptyset) \in \cL_0\inter\cL_2^{0}$ and it is not hard to see that $(\mathit{\bf Path}(\{v_1, \ldots, v_n\}), L, R, \emptyset) \in \cL_3^{2}$, because the antecedent of every issued grant is $\true$.

For part (b), we show that we can reduce the 3-satisfiability problem to the problem of determining whether
$\XProcTwo(\cc, L, R, \scc) = \true$, for $(\cc,L,R, \scc)\in \cL_0\inter\cL_1\cap\cL_3^2$.  Let $f = c_1\land\ldots\land c_n$ be a formula in propositional logic, where each $c_i$ is a clause with three disjuncts.  Let $q_1, \ldots, q_m$ be the primitive propositions mentioned in $f$.  We want to determine if $f$ is satisfiable.

\begin{sloppypar}
To encode the problem as an XrML query, suppose that $p_1, \ldots, p_n, p_t, p_f$ are distinct primitive principals, $\pred{Pr}$ is a property, and $x_1, \ldots, x_m$ are distinct variables of sort $\Princ$.  Let $g_0$ be a fixed closed grant.  Given principals $t_1, \ldots, t_m$, we define grants $g_1(t_1), \ldots, g_m(t_1, \ldots, t_m)$ inductively
as follows: $g_1(t_1)$ is the grant $\true \imp \Permitted(t_i, \issue, g_0)$ and, for $i = 2, \ldots, m$, $g_i$ is
the grant $\true \imp \Permitted(t_i, \issue, g_{i-1}(t_1, \ldots, t_{i-1}))$.  Let $\cc(t_1, \ldots, t_m)$ be the conclusion $\Permitted(t_m, \issue, g_{m-1}(t_1, \ldots, t_{m-1}))$.  For ease of exposition, let $\cc'$ be the conclusion $\cc(x_1, \ldots, x_m)$.  Let
$
L = \{(p_i, \forall x_1\ldots\forall x_m(e'[x_j/p_t])) \mid \mbox{$q_j$ is a disjunct of $c_i$}\} \union
\{(p_i, \forall x_1\ldots\forall x_m(e'[x_j/p_f])) \mid \mbox{$\neg q_j$ is a disjunct of $c_i$}\}
$
and let $R = \{\forall x_1 \ldots \forall x_m ((\bigwedge_{i = 1, \ldots, n}\Said(p_i, \cc'))\imp
\pred{Pr}(p_t)\}$.  We claim that $f$ is satisfiable iff
$\XProcTwo(\pred{Pr}(p_t), L, R, \emptyset) = \true$.  Note that
$(\pred{P}(p_t),L,R, \emptyset)\in \cL_1 \inter \cL_0\inter\cL_3^2$, since none of the grants mention
a variable of sort $\Rsrc$, the $\union$ operator is not mentioned in the query, and the antecedent of
every issued grant is $\true$.

To prove the claim, first note that $\XProcTwo(\pred{Pr}(p_t), L, R, \emptyset) = \true$ iff
$\bigwedge_{i = 1, \ldots, n}\Said(p_i, \cc')\sigma$ holds for some substitution $\sigma$.  It is not hard to see that if $\sigma$ exists, then $f$ is satisfied by the truth assignment that sets $q_i = \true$ if $\sigma$ sets $x_i$ to $p_t$, and sets $q_i$ to $\false$ otherwise.  Similarly, if $f$ is satisfied by a truth assignment $A$, then $\bigwedge_{i = 1, \ldots, n}\Said(p_i, \cc')\sigma$ holds for the substitution $\sigma$ that replaces $x_i$ by $p_t$ if $A$ assigns $x_i$ to $\true$, and replaces $x_i$ by $p_f$ otherwise.
\end{sloppypar}

For part (c), we show that we can reduce the 3-satisfiability problem to
the problem of determining whether $\XProcTwo(\cc, L, R, \scc) = \true$,
for $(\cc, L, R, \scc)\in \cL_0\inter\cL_1\cap\cL_2^{0}$.  As in part (b), let $f$
be the 3-CNF formula $c_1 \land \ldots \land c_n$, whose primitive
propositions are $q_1, \ldots, q_m$.  Define the condition $e(t_1,
\ldots, t_m)$ as in part (b); again, take $e'$ to be an abbreviation
for $e(x_1, \ldots, x_m)$.  Let $p_1', \ldots, p_m'$ be fresh
principals, distinct from $p_1, \ldots, p_n, p_f, p_t$.  We claim that
$f$ is satisfied iff $\XProcTwo(\cc(p'_1, \ldots, p'_m), L,  R,
\emptyset) = \true$, where

\[
\begin{array}{lll}
L &=& \{(p_i, \forall x_1 \ldots \forall x_m(\Said(p_{i+1},\cc'[x_j/p_t])\imp \cc'[x_j/p]))\mid \\
&&\vtab\vtab q_j \mbox{ is a disjunct of $c_i$, } p \ne p_f, i = 1, \ldots, n-1\}\\
&&\union \{(p_i, \forall x_1 \ldots \forall x_m(\Said(p_{i+1},\cc'[x_j/p_f])\imp \cc'[x_j/p]))\mid\\
&&\vtab \neg q_j \mbox{ is a disjunct of $c_i$, } p \ne p_t, i = 1, \ldots, n-1\}\\
&&\union \{(p_n, \forall x_1 \ldots \forall x_m(\cc'[x_j/p]))\mid\\
&&\vtab q_j \mbox{ is a disjunct of $c_n$ and } p \ne p_f, \mbox{ or }
\neg q_j \mbox{ is a disjunct of $c_n$ and } p \ne p_t\}\\
R &=& \{\Said(p_1, \cc(p'_1, \ldots, p'_m))\imp e(p'_1, \ldots, p'_m)\}.
\end{array}
\]

If $t_1, \ldots, t_m$ are variable-free principals, let $A(t_1, \ldots, t_m)$ be the set of all truth assignments to $q_1,\ldots, q_m$ such that $q_i$ is assigned $\true$ if $t_i = p_t$ and $q_i$ is assigned
$\false$ if $t_i = p_f$, for $i = 1,\ldots, m$.  (If $t_i \notin \{p_t, p_f\}$, then there are no constraints on $q_i$.)
Let $A_i(t_1, \ldots, t_m)$ be the set of all truth assignments to
$q_1,\ldots, q_m$ under which $c_i\land\ldots\land c_n$ is $\true$.  We
show by induction on $n-i$ that $A_i(t_1, \ldots, t_m)$ is nonempty iff
$\Said(p_i,\cc(t_1, \ldots, t_m))$ holds.  If $n-i =
0$, then $i = n$.  It is easy to see that $A_i(t_1, \ldots, t_m)$ is
nonempty iff, for some $j = 1, \ldots, m$, either $q_j$ is a disjunct of
$c_n$ and $t_j \ne p_f$, or $\neg q_j$ is a disjunct of $c_n$ and $t_j
\ne p_t$.  For the inductive step, suppose that $n-i > 0$.  Clearly,
$A_i(t_1, \ldots, t_m)$ is nonempty iff there is an assignment in
$A_{i-1}(t_1, \ldots, t_m)$ under which $c_i$ is $\true$.  If there is
at least one such assignment, then $A_{i-1}(t_1', \ldots, t_m')$ is
nonempty, where $t_1', \ldots, t_m'$ are variable-free principals such
that, for some $j \in \{1, \ldots, m\}$ and for all $i \ne j$, $t_i' =
t_i$ and either $q_j$ is a disjunct of $c_i$, $t_j \ne p_f$, and $t_j' =
p_t$, or $\neg q_j$ is a disjunct of $c_i$, $t_j \ne p_t$, and $t_j' =
p_f$.  It follows from the induction hypothesis that $\Said(p_{i-1},
e(t_1', \ldots, t_m'))$ holds and it follows from
$L$ that $\Said(p_{i}, e(t_1, \ldots, t_m))$ holds as well.  If there is no assignment in
$A_{i-1}(t_1, \ldots, t_m)$ under which $c_i$ is $\true$ then, for every disjunct $q_j$ in
$c_i$, $t_i = p_f$ and, for every disjunct $\neg q_j$ in $c_i$, $t_j = p_t$.  It follows that
$A_{i}(t_1, \ldots, t_m) = \emptyset$ and $\Said(p_i, e(t_1, \ldots, t_m))$ does not hold.

The desired result now follows quickly.  It is  easy to see that $\XProcTwo(\cc, L, R, \emptyset) = \true$ iff $\Said(p_1, \cc(p'_1, \ldots, p'_m))$ holds.  Since none of $p_1', \ldots, p_m'$ is $p_f$ or $p_t$, by definition, $A(p_1', \ldots, p_m')$ consists of all truth assignments.  Thus, by the induction argument, it follows that $\XProcTwo(\cc, L, R, \emptyset) = \true$ iff $f= c_1 \land \ldots \land c_n$ is satisfiable.  Moreover, it is easy to see that $(\cc, L, R, \emptyset) \in \cL_0\inter\cL_1 \inter \cL_2^{0}$, because
the query does not mention union and, for every variable $x$ mentioned in a grant $g$ that is in
$R \union L$, $x$ is mentioned in the conclusion of $g$.
\eprf

We next prove Theorem~\ref{t:NPHardAll1}, which considers the complexity
of determining whether $\XProcTwo(\cc, L, R, \scc)$ returns $\true$ for
$(\cc, L, R, \scc) \in \cL_0\inter\cL_1\inter\cL_2^n\inter\cL_3^h$.
In the statement of the theorem, we viewed $n$ and $h$ as constants.
In our proof, we treat them as parameters, so as to bring out their role.

\begin{figure*}[tb]
\begin{center}
\rotatebox{270}{\includegraphics[width=2.5in, height=4.5in]{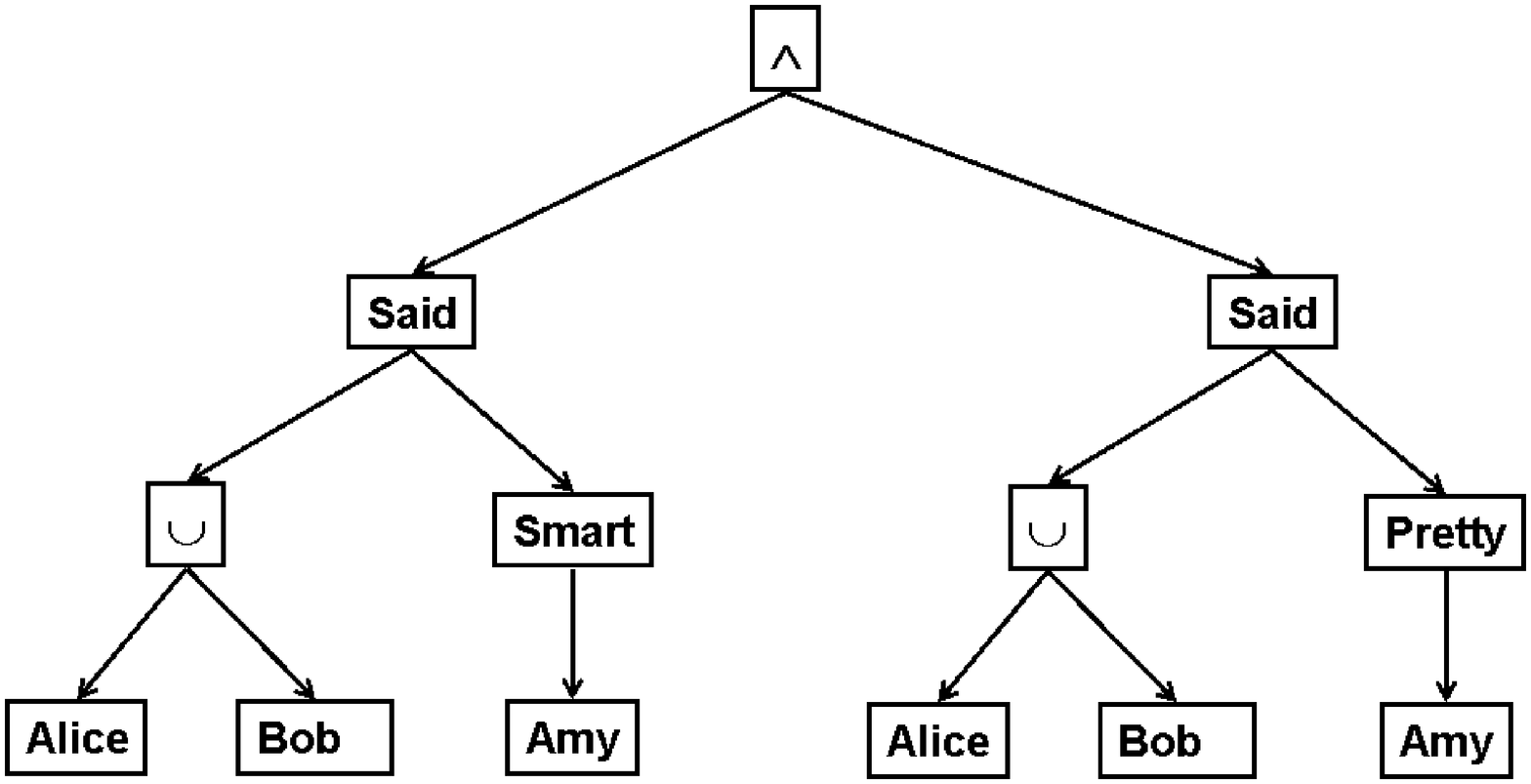}}
\end{center}
\caption{A tree representing $\Said(\{\const{Alice},
\const{Bob}\}, \pred{Smart}(\const{Amy}))\land \Said(\{\const{Alice},
\const{Bob}\}, \pred{Pretty}(\const{Amy}))$}
\label{tree}
\end{figure*}

\begin{figure*}[tb]
\begin{center}
\rotatebox{270}{\includegraphics[width=2.5in, height=4.5in]{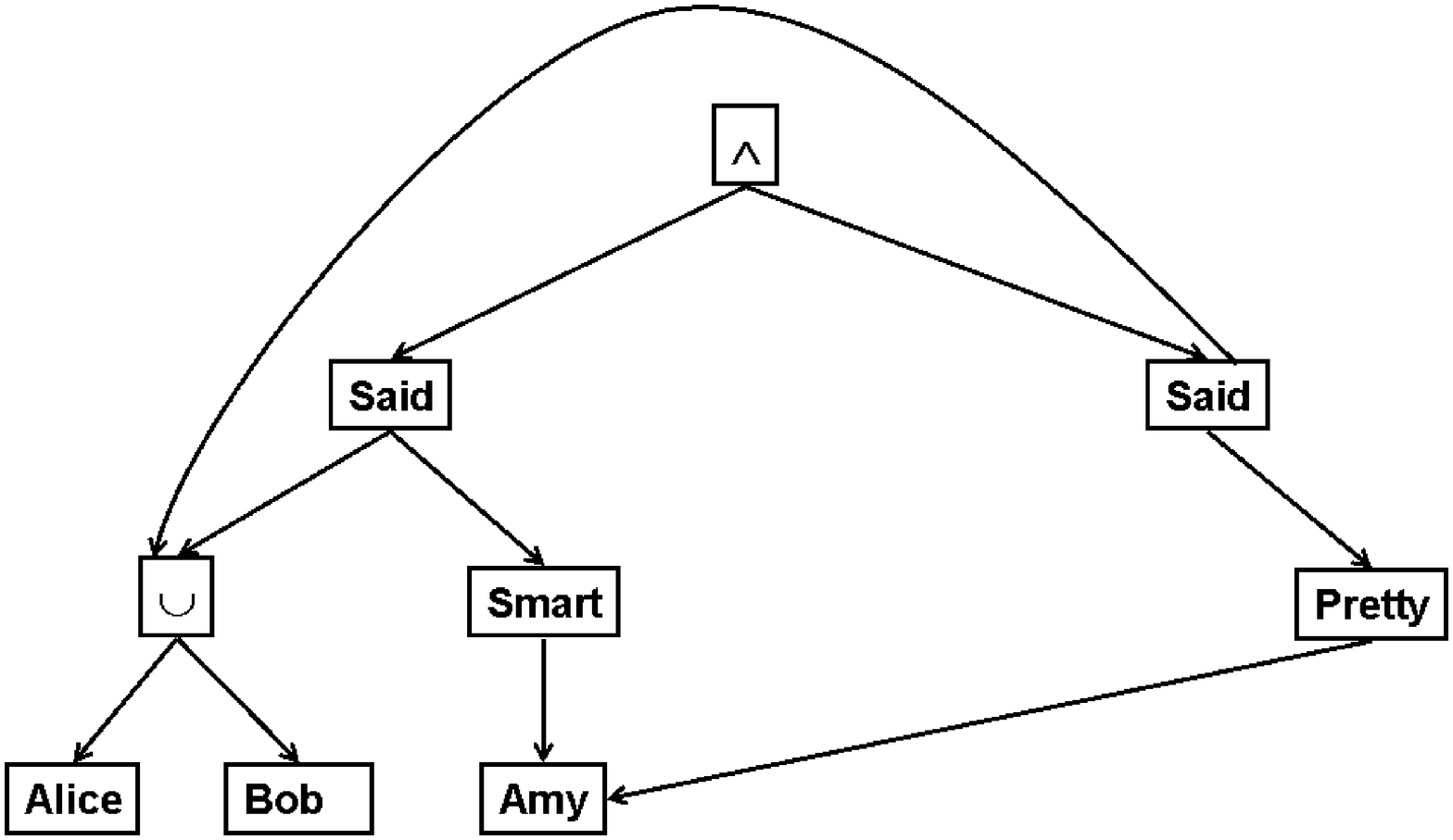}}
\end{center}
\caption{A dag representing $\Said(\{\const{Alice},
\const{Bob}\}, \pred{Smart}(\const{Amy}))\land \Said(\{\const{Alice},
\const{Bob}\}, \pred{Pretty}(\const{Amy}))$}
\label{dag}
\end{figure*}

To prove the theorem we need three preliminary lemmas.  The first uses the fact that, for every
condition $\cd$, there is a dag (directed acyclic graph) $G_d$ such that $G_{\cd}$ represents
$\cd$ and $G_{\cd}$ is no larger than $\cd$.  To make this precise, recall that $\len{s}$ is the
length of string $s$ when viewed as a string of symbols.  For ease of exposition, we assume that
each pair of parenthesis and set braces has length 2, and each comma has length 1.  For a graph
$G(V,E)$, let $\len{G} = \card{V} + \card{E}$.  It is easy to see that a condition $\cd$ can be
represented as a tree $T_d$, where $\len{T_d}\le\len{\cd}$.  For example, we can represent the
condition $\cd = \Said(\{\const{Alice}, \const{Bob}\}, \pred{Smart}(\const{Amy}))\land
\Said(\{\const{Alice}, \const{Bob}\}, \pred{Pretty}(\const{Amy}))$ as the tree $T_d$ shown in
Figure~\ref{tree}.
Note that $\len{d} = 27$ and, because the tree has $13$ nodes and $12$ edges, $\len{T_d} = 25$.
By ``merging'' identical subtrees, we can create a dag representation of $\cd$ that can be
substantially smaller than $\len{\cd}$.  Continuing our example, the dag $D_\cd$
in
Figure~\ref{dag} represents the condition
$\Said(\{\const{Alice}, \const{Bob}\}, \pred{Smart}(\const{Amy}))\land \Said(\{\const{Alice},
\const{Bob}\}, \pred{Pretty}(\const{Amy}))$ and $\len{D_\cd} = 19$.

\lem\label{l:boundD}
Suppose that $T$ is the call tree for an execution of $\CMetTwo(\cd, L, \emptyset)$;
every license in $L$ is restrained; the $\union$ operator is not mentioned in $\cd$
or in a grant in $L$; and $v$ is a node in $T$ with label $\CMetTwo(\cd', L, S)$.
If $G_{\cd}$ is a dag representing $\cd$, then there exists a dag $G_{\cd'}$
representing $\cd'$ such that $\len{G_{\cd'}} \le h\len{L} + \len{G_{\cd}}$, where
$h$ is the height of $T$.
\elem
\prf
Because $v$ is a node in $T$, there is a path $v_0, \ldots, v_k$ in $T$ such that
$v_0$ is the root of $T$ and $v_k = v$.  We prove by induction on $k$ that there
is a dag $G_{\cd'}$ representing $\cd'$ such that
$\len{G_{\cd'}} \le k\len{L} + \len{G_{\cd}}$.  Since $k \le h$ by assumption, it
easily follows that $\len{G_{\cd'}} \le h\len{L} + \len{G_{\cd}}$.

If $k=0$, then $v$ is the root of $T$, so $\cd' = \cd$.  If $k>0$, then $v$ is the
child of a node $v_{k-1}$.  Let $\CMetTwo(\cd'', L, S')$ be the label of $v_{k-1}$.
The proof is by cases on the structure of $\cd''$.  It follows from the description
of $\CMetTwo$ that $\cd''$ is not $\true$ because $\cd''$ is not a leaf in $T$.  If
$\cd''$ is a conjunction, then $\cd'$ is a conjunct of $\cd''$.  So the space needed
to represent $\cd'$ is less than the space needed to represent $\cd''$, thus the
result follows easily from the induction hypothesis.  Finally, if $\cd''$ has the
form $\Said(p, \cc)$, then it follows from the description of $\CMetTwo$ that there
is a license $(p,g)\in L$, where $g = \forall x_1\ldots\forall x_m(\cd_g\imp\cc_g)$,
and a closed substitution $\sigma$ such that $\cd' = \cd_g\sigma$ and $\cc_g\sigma =
\cc$.  A dag representing $\cd_g\sigma$ (i.e., $\cd'$) can be obtained by taking a
dag representing $\cd_g$ and replacing every variable $x$ by a dag representing
$\sigma(x)$.  Because every grant in $L$ is restrained, $g$ is restrained, so
$\sigma$ assigns every variable of sort $\Rsrc$ mentioned in $\cd_g$ to a term in
$\cc$.  Since $\sigma(x)$ is a subterm of $e$ or a primitive principal, given
a dag $G_{\cd_g}$ representing $\cd_g$ and a dag $G_{\cc}$ representing $\cc$, we
can construct a dag $G_{\cd'}$ representing $\cd'$ such that
$\len{G_{\cd'}} \le \len{G_{\cd_g}} + \len{G_{\cc}}$.  Since, for every condition
$\cd$, there is a tree representation of $\cd$ whose size is at most $\len{\cd}$,
there is a dag $G_{\cd_g}$ representing $\cd_g$ such that
$\len{G_{\cd_g}} \le \len{\cd_g}$.  Because $\cd_g$ is the antecedent of a grant in
$L$, $\len{\cd_g}<\len{L}$ so it follows that $\len{G_{\cd_g}} < L$.  Because $e$
is a subterm of $d'' = \Said(p,e)$, and by the induction hypothesis, there is a dag
$G_{d''}$ representing $d''$ such that $\len{G_{d''}} \le (k-1)\len{L} + \len{G_{\cd}}$,
there is surely a dag $G_e$ representing $e$ such that
$\len{G_e} \le (k-1)\len{L} + \len{G_{\cd}}$.  Putting this all together, it follows
that there is a dag $G_{d'}$ representing $d'$ such that
$\len{G_{d'}} \le k\len{L} + \len{G_{\cd}}$.
\eprf

\begin{sloppypar}
\lem\label{complexityestimate}
If $\CMetTwo(\cd, L, \emptyset)$ is $h$-bounded, the $\union$ operator is not mentioned in $\cd$ or
in a grant in $L$, $L$ is both restrained and $n$-restricted, and $G_{\cd}$ is a dag representing $\cd$,
then the output of $\CMetTwo(\cd, L, \emptyset)$ can be determined in time
$$
O(\max(\len{G_d}, \len{L}\len{P_0}^n) (\len{L}\len{P_0}^n)^{h-2}
(\len{L}\len{P_0}^n + (h\len{L} + \len{G_{\cd}})(h + \len{L}))).
$$
\elem
\prf
Let $T$ be the call tree for an execution of $\CMetTwo(\cd, L, \emptyset)$.
Our goal is to compute the truth value associated with the root of $T$, since
that truth value is the output of $\CMetTwo(\cd, L,\emptyset)$.

It is clear that once we have written the call tree, computing the truth value
of the root can be done in time linear in the number of nodes in the tree.  The
obvious way to construct the tree is to start at the root and, for each node $v$,
construct the successors of $v$ (if there are any).  In constructing the call
tree, we assume that the condition $\cd'$ and the elements of the set $S$ in a
node labeled $\CMetTwo(\cd', L, S)$ are described using the dags of
Lemma~\ref{l:boundD}.  Consider a node $v$ in $T$ that is labeled
$\CMetTwo(\cd', L, S)$ and is neither the root nor a leaf.  Since $v$ is not a
leaf, $\cd' \ne \true$.  If $\cd'$ is a conjunction, then a bound on the number
of conjuncts (and hence on the successors of the node) is $\len{L}$ since $\cd'$
is of the form $\cd_g\sigma$, where $\cd_g$ is the antecedent of a grant $g$ that
is in $L$, and $\sigma$ is a closed substitution.  It is easy to see that $\cd_g$,
and hence $\cd_g\sigma$, has at most $\len{L}$ conjuncts, and these can be
computed in time $O(\len{L})$.

Suppose that $\cd'$ is of the form $\Said(p,\cc)$.  If $\cd'\in S$, then $v$ is a leaf.
Since the height of $T$ is at most $h$, $S$ has at most $h$ elements.  It follows from
Lemma~\ref{l:boundD} that each of these elements can be represented using a dag of size
at most $h\len{L} + \len{G_{\cd}}$, so checking whether $\Said(p,\cc) \in S$ can be
done in time $O(h^2\len{L} + h \len{G_{\cd}})$.  If $\cd' \notin S$, then each child of
$v$ has the form $\cd_g\sigma$, where $g= \forall x_1\ldots\forall x_i(\cd_g\imp\cc_g)$
is a grant in $L$ and $\sigma$ is a closed substitution such that $\cc_g\sigma = \cc$.
Since every grant in $L$ is restrained and $n$-restricted, $\cd_g$ mentions at most $n$
variables that are not mentioned in $\cc_g$ and each of these variables is of sort
$\Princ$.  Since $\cd$ and the grants in $L$ do not mention the $\union$ operator and
$\card{P_0} = \len{P_0}$, there are at most $\len{P_0}$ substitutions for each variable
and thus $\len{P_0}^n$ possible substitutions $\sigma$.  Finding $\sigma(x)$ for all of
the variables $x$ that are mentioned in $\cc_g$ takes time linear in the size of the dag
representing $\cc$ (since $\cc_g\sigma = \cc$).  Clearly the dag representing $\cc$ has
size less than that representing $\cd' = \Said(p,\cc)$. By Lemma~\ref{l:boundD}, the
latter dag has size at most $h\len{L}  + \len{G_{\cd}}$.  Since $\card{L}\le \len{L}$,
there are at most $|L|\len{P_0}^n$ children of $v$ and computing what they are takes
time $O(\len{L}\len{P_0}^n + (h\len{L} + \len{G_{\cd}})(h + \len{L}))$.

Similarly, the root of $T$ has at most $ \max(\len{G_d}, \len{L}\len{P_0}^n) $ children since the root
has zero children if $\cd = \true$, less than $\len{G_d}$ children if $\cd$ is a conjunction, and at most
$\len{L}\len{P_0}^n$ children if $\cd$ is a $\Said$ condition.  The children of the root can be computed
in time $O(\len{G_d})$ if $\cd$ is a conjunction and in time $O(\len{L}\len{P_0}^n + \len{G_d}\len{L})$ if
$\cd$ is a $\Said$ condition.  This follows from the reasoning given for the case when the node is neither
the root nor a leaf modified to account for the fact that $\cd \not\in S$, since $S =\emptyset$, and there
is a dag representation of $\cd$ that has length $\len{G_d}$.

To determine the number of non-leaf nodes of $T$, observe that, if the root of $T$ has $n$ children and
each subtree of $T$ has at most $m$ non-leaf nodes, then $T$ has at most $1 +nm$ non-leaf nodes.  It
follows that $T$ has at most $1 +2\max(\len{G_d}, \len{L}\len{P_0}^n) (\len{L}\len{P_0}^n)^{h-2}$ non-leaf
nodes, since a tree with outdegree at most $c$ and height $h$ has $c^h/(c-1) \le 2c^{h-1}$ non-leaf
nodes.  Thus, it takes time
$$
O(\max(\len{G_d}, \len{L}\len{P_0}^n) (\len{L}\len{P_0}^n)^{h-2}
(\len{L}\len{P_0}^n + (h\len{L} + \len{G_{\cd}})(h + \len{L})))
$$
to compute the children of the $\max(\len{G_d},\len{L}\len{P_0}^n) (\len{L}\len{P_0}^n)^{h-2}$
non-leaf nodes other than the root.  Since this time dominates the time to compute the children of the
root, it is also the time required to compute $T$.

Once $T$ is constructed, the truth value of its root can be computed in time linear in the number of
nodes of $T$.  Thus, $\CMetTwo(\cd, L, \emptyset)$ can be computed in time
$$
O(\max(\len{G_d}, \len{L}\len{P_0}^n) (\len{L}\len{P_0}^n)^{h-2}
(\len{L}\len{P_0}^n + (h\len{L} + \len{G_{\cd}})(h + \len{L}))).
$$
\eprf
\end{sloppypar}

\begin{sloppypar}
\lem\label{l:ComputeAuth2}
Suppose that $(\cc, L, R, \scc)$ is a query in $\cL_0\cap\cL_1\cap\cL_2^n\cap\cL_3^h$ such that
$\cc\not\in\scc$ and $D$ is the output of $\XATwo(\cc, L, R, \scc)$.  Then
\begin{itemize}
\item[(a)] $\card{D}$ is at most $\card{P_0}^n(\card{R}+\card{L})$;
\item[(b)] if $\cd$ is a closed condition in $D$, then there is a dag $G_{\cd}$ representing
$\cd$ such that $\len{G_{\cd}}\le \len{R}+\len{L}+\len{\cc}$; and
\item[(c)] $D$ can be computed in time
$
O(\len{L}\len{E\union\{\cc\}} + \len{L}^2\log(\len{R} + 1) + \len{L}^2(\len{L}\len{P_0}^n)^{h+1}h^2).
$
\end{itemize}
\elem
\prf
Let $\EX$ be an execution of $\XProcTwo(\cc, L, R, \scc)$ and let $G  = G(\cc, L, R, \scc, \EX)$.

For part (a), by Theorem~\ref{t:generalizeCorrect1}(b), if $\cc \notin E$, then
\begin{equation}\label{eq1-xrml}
D = \begin{prog}
  \{\cd \mid \mbox{for some grant }\forall x_1\ldots\forall x_m(\cd_g\imp\cc_g)\in G
\mbox{ and closed substitution }\sigma, \\
\cd_g\sigma= \cd\mbox{ and } \cc_g\sigma = \cc\}.\end{prog}
\end{equation}
Since every grant in $G$ is either in $R$ or $L$, $\card{G} \le \card{R}+\card{L}$.  Moreover,
because $(\cc, L, R, \scc)\in \cL_0\cap\cL_2^n$, for every grant
$g = \forall x_1\ldots\forall x_m(\cd_g\imp\cc_g) \in G$, there are at most $n$ variables
mentioned in $\cd_g$ that are not mentioned in $\cc_g$, and each of these variables is of sort
$\Princ$.  As in the proof of Lemma~\ref{complexityestimate}, it follows that there are at most
$\card{P_0}^n$ substitutions of variables in $g$ to closed terms such that $\cc_g\sigma = \cc$
because $(\cc, L, R, \scc)\in \cL_1$.  Part (a) follows immediately.

For part (b), let $\cd$ be a closed condition in $D$.  By (1), $\cd = \cd_g\sigma$, where
$\cd_g$ is the antecedent of a grant $g \in G$ and $\sigma$ is a closed substitution.  By the
proof of part (a), $\sigma$ assigns every variable in $\cd_g$ to a term in $\cc$ or to a
principal in $P_0$.  Given dags $G_{\cc}$ and $G_{\cd_g }$ representing $\cc$ and $\cd_g$,
respectively, we can obtain a dag $G_{\cd}$ representing $\cd$ by replacing every variable in
$G_{\cd_g}$ by either a subgraph of $G_{\cc}$ or by some $p\in P_0$.  So there is a dag
$G_{\cd}$ representing $\cd$ such that $\len{G_{\cd}} \le \len{G_{\cd_g}}+\len{G_{\cc}}$.
Recall that, for every string $s$, there is a dag $G_{s}$ representing $s$ such that
$\len{G_{s}} \le \len{s}$.  So there is a dag $G_{\cd}$ representing $\cd$ such that
$\len{G_{\cd}}\le \len{\cd_g} + \len{\cc}$.  Since $\cd_g$ is the antecedent of a grant in $G$
and every grant in $G$ is a grant in $R$ or $L$, $\len{\cd_g} < \len{R}+\len{L}$, and we are
done.

For part (c), by (\ref{eq1-xrml}), we can compute $D$ by (i) checking whether $\cc \in \scc$;
(ii) computing $G$; and (iii) for each grant
$g = \forall x_1\ldots\forall x_m(\cd_g\imp\cc_g) \in G$, computing
$D_g = \{\cd \mid \mbox{for some closed substitution }\sigma, \cd_g\sigma = \cd\mbox{ and }
\cc_g\sigma = \cc\}$.  (Observe that these are the same steps taken in $\XATwo$; however, our
approach computes $G$ more efficiently.)  Step (i) takes time $O(\len{\scc})$.  We show below
that $G$ can be completed in time
$O(\len{L}^h\len{P_0}^n(2^{h-1} + \len{L}^2\len{P_0}^{n(h-1)})(\len{P_0}^n
+ h^2 + h\len{L}) + \len{L}^2 \log(\len{R}+1) + \len{L}(\len{E} + \len{e})).$
For step (iii), essentially the same arguments as those used in Lemma~\ref{complexityestimate}
show that, given grant $g \in G$, $D_g$ can be computed in time
$O(|\cc| + |\cc_g| + \len{P_0}^n|\cd_g|)$.  So, $\{D_g\mid g\in G\}$ can be computed in time
$O(|G|(|\cc| + \len{P_0}^n))$.  Since $\len{G}\le \len{R}+\len{L}$, the total time needed to
compute $D$ is
$O(\len{E} + \len{L}^h\len{P_0}^n(2^{h-1} + \len{L}^2\len{P_0}^{n(h-1)})(\len{P_0}^n + h^2 +
h\len{L}) + \len{L}^2 \log(\len{R}+1) + \len{L}(\len{E} + \len{e}) + \len{R}(\len{\cc} + \len{P_0}^n)).$

For step (ii), let $A  = A(\cc, L, R, \scc, \EX)$.  For all integers $k \ge 0$, define the set $G'_k$
of grants inductively as follows: $G'_0 = R$ and, for $i>0$,
$G'_i = R\union \{g\mid\mbox{for some principal $p$, } (p,g)\in L \mbox{ and }
\bigwedge_{g'\in G'_{i-1}}\transwithE{g'}{L}{A}{\emptyset}{(E\union\{\cc\})}\rimp
\Permitted(p, \issue, c_g)\mbox{ is acceptably valid}\}$.  We claim that $G'_{\card{L}} = G$.

To show that $G'_{\card{L}} \subseteq G$, we prove by induction that $G'_i\subseteq G$ for all $i\ge 0$.
The base case is immediate because $G'_0 = R$.  For the inductive step, it suffices to show that, if
there is a license $(p,g)\in L$ and a subset $G'\subseteq G$ such that
$\bigwedge_{g'\in G'}\transwithE{g'}{L}{A}{\emptyset}{(E\union\{\cc\})}\rimp\Permitted(p,\issue, c_g)$
is acceptably valid, then $g\in G$.  Let
$\phi = ((\bigwedge_{\ell\in L}\transwithE{\ell}{L}{A}{\emptyset}{(\scc\union\{\cc\})})\land
(\bigwedge_{g\in R}\transwithE{g}{L}{A}{\emptyset}{(\scc\union\{\cc\})}))$.
Because $(p,g)\in L$, it is immediate from the definition of $G$ that $g\in G$ if
$\phi\rimp\Permitted(p, \issue, c_g)$ is acceptably valid.  Because $G'\subseteq G$, every grant
$g'\in G'$ is either in $R$ or there is a principal $p'$ such that $(p', g')\in L$ and
$\phi\rimp\Permitted(p', \issue, c_{g'})$ is acceptably valid.  It follows that
$\phi\rimp\bigwedge_{g'\in G'}\transwithE{g'}{L}{A}{\emptyset}{(E\union\{\cc\})}$ is acceptably valid.
Since
$\bigwedge_{g'\in G'}\transwithE{g'}{L}{A}{\emptyset}{(E\union\{\cc\})}\rimp\Permitted(p,\issue, c_g)$
is acceptably valid, $\phi\rimp\Permitted(p, \issue, c_g)$ is acceptably valid.

To show that $G\subseteq G'_{\card{L}}$, we first observe that, for all $i$, $G'_i \subseteq G'_{i+1}$
and, if $G'_i = G'_{i+1}$, then $G'_i = G'_{i+j}$ for all $j > 0$.  Since $G'_0 = R$ and
$G'_i \subseteq R\union\{g\mid\mbox{for some principal $p$, }(p, g)\in L\}$, it follows that
$G'_{\card{L}} = G'_{\card{L} + 1}$.  To show that $G\subseteq G'_{\card{L}}$, it suffices to show that
for all licenses $(p,g)\in L$ such that $\phi\rimp\Permitted(p, \issue, c_g)$ is acceptably valid,
$g\in G'_{\card{L}}$.  Suppose by way of contradiction that there is a license $(p,g)\in L$ such that
$\phi\rimp\Permitted(p, \issue, c_g)$ is acceptably valid and $g\not\in G'_{\card{L}}$.  Let
$\phi' = \bigwedge_{g'\in G'_{\card{L}}}\transwithE{g'}{L}{A}{\emptyset}{(E\union\{\cc\})}$.  Since
$G'_{\card{L}} = G'_{\card{L}+1}$, the grant $g \not\in G'_{\card{L}+1}$ so, by the definition of
$G'_{\card{L}+1}$, the formula $\phi' \rimp\Permitted(p,\issue, c_g)$ is not acceptably valid.  It
follows that there is an acceptable model $m$ that satisfies $\phi'\land\neg\Permitted(p,\issue, c_g)$
and is ``most forbidding'' in the sense that, for all principals $p'$ and grants $g'$, either $m$ does
not satisfy $\Permitted(p', \issue, c_{g'})$ or the model $m'$ that does not satisfy
$\Permitted(p', \issue, c_{g'})$ and is otherwise identical to $m$ does not satisfy $\phi'$.  Since $m$ satisfies $\neg\Permitted(p,\issue, c_g)$ and $\phi\rimp\Permitted(p, \issue, c_g)$ is acceptably valid,
$m$ does not satisfy $\phi$.  Because $R\subseteq G'_{\card{L}}$ and $m$ satisfies $\phi'$, $m$
satisfies $\bigwedge_{g'\in R}\transwithE{g'}{L}{A}{\emptyset}{(E\union\{\cc\})}$.  So, there is a
license $(p',g')\in L$ such that $m$ does not satisfy
$\transwithE{(p',g')}{L}{A}{\emptyset}{(E\union\{\cc\})}$.  If
$\Permitted(p', \issue, g')\in \scc\union\{\cc\}$, then
$\transwithE{(p', g')}{L}{A}{\emptyset}{(\scc\union\{\cc\})} = \true$,
so $m$ satisfies $\transwithE{(p',g')}{L}{A}{\emptyset}{(E\union\{\cc\})}$.  Thus,
$\Permitted(p', \issue, g')\not\in \scc\union\{\cc\}$.  But then
$\transwithE{(p', g')}{L}{A}{\emptyset}{(\scc\union\{\cc\})} =
\Permitted(p', \issue, c_{g'})\rimp\transwithE{g'}{L}{A}{\emptyset}{(\scc\union\{\cc\})}$.
Since $m$ does not satisfy this formula, $m$ satisfies $\Permitted(p', \issue, c_{g'})$.  By the
construction of $m$, the model $m'$ that does not satisfy $\Permitted(p', \issue, c_{g'})$ and is
otherwise identical to $m$ does not satisfy $\phi'$.  So there is a grant
$g'' = \forall x_1\ldots \forall x_n(\cd_{g''} \imp \cc_{g''})\in G'_{\card{L}}$ such that $m'$
does not satisfy $\transwithE{g''}{L}{A}{\emptyset}{(\scc\union\{\cc\})}$.  Because $m$ satisfies
$\transwithE{g''}{L}{A}{\emptyset}{(\scc\union\{\cc\})}$ and the two models $m$ and $m'$ differ
only in their interpretation of $\Permitted(p', \issue, c_{g'})$, it follows from the translation
of $g''$ that there is a substitution $\sigma$ such that
$\cc_{g''}\sigma = \Permitted(p', \issue, g')$, $\cc_{g''}\sigma\not\in \scc\union\{\cc\}$, and
$\transwithE{\cd_{g''}\sigma}{L}{A}{\emptyset}{(\scc\union\{\cc\})}$ is valid.  So
$\transwithE{g''}{L}{A}{\emptyset}{(\scc\union\{\cc\})}\rimp\Permitted(p',\issue, c_{g'})$ is
acceptably valid. Since $g'' \in G'_{\card{L}}$, $\phi'\rimp\Permitted(p', \issue, c_{g'})$ is
acceptably valid, $g'\in G'_{\card{L}+1}$.  Because $G'_{\card{L}+1} = G'_{\card{L}}$, the grant
$g' \in G'_{\card{L}}$ and, since $m$ satisfies $\phi'$, $m$ satisfies
$\transwithE{g'}{L}{A}{\emptyset}{(\scc\union\{\cc\})}$.  So $m$ satisfies
$\transwithE{(p', g')}{L}{A}{\emptyset}{(\scc\union\{\cc\})}$, which contradicts the assumptions.

We next consider the complexity of computing $G = G_{\card{L}}'$.  Let
$L' = \{(p,g) \in L \mid \Permitted(p, \issue, g)\not\in E\union\{\cc\} \}$.  Clearly, we can
compute $L'$ in time $c_0\len{L}\len{\scc\union \{\cc\}}$ for some constant $c_0$.  For all
$k > 1$, let $L'_k = \{(p,g) \in L' \mid g \notin G'_k\}$ and let $G''_{k} = G'_{k} - G'_{k-1}$.
We plan to compute $G'_k$ inductively,  It will be useful in the induction to represent the
elements of $G'_k$ in a \emph{splay tree}.  (Recall that a splay tree is a form of binary search
tree such that $k$ insertions and searches can be done in a tree with at most $n$ nodes in time
$O(k \log n)$ \cite{ST}.)  If $G'_k$ is represented as a splay tree, then we can compute $L'_k$
in time $O(\len{L} \log{(\len{L} + \len{R}}))$ (since $G_k' \subseteq L \union R$).

For $0 < k < \card{L}$,
\[
\begin{array}{ll}
G''_{k+1} &= \{g\mid\mbox{for some principal $p$, } (p,g)\in L'_k \mbox{ and }\\
&\qquad \bigwedge_{g'\in G''_{k}} \transwithE{g'}{L}{A}{\emptyset}{(E\union\{\cc\})}\rimp
\Permitted(p,\issue, c_g)\mbox{ is acceptably valid}\}.
\end{array}
\]
By Lemma~\ref{l7},
\[
G_{k+1}'' = \union_{(p,g) \in L_k'} \union_{g' \in G_k''}\{g\mid \transwithE{g'}{L}{A}{\emptyset}
{(E\union\{\cc\})}\rimp\Permitted(p,\issue, c_g)\mbox{ is acceptably valid}\}.
\]
Moreover, it follows from Lemma~\ref{l7} that, for $(p,g) \in L'$,
$\transwithE{g'}{L}{A}{\emptyset}{(E\union\{\cc\})}\rimp\Permitted(p,\issue, c_g)$  is acceptably
valid iff the formula $d_{g'} \sigma$ is valid for some $A$-closed substitution $\sigma$ such that
$e_{g'} \sigma = \Permitted(p,\issue,c_g)$, where
$g' = \forall x_1 \ldots \forall x_n (d_{g'} \rimp e_{g'})$.  Given $(p,g) \in L'$ with
$g \notin G_k'$ and $g' \in G_{k}''$, we can clearly check in time $c_1(|e_{g'}| + |(p,g)|)$ if
there exists an $A$-closed substitution $\sigma$ such that $e_{g'}\sigma = \Permitted(p,\issue,g)$,
where $c_1$ is a constant independent of $k$.  If so, as in part (a), there are at most
$\card{P_0}^n$ distinct formulas of the form $d_{g'}\sigma$ (since there are at most $\card{P_0}^n$
possible substitutions for the free variables in $d_{g'}$).  It follows from
Theorem~\ref{t:generalizeCorrect1}(a) that
$\transwithE{\cd_{g'}\sigma}{L}{A}{\emptyset}{(\scc\union\{\cc\})}$ is valid iff
$\CMetTwo(\cd_{g'}\sigma, L, \emptyset) = \true$.  We show shortly that there is an execution of
$\XProcTwo(\cc, L, R, \scc)$ that calls $\CMetTwo(\cd_{g'}\sigma, L, \emptyset)$, so
$\CMetTwo(\cd_{g'}\sigma, L, \emptyset)$ is $h$-bounded.  It follows from
Lemma~\ref{complexityestimate} that we can determine if
$\CMetTwo(\cd_{g'}\sigma, L, \emptyset) = \true$ in time
$
c_2\max(\len{G_{\cd_{g'}\sigma}}, \len{L}\len{P_0}^n) (\len{L}\len{P_0}^n)^{h-2}
(\len{L}\len{P_0}^n + (h\len{L} + \len{G_{\cd_{g'}\sigma}})(h + \len{L})),
$
where $c_2$ is a constant independent of $k$ and $G_{\cd_{g'}\sigma}$ is a dag representing
$\cd_{g'}\sigma$.  As in the proof of part (b), we can obtain $G_{\cd_{g'}\sigma}$ from a dag
$G_{\cd_{g'}}$ representing $\cd_{g'}$ by replacing every variable with a principal in $P_0$ or a
resource mentioned in $\Permitted(p, \issue, g)$.  So there is a dag $G_{\cd_{g'}\sigma}$
representing $\cd_{g'}\sigma$ such that $\len{G_{\cd_{g'}\sigma}} < \len{\cd_{g'}} + \len{g}$.
Repeating this process for each of the at most $\len{P_0}^n$ formulas $\cd_{g'}\sigma$, it follows
that we can check if
$\transwithE{g'}{L}{A}{\emptyset}{(E\union\{\cc\})}\rimp\Permitted(p, \issue, c_g)$  is acceptably
valid in time
$c_2\len{P_0}^n\max(\len{\cd_{g'}}+\len{g}, \len{L}\len{P_0}^n) (\len{L}\len{P_0}^n)^{h-2}
(\len{L}\len{P_0}^n + (h\len{L} + \len{\cd_{g'}}+\len{g})(h + \len{L})).$

Assuming we have already computed $L'_k$ and $G''_k$, we can repeat the process above for all
$g' \in G''_k$ and $(p,g) \in L'_k$.  It is not hard to show that we can compute $G''_{k+1}$ in time
$$
\begin{array}{ll}
&\sum_{g' \in G''_k} \sum_{(p,g) \in L'_k} c_1(\len{\cc_{g'}} + \len{(p,g)}) +
\\& c_2\len{P_0}^n\max(\len{\cd_{g'}}+\len{g}, \len{L}\len{P_0}^n) (\len{L}\len{P_0}^n)^{h-2}
(\len{L}\len{P_0}^n + (h\len{L} + \len{\cd_{g'}}+\len{g})(h + \len{L}))\\
\le &2c_1\len{G''_{k}}\len{L} + c_2\len{P_0}^n(\len{L}\len{P_0}^n)^{h-2}(h + \len{L}) \cdot\\
& \sum_{g' \in G''_k} \sum_{(p,g) \in L} (\len{\cd_{g'}}+\len{g} + \len{L}\len{P_0}^n)
(\len{L}\len{P_0}^n + h\len{L} + \len{\cd_{g'}}+\len{g})\\
\le &2c_1\len{G''_{k}}\len{L} + c_2\len{P_0}^n(\len{L}\len{P_0}^n)^{h-2}(h + \len{L})
2\len{G''_k}\len{L}^2 \len{P_0}^n(\len{L}\len{P_0}^n + h \len{L} + \len{G''_k} + \len{L})\\
\le &2c_1\len{G''_{k}}\len{L} + 2c_2\len{G''_{k}}(\len{L}\len{P_0}^n)^h(h + \len{L})
(\len{L}\len{P_0}^n + h \len{L} + \len{G''_k} + \len{L})\\
\le &c_3 \len{G''_{k}}(\len{L}\len{P_0}^n)^h(h + \len{L})(\len{L}\len{P_0}^n + h \len{L} +
\len{G''_k} + \len{L})
\end{array}
$$
for some constant $c_3$.  We can then build the splay tree for $G_{k+1}'$ by inserting the grants
in $G_k''$ into the splay tree for $G_{k}'$; this can be done in time
$O(\len{G_k''} \log(\len{L} + \len{R}))$.

Since $\union_{k=1}^{\len{L}} G_k'' \subseteq L$, the total time to compute $G_1'', \ldots, G_k''$
(ignoring the time to compute the sets $L'$ and $L'_k$, and to build the splay trees for $G'_k$) is
at most
$$
\begin{array}{ll}
c_4\len{L}^2(\len{L}\len{P_0}^n)^{h+1}h^2
\end{array}
$$
for some constant $c_4$; i.e., it is $O(\len{L}^2(\len{L}\len{P_0}^n)^{h+1}h^2)$.

Now taking into account the complexity of computing $L'$ and $L'_k$ and to build the splay trees, and
using the observation that $\log(a+b) \le \log(a+1) + \log(b+1)$, we get that the complexity for
computing $G$ is
$$
O(\len{L}\len{E\union\{\cc\}} + \len{L}^2\log(\len{R} + 1) + \len{L}^2(\len{L}\len{P_0}^n)^{h+1}h^2).
$$

It remains to show that if $g'= \forall x_1\ldots\forall x_n(\cd_{g'}\imp\cc_{g'})\in G_k' - G_{k-1}'$,
$(p,g)\in L'$ with $g \notin G_k'$, and $\cc_{g'}\sigma = \Permitted(p, \issue, g)$ and $A$-closed
substitution $\sigma$, then there is an execution $\EX$ of $\XProcTwo(\cc, L, R, \scc)$ that calls
$\CMetTwo(\cd_{g'}\sigma, L, \emptyset)$.  Because $\cc\not\in\scc$ by assumption,
$\XProcTwo(\cc, L, R, \scc)$ calls $\XATwo(\cc, L, R, \scc)$, which calls
$\XProcTwo(\Permitted(p, \issue, g), L, R, \scc\union \{\cc\})$, which calls
$\XATwo(\Permitted(p, \issue, gR), L, R, \scc\union \{\cc\})$.  Since $(p, g)\in L'$,
$\Permitted(p, \issue, g)\not\in\scc\union \{\cc\}$.  It follows that
$\XATwo(\Permitted(p, \issue, g), L, R, \scc\union \{\cc\})$ computes
$G(\Permitted(p, \issue, g), L, R, \scc\union \{\cc\}, \EX)$ and, if
$g' \in G(\Permitted(p, \issue, g), L, R, \scc\union \{\cc\}, \EX)$, then
$\XATwo(\Permitted(p, \issue, g), L, R, \scc\union \{\cc\})$ returns a set $D$ that includes
$\cd_{g'}\sigma$.  After $\XATwo(\Permitted(p, \issue, g), L, R, \scc\union \{\cc\})$ returns $D$, it
is easy to see that some execution of $\XProcTwo(\Permitted(p, \issue, g), L, R, \scc\union\{\cc\})$
calls $\CMetTwo(\cd_{g'}\sigma, L, \emptyset)$.  So, in short, it suffices to show that
$g' \in G(\Permitted(p, \issue, g), L, R, \scc\union \{\cc\}, \EX)$.  The proof is by induction on $k$.
If $k = 0$, then $g'\in R\subseteq G(\Permitted(p, \issue, g), L, R, \scc\union \{\cc\}, \EX)$.  If
$k > 0$ then, by the induction hypothesis,
$G'_{k-1} \subseteq G(\Permitted(p, \issue, g), L, R, \scc\union \{\cc\}, \EX)$, so
$
\bigwedge_{\ell\in L}\transwithE{\ell}{L}{A}{\emptyset}{\scc\union \{\cc\union \Permitted(p, \issue, g)\}}
\land \bigwedge_{g'''\in R}\transwithE{g'''}{L}{A}{\emptyset}{\scc\union \{\cc\union \Permitted(p, \issue, g)\}}
\rimp \bigwedge{g''\in G'_{k-1}}\transwithE{g''}{L}{A}{\emptyset}{\scc\union \{\cc\union
\Permitted(p, \issue, g)\}}
$
is acceptably valid.  Since $g'\in G'_k - G'_{k-1}$, there is a grant $g''\in G'_{k-1}$ and a principal
$p'$ such that $(p', g')\in L$ and $\transwithE{g''}{L}{A}{\emptyset}{\scc\union \{\cc\}}\rimp
\Permitted(p', \issue, g')$ is acceptably valid.  Because $g'\in G'_k$ and $g \not\in G'_k$, $g\ne g'$
and, thus, it follows from the translation that
$\transwithE{g''}{L}{A}{\emptyset}{\scc\union \{\cc\union \Permitted(p, \issue, g)\}}\rimp
\Permitted(p', \issue, g')$ is acceptably valid.  Putting the pieces together, there is a principal
$p'$ such that $(p', g')\in L$ and
$\bigwedge_{\ell\in L}\transwithE{\ell}{L}{A}{\emptyset}{\scc\union
\{\cc\union \Permitted(p, \issue, g)\}}
\land \bigwedge_{g'''\in R}\transwithE{g'''}{L}{A}{\emptyset}{\scc\union \{\cc\union
\Permitted(p, \issue, g)\}}\rimp \Permitted(p', \issue, g')$
is acceptably valid, so $g' \in  G(\Permitted(p, \issue, g), L, R, \scc\union \{\cc\}, \EX)$.
\eprf
\end{sloppypar}

We are now ready to prove Theorem~\ref{t:NPHardAll1}.
\othm{t:NPHardAll1}
For fixed $n$ and $h$, if $(\cc, L, R, \scc) \in\cL_0\cap\cL_1\cap\cL_2^{\numP}\cap\cL_3^{\lenC}$,
then determining whether $\XProcTwo(\cc, L, R, \scc)$ returns $\true$ takes time
$O(\len{L}\len{E} + (\len{R} + \len{L})(\len{L}^{h-1}(\len{L} + \len{R}+ \len{\cc})^2))$.
\eothm
\begin{sloppypar}
\prf
Let $D$ be the output of $\XATwo(\cc, L, R, \scc)$.  It is immediate from the description of
$\XProcTwo$ that $\XProcTwo(\cc, L, R, \scc) = \true$ iff there is some condition $\cd\in D$
such that $\CMetTwo(\cd, L, \emptyset) = \true$.  So the output of $\XProcTwo(\cc, L, R, \scc)$
can be determined in time $T + \card{D}T'$, where $T$ is the time needed to compute $D$ and
$T'$ is the time needed to determine the output of $\CMetTwo(\cd, L, \emptyset)$ for a condition
$\cd\in D$.  By Lemma~\ref{l:ComputeAuth2}(c),
$\CMetTwo(\cd, L, \emptyset)$ for a condition $\cd\in D$.  By Lemma~\ref{l:ComputeAuth2}(c),
$
T = c_1(\len{L}\len{E\union\{\cc\}} + \len{L}^2\log(\len{R} + 1) + \len{L}^2(\len{L}\len{P_0}^n)^{h+1}h^2)
$
for some constant $c_1$.  If $n$ and $h$ are treated as constants, then
$T = c'_1(\len{L}\len{E\union\{\cc\}} + \len{L}^2\log(\len{R} + 1) + \len{L}^{h+3})$ for some constant $c'_1$;
i.e., $T$ is $O(\len{L}\len{E\union\{\cc\}} + \len{L}^2\len{R} + \len{L}^{h+3})$.

By Lemma~\ref{l:ComputeAuth2}(a), $\card{D} \le \card{P_0}^n(\card{R}+\card{L})$.
By Lemma~\ref{complexityestimate}, $T'$ is at most
$c_2(\len{G_d}+\len{L}\len{P_0}^n)(\len{L}\len{P_0}^n)^{h-2}(\len{L}\len{P_0}^n + (h\len{L} +\len{G_d})
(h + \len{L}))$, for some constant $c_2$.  If $n$ and $h$ are treated as constants, then there is a
constant $c'_2$ such that $T'$ is at most
$$
\begin{array}{ll}
& c'_2(\len{G_d}+\len{L})\len{L}^{h-2}(\len{L} + (\len{L} +\len{G_d})\len{L})\\
= & c'_2\len{L}^{h-1}(\len{G_d}+\len{L})(1 + (\len{L} +\len{G_d}))\\
= & c'_2\len{L}^{h-1}(\len{G_d}+\len{L})(2(\len{G_d}+\len{L}))\\
\le & 2c'_2\len{L}^{h-1}(\len{G_d}+\len{L})^2.
\end{array}
$$
Since, by Lemma~\ref{l:ComputeAuth2}(b), $\len{G_d} \le \len{R}+\len{L}+\len{\cc}$, it follows that
$T' \le 2c'_2\len{L}^{h-1}(2\len{L} + \len{R} + \len{\cc})^2$,
i.e., $O(\len{L}^{h-1}(\len{L} + \len{R} + \len{\cc})^2)$.

Since $\card{D} \le \card{P_0}^n(\card{R}+\card{L}) \le \len{P_0}^n(\len{R} + \len{L})$, a
straightforward computation shows that $T + \card{D}T'$, the time needed to determine whether
$\XProcTwo(\cc, L, R, \scc)$ returns $\true$, is
$O(\len{L}\len{E} + (\len{R} + \len{L})(\len{L}^{h-1}(\len{L} + \len{R}
+ \len{\cc})^2))$.
\eprf
\end{sloppypar}

\othm{t:ext1}
Let $(\cc, L, R, \scc)$ be a tuple in $\cL_0\cap\cL_1\cap\cL_2^0\cap\cL_3^2$ extended to
include negated $\Said$ conditions and negated conclusions.  The problem of deciding whether
$$
\bigwedge_{\ell\in L} \tranwithE{\ell} \land \bigwedge_{g\in R} \tranwithE{g}\rimp\tranwithE{\cc}
$$
is valid is NP-hard.  This result holds even if $\cc$, all of the licenses in $L$, and all
of the conclusions in $\scc$ are in XrML, all but one of the grants in $R$ is in XrML, and
the one grant that is in in XrML$^\neg$ -- XrML is of the form
$\forall x_1\ldots\forall x_n(\neg \cc)$.
\eothm
\prf
The proof is by reduction of the 3-satisfiability problem.  The reduction is identical to the
reduction given in the proof for the case of $\cL_0\inter\cL_1 \inter \cL_3^2$ in
Theorem~\ref{t:NPHardAll}, except that
$R = \{\forall x_1 \ldots\forall x_m((\bigwedge_{i= 1, \ldots, n}\Said(p_i, \cc'))\imp\cc'), \neg\cc'\}$.
To show that $\XProcTwo(\cc, L, R, \emptyset) = \true$ iff $f$ is valid, we observe that
$\XProcTwo(\cc, L, R, \emptyset) = \true$ iff $L$ and $R$ imply $\false$, which occurs iff
$\bigwedge_{i= 1, \ldots, n}\Said(p_i, \cc')\sigma$ holds for some substitution $\sigma$.  The rest of
the argument proceeds as in the proof of Theorem~\ref{t:NPHardAll}.
\eprf

\bibliographystyle{acmtrans}
\bibliography{vickyw}

\end{document}